\documentclass[preprint]{elsarticle}
\graphicspath{{./img/}}
\DeclareGraphicsExtensions{.pdf,.jpeg}

\usepackage{amsmath}
\usepackage{trfsigns} 
\usepackage{lineno}   
\usepackage{setspace} 
\usepackage[hidelinks=true]{hyperref} 

\begin{document}

\begin{frontmatter}
\journal{Nuclear Instruments and Methods in Physics Research A}
\title{Digital high-pass filter deconvolution by means of an infinite impulse response filter}

\author[HZDR]{P.~F\"odisch\corref{cor1}}
\ead{p.foedisch@hzdr.de}
\cortext[cor1]{Corresponding author}
\author[HZDR,HTW]{J.~Wohsmann}
\author[HZDR]{B.~Lange}
\author[HTW]{J.~Sch\"onherr}
\author[Oncoray,HZDRR,DKFZ]{W.~Enghardt}
\author[HZDR,HTW]{P.~Kaever}

\address[HZDR]{Helmholtz-Zentrum Dresden - Rossendorf, Department of Research Technology,\\Bautzner Landstr. 400, 01328 Dresden, Germany}
\address[HTW]{Dresden University of Applied Sciences, Faculty of Electrical Engineering,\\Friedrich-List-Platz 1, 01069 Dresden}
\address[Oncoray]{OncoRay - National Center for Radiation Research in Oncology, Faculty of Medicine and University Hospital Carl Gustav Carus, Technische Universit\"at Dresden,\\Fetscherstr. 74, PF 41, 01307 Dresden, Germany}
\address[HZDRR]{Helmholtz-Zentrum Dresden - Rossendorf, Institute of Radiooncology,\\Bautzner Landstr. 400, 01328 Dresden, Germany}
\address[DKFZ]{German Cancer Consortium (DKTK) and German Cancer Research Center (DKFZ),\\Im Neuenheimer Feld 280, 69120 Heidelberg, Germany}

\begin{abstract}
In the application of semiconductor detectors, the charge-sensitive amplifier is widely used in front-end electronics. The output signal is shaped by a typical exponential decay. Depending on the feedback network, this type of front-end electronics suffers from the ballistic deficit problem, or an increased rate of pulse pile-ups. Moreover, spectroscopy applications require a correction of the pulse-height, while a shortened pulse-width is desirable for high-throughput applications. For both objectives, digital deconvolution of the exponential decay is convenient. With a general method and the signals of our custom charge-sensitive amplifier for cadmium zinc telluride detectors, we show how the transfer function of an amplifier is adapted to an infinite impulse response (IIR) filter. This paper investigates different design methods for an IIR filter in the discrete-time domain and verifies the obtained filter coefficients with respect to the equivalent continuous-time frequency response. Finally, the exponential decay is shaped to a step-like output signal that is exploited by a forward-looking pulse processing.
\end{abstract}

\end{frontmatter}
\section{Introduction}
\label{sec_intro}
\begin{figure}[ht]
\centering
\includegraphics[width=0.45\textwidth]{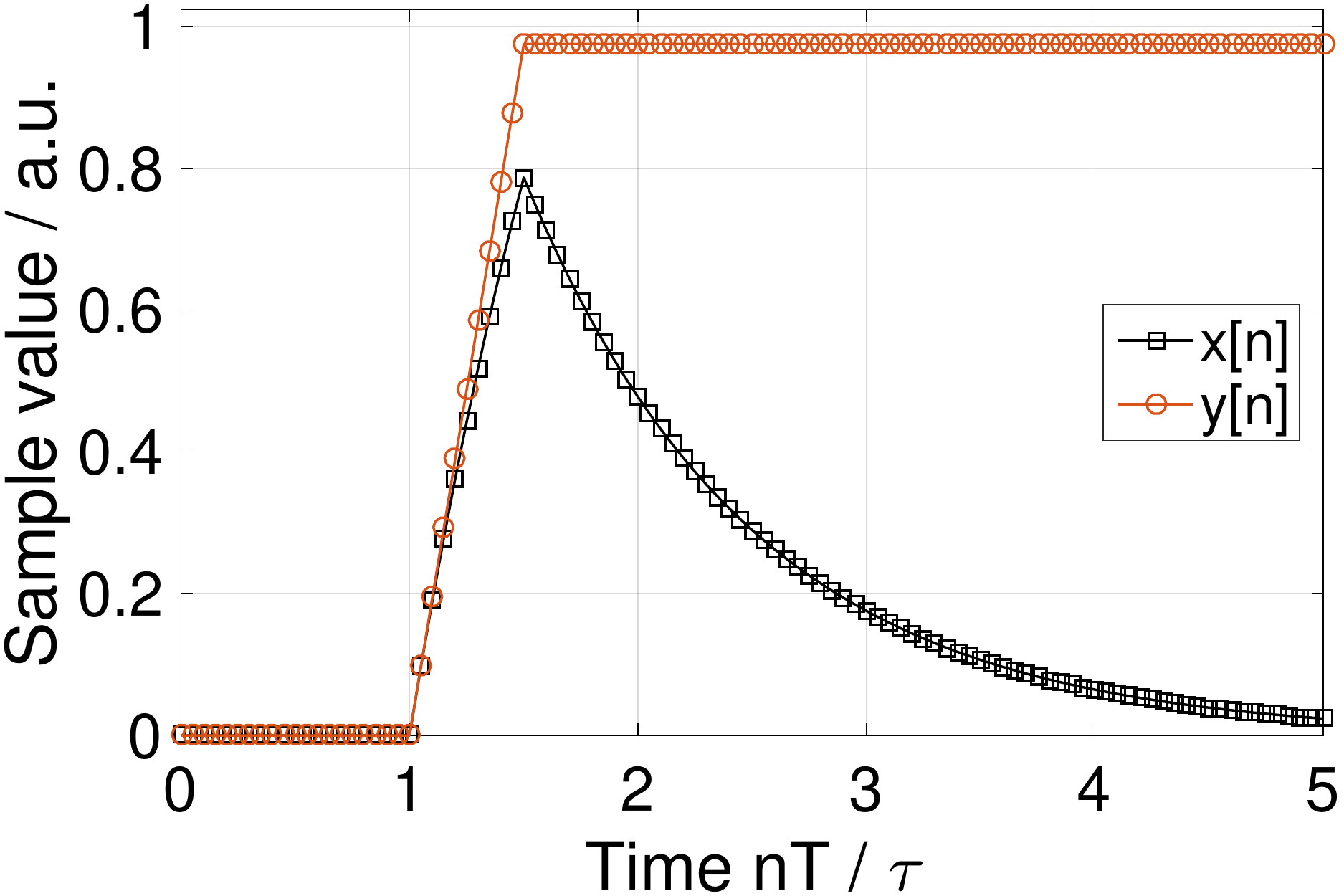}
\caption{A sampled output signal $x[n]$ of an amplifier with a typical exponential decay at sampling interval T (simulated). The decay has the time constant $\tau = 20 T$, and the parameter $k$ of the deconvolution is $k = 1 - {\mathrm{e}}^\frac{-T}{\tau}$. This results in a step-like signal $y[n]$ with a corrected amplitude due to the ballistic deficit.}
\label{fig_expdecay}
\end{figure}

A gamma-ray detector system based on a semiconductor detector such as cadmium zinc telluride (CdZnTe, CZT) usually consists of the detector crystal, the analog readout electronics for the amplification of the detector signal, and a pulse processing unit. Nowadays, the pulse processing is mainly integrated by an application-specific integrated circuit or by a digital circuit in a field-programmable gate array (FPGA). As we recently showed~\protect\cite{foedisch_csa}, the front-end electronics can be appropriately implemented with a charge-sensitive amplifier with a continuous reset through an $RC$ feedback circuit. This type of amplifier discharges the integrated detector current from the feedback capacitor $C$ with the resistor $R$. Thus the typical signal shape with an exponential decay is seen at the output of the charge-sensitive amplifier. A well-known problem of the charge-sensitive amplifier with $RC$ feedback is the ballistic deficit. This is caused by continuous discharge of the feedback capacitor, even though the current of the detector is integrated. If the ratio of the $RC$ time constant over the integration time decreases, the ballistic deficit dominates the measured peak amplitude~\protect\cite{foedisch_csa}. To eliminate this effect and to reconstruct the initial charge by a deconvolution of the exponential decay, Stein et al. presented the Moving Window Deconvolution (MWD)~\protect\cite{stein1}-\protect\cite{stein3}. 
Later, Jordanov et al.~\protect\cite{jordanov1}-\protect\cite{jordanov3} described the same approach~\protect\cite{stein3}. However, both algorithms are derived by an analysis of the exponential decay in the time domain. As a result, a deconvolution of the exponential decay in the discrete-time domain is described by~\protect\cite{stein3}

\begin{equation}
\label{eqn_deconv}
y[n] = x[n] + k \sum_{i=-\infty}^{n-1} {x[i]} \, , 
\end{equation}
where $y[n]$ is the deconvolution of the signal $x[n]$, which is the value of the continuous signal $x(t)$ at the discrete time $t = n T$ with the sampling interval $T$. The value of $k$ is set to $(1-k')$, where $k'={\mathrm{e}}^\frac{-T}{\tau}$ is "the decay constant of the preamplifier transfer function for one sampling interval"~\protect\cite{stein1}. The authors proposed an alternative value $k = \frac{T}{\tau}$ in \cite{stein3} assuming $\tau \gg T$. Both parameters for Eq.~\protect\eqref{eqn_deconv} transform the exponential decay of the signal into a step-like signal, as shown in Fig.~\protect\ref{fig_expdecay}.

The discrete-time signal $x[n]$ shown in Fig.~\protect\ref{fig_expdecay} corresponds to an output signal of an ideal charge-sensitive amplifier with a feedback resistor $R$ and feedback capacitance $C$. For a rectangular shaped input current pulse with amplitude $I$, where the current flows in the time interval from $t_a$ to $t_b$, the continuous-time output signal $x(t)$ of the amplifier is calculated by

\begin{equation}
x(t) = I R \left[ \left( 1 - {\mathrm{e}}^{\frac{t_a - t}{\tau}} \right) {\theta}{(t - t_a)}  - \left( 1 - {\mathrm{e}}^{\frac{t_b - t}{\tau}} \right){\theta}{(t - t_b)} \right] \, .
\end{equation}
Here $\theta(t)$ is the Heaviside step function, and $t_b > t_a > 0$. Regarding Stein's approach for the MWD, the presented deconvolution is calculated by the recursive representation of Eq.~\protect\eqref{eqn_deconv}, which is derived by

\begin{align}
y[n]                                 &=y[n-1] + d\\
x[n]+k \sum_{i=-\infty}^{n-1} {x[i]} &=x[n-1] + k \sum_{i=-\infty}^{n-2} {x[i]} + d\\
d                                    &=x[n]   + \left( k - 1\right) x[n-1]\\
y[n]                                 &=y[n-1] + x[n] + \left( k - 1 \right) x[n-1]\label{eqn_deconv_rec}
\end{align}
According to the time-shifting property of the z-transformation~\protect\cite{oppenheim}

\begin{equation}
\label{eqn_ztime}
x[n-k] \stackrel{\mathcal{Z}}{\longleftrightarrow} z^{-k} X(z) \, ,
\end{equation}
Eq.~\protect\eqref{eqn_deconv_rec} can be rewritten as

\begin{equation}
\label{eqn_iirh}
\frac{Y(z)}{X(z)} = \frac{1 + (k-1)z^{-1}}{1 - z^{-1}} \, .
\end{equation}
It is obvious that Eq.~\protect\eqref{eqn_iirh} is an equivalent of the generalized transfer function of an infinite impulse response (IIR) filter, which is defined as~\protect\cite{oppenheim}

\begin{equation}
\label{eqn_iir_transfer}
H(z) = \frac{\sum_{k=0}^{M} {b_k z^{-k}}}{1 - \sum_{k=1}^{N} {a_k z^{-k}}}
\end{equation}
For further clarification, the fundamental operation of the pulse shapers described by Stein et al.~\protect\cite{stein1} or by Jordanov et al.~\protect\cite{jordanov1} is a deconvolution of the transfer function of the amplifier and can be substituted by an IIR filter. Recently, Jordanov presented an unfolding-synthesis technique \cite{jordanov4} that also demands an accurate deconvolution of the amplifier transfer function. Both constructed their digital algorithms intuitively by an extensive analysis of the signals in the time domain. By doing so, they neglected the established design methods for digital filters. Consequently, we will show a further analysis of the amplifier transfer function in the $s$-domain (frequency domain of continuous-time signals) and design the corresponding digital filter for the deconvolution in the $z$-domain (equivalent frequency domain of discrete-time signals). Finally, we will verify our algorithms with the signals of a CZT detector in conjunction with a charge-sensitive amplifier.

\section{Discrete-time inverse amplifier transfer function}
The charge-to-voltage transfer function of an ideal charge-sensitive amplifier with an $RC$ feedback network and the voltage $v_\mathrm{O}$ at its output is given by~\protect\cite{foedisch_csa}

\begin{equation}
H(s) = \frac{v_\mathrm{O}}{Q} = \frac{s R}{1 + s R C} \, .
\end{equation}
By normalizing the charge $Q$ to the feedback capacitance $C$ with $\frac{Q}{C} = v_\mathrm{Q}$, the transfer function $H_\mathrm{Q}$ becomes

\begin{equation}
\label{eqn_hq}
H_\mathrm{Q} \left( s \right) = \frac{v_\mathrm{O}}{v_\mathrm{Q}} = \frac{s \tau}{1 + s \tau} \, ,
\end{equation}
where $\tau=R C$ is the characteristic time constant of the charge-sensitive amplifier. The transfer function is identical to that of a first-order high-pass filter. Therefore, the signal seen at the output of the amplifier is a convolution of the charge input signal and a high-pass filter. Moreover, as we want to reconstruct the input signal, the deconvolution of the high-pass filter is realized with the inverse transfer function of the amplifier. A deconvolution in the discrete-time domain requires an adequate approximation of ${H_\mathrm{Q}}^{-1}$  in the $z$-domain. Because the design methods for discrete-time filters that transform continuous-time filters are numerous, we will focus our investigations on a set of established methods and will test the accuracy of the transformation from the $s$-plane to the $z$-plane.
At first, with the corresponding Laplace-transformation of the difference quotient of the continuous-time signal $x(t)$ 

\begin{equation}
s X \Laplace \frac{dx(t)}{dt} = \lim_{h \to 0}{\frac{x(t+h) - x(t)}{h}} \, ,
\end{equation}
the equivalent difference quotient for the discrete-time signal $x(nT)$ is

\begin{align}
\lim_{h \to T}{\frac{x(nT+h) - x(nT)}{h}} &=\frac{x(nT+T) -x(nT)}{T}\nonumber\\
                                          &=\frac{x[n+1] - x[n]}{T}\label{eqn_forwarddiffn}
\end{align}
By using Eqs.~\protect\eqref{eqn_ztime} and \protect\eqref{eqn_forwarddiffn}, the transformation of the $s$-domain to the $z$-domain is therefore derived by

\begin{equation}
\label{eqn_forwarddiff}
s \longrightarrow \frac{z-1}{T} \, ,
\end{equation}
which is referred to as the forward difference method. In the same way, but by setting the difference quotient to ${\frac{x(t)-x(t-h)}{h}}$, the corresponding backward difference is defined by

\begin{equation}
\label{eqn_backwarddiff}
s \longrightarrow \frac{z-1}{z T}
\end{equation}
It is clear that these design methods replace the continuous-time differentials with a discrete-time difference. The exact relation of $s$ and $z$ in the context of the z-transformation is given by

\begin{equation}
\label{eqn_zs}
z = {\mathrm{e}}^{sT} \Longleftrightarrow s = \frac{1}{T}\mathrm{ln}\left( z \right)\, ,
\end{equation}
which cannot be used for the expression of a discrete series of samples with respect to Eq.~\protect\eqref{eqn_ztime}. Therefore, another substitution of $s$ is derived by solving the differential equation corresponding to $H(s)$ by the approximation of an integral with the trapezoidal rule~\protect\cite{oppenheim}, \protect\cite{proakis}. A replacement of $s$ with

\begin{equation}
\label{eqn_blt}
s \longrightarrow \frac{2}{T}\frac{z-1}{z+1} \, ,
\end{equation}
is referred to as a bilinear transformation. Besides the approximation of $s$ with a suitable expression in terms of $z$, there exist further design techniques based on the mapping of the zeros and poles of the transfer function in the $s$-plane directly into zeros and poles in the $z$-plane \cite{proakis}. The matched-z transformation, as described in~\protect\cite{proakis}, maps the zeros $z_k$ and poles $p_k$ of the continuous-time transfer function $H(s)$ to the discrete-time transfer function $H(z)$ with the relation

\begin{equation}
\label{eqn_matchedz}
H(s) = \frac{\displaystyle \prod_{k=1}^{M} {s - z_k}}{\displaystyle \prod_{k=1}^{N} {s - p_k}} \longrightarrow \frac{\displaystyle \prod_{k=1}^{M} {z - {\mathrm{e}}^{z_k T}}}{\displaystyle \prod_{k=1}^{N} {z - {\mathrm{e}}^{p_k T}}} = H( z )\, , 
\end{equation}
where $T$ is the sampling interval.
After all of the above, the transfer functions in the $z$-domain used to deconvolute the continuous-time amplifier transfer function  $H_\mathrm{Q}$ of Eq.~\protect\eqref{eqn_hq} are summarized in tab.~\protect\ref{tab_transforms} (detailed results are illustrated in ~\protect\ref{sec_app_iir_comparison}).

\begin{figure}[ht]
\centering
\includegraphics[width=0.45\textwidth]{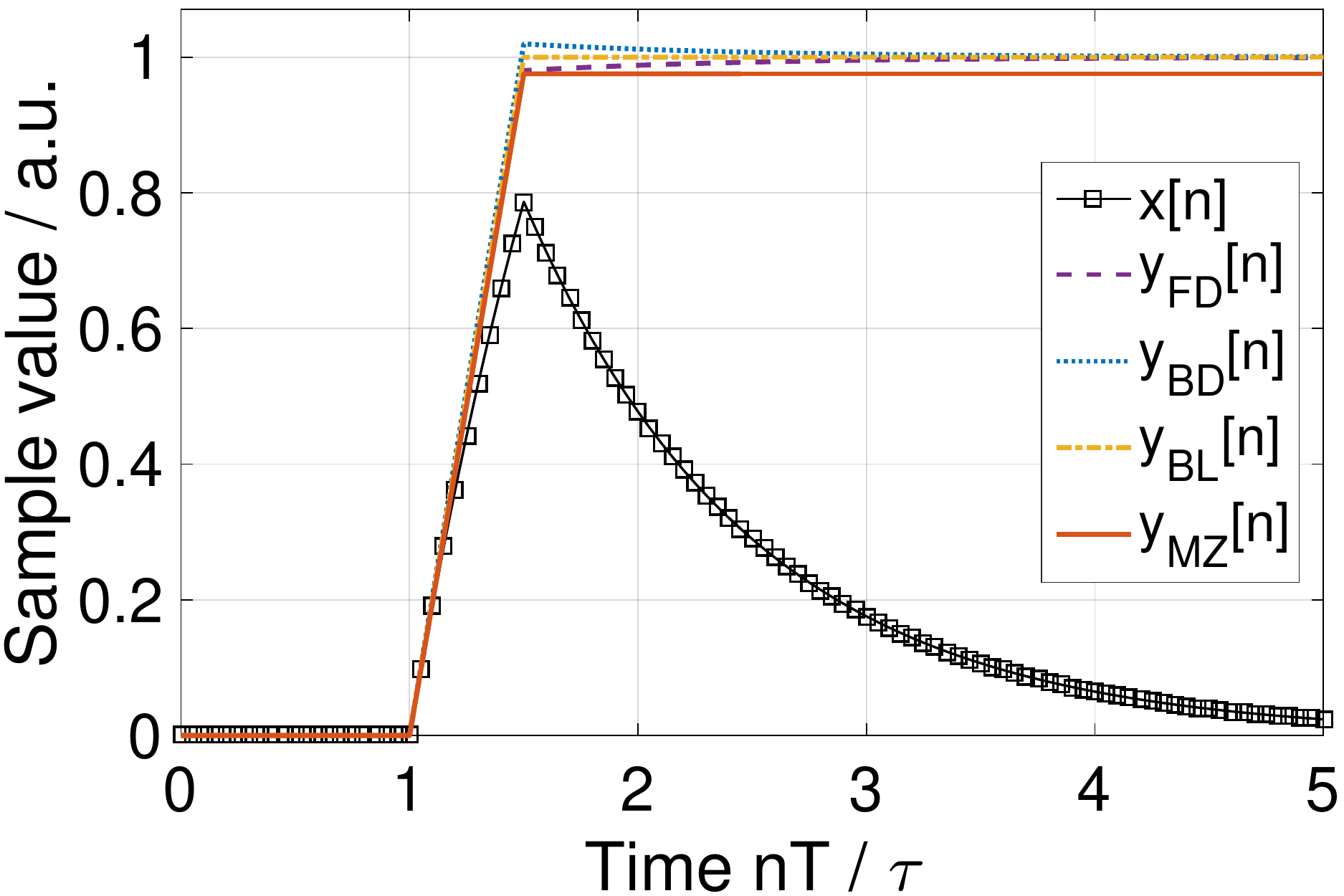}
\caption{A comparison of the results obtained by different transfer functions for the deconvolution of the signal $x[n]$. The filter based on the forward difference method ($y_\mathrm{FD}$) produces a little undershoot, whereas the backward difference ($y_\mathrm{BD}$) method shows an overshoot in the analogy. Further, the filters based on the bilinear ($y_\mathrm{BL}$) or matched-z ($y_\mathrm{MZ}$) transformations seem to have a flat response with respect to the exponential decay of the input signal.}
\label{fig_expdecay_comp}
\end{figure}

\begin{table}[ht]
\centering
\caption{Summary of the investigated design methods for an infinite impulse response filter. All methods rely on an approximation of the continuous-time transfer function in the $z$-domain.}
\label{tab_transforms}
\begin{tabular}{|l|l|}
\hline
{Design method} & {Transfer function}\\
\hline
{Forward diff. (Eq.~\eqref{eqn_forwarddiff})} & {$\displaystyle H_\mathrm{FD} (z) = \frac{1 + \left( \frac{T}{\tau} - 1 \right) {z}^{-1}}{1 - {z}^{-1}}$}\\
\hline
{Backward diff. (Eq.~\eqref{eqn_backwarddiff})} & {$\displaystyle H_\mathrm{BD} (z) = \frac{\left( \frac{T}{\tau} + 1 \right) - {z}^{-1}}{1 - {z}^{-1}}$}\\
\hline
{Bilinear transf. (Eq.~\eqref{eqn_blt})} & {$\displaystyle H_\mathrm{BL} (z) = \frac{\left( \frac{T}{2 \tau} + 1 \right) + \left( \frac{T}{2 \tau} - 1 \right){z}^{-1}}{1 - {z}^{-1}}$}\\
\hline
{Matched-z transf. (Eq.~\eqref{eqn_matchedz})} & {$\displaystyle H_\mathrm{MZ} (z) = \frac{1 - {\mathrm{e}}^{-\frac{T}{\tau}}{z}^{-1}}{1 - {z}^{-1}}$}\\
\hline
\end{tabular}
\end{table}
The transfer functions shown in tab.~\protect\ref{tab_transforms} are all identical in the limit $T \ll \tau$. But with real constraints, where $\tau$ is chosen to be as small as possible to meet the requirements of the application, and $T$ is chosen to be as large as possible in the range of the Nyquist frequency, the results of the presented IIR filters are slightly different. After an analysis of the filter responses in the time-domain with the example shown from Fig.~\protect\ref{fig_expdecay}, it is apparent that the bilinear transformation performs best for the specified values, as the voltage step is expected to have a unity value without the presence of the ballistic deficit effect (Fig.~\protect\ref{fig_expdecay_comp}).

As the methods based on the difference quotient obviously result in distorted output signals, these approaches will be neglected for our application. Moreover, the results of the matched-z transformation are improved if this method is extended by an additional matched gain. The discrete-time transfer function $H_\mathrm{MZ}$ is adjusted by the gain correction constant $G$, so that

\begin{equation}
G = \frac{\left| H(\mathrm{j} \omega) \right|}{\left| H_\mathrm{MZ}({\mathrm{e}}^{\mathrm{j}  \omega T}) \right|}
\end{equation}
matches the ratio of the magnitudes of the frequency responses at a specific frequency $\omega$. For the best gain matching of the inverse high-pass filter regarding the transfer function in the $s$-domain, $\omega$ is set to the characteristic $3\,\mathrm{dB}$ corner frequency $\frac{1}{\tau}$, as carried out by a comparison of the different values for $\omega$. This is shown in~\protect\ref{sec_app_gainT} and~\protect\ref{sec_app_gainTAU}. Consequently, the optimum IIR filter coefficients for the ideal inverse amplifier transfer function are

\begin{align}
b_n &=\left\{G\left(\frac{1}{\tau}\right), G\left(\frac{1}{\tau}\right){\mathrm{e}}^{\frac{- \tau}{T}} \right\}\\
a_n &=\left\{1, -1\right\}
\end{align}
where the specific gain matching constant $G(\omega)$ for the described system is calculated as

\begin{equation}
G(\omega) = \frac{2 \left| {\mathrm{sin}}{\left( \frac{\omega T}{2}\right)} \right| \sqrt{{\left(\omega T\right)}^{2} + 1}}{\omega \tau \sqrt{{\mathrm{e}}^{\frac{-2 T}{\tau}} - 2 {\mathrm{e}}^{\frac{-T}{\tau}} {\mathrm{cos}}{\left( \omega T \right)} + 1}}
\end{equation}

\section{Application to measured signals}
\begin{figure}[ht]
\centering
\includegraphics[width=0.50\textwidth]{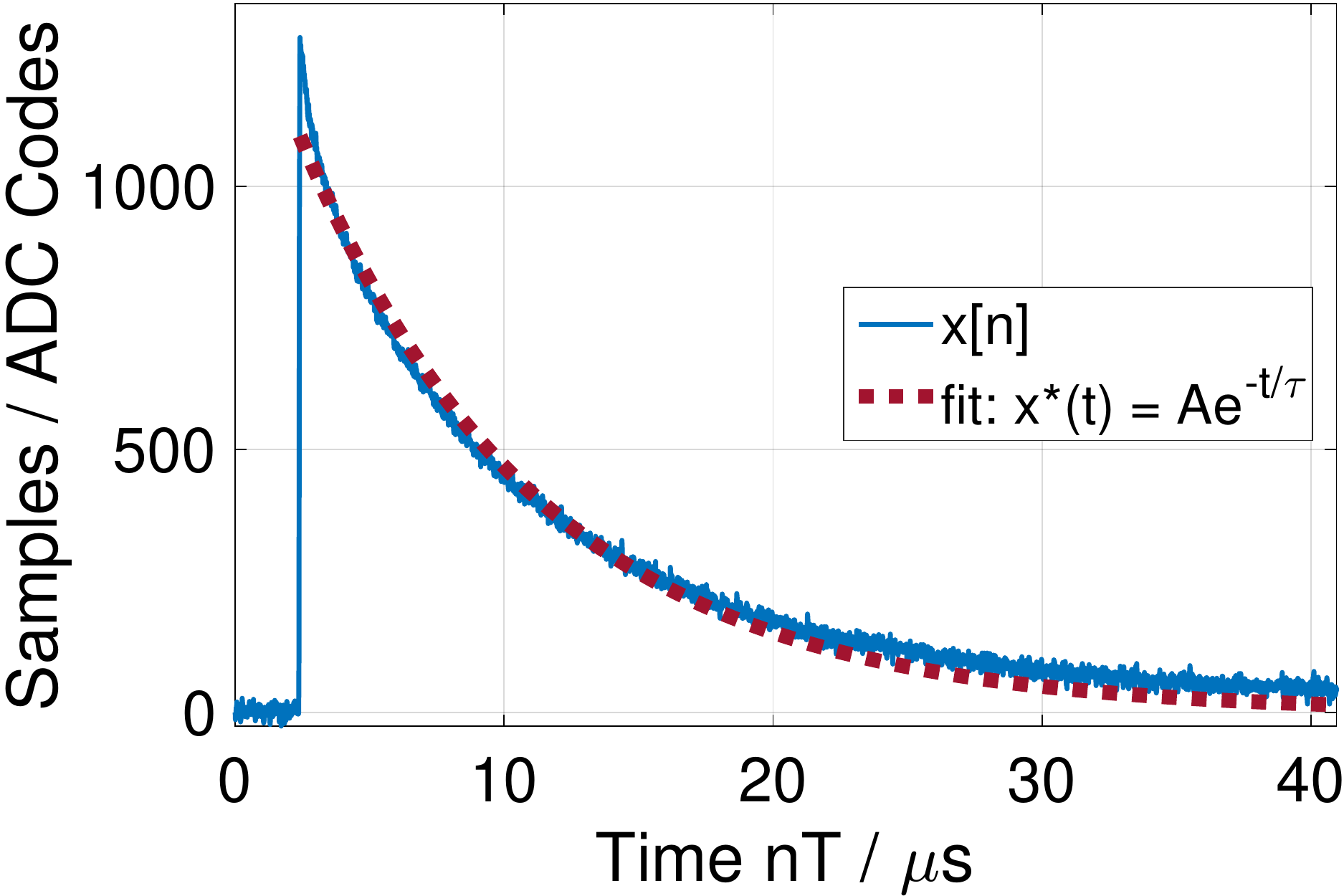}
\caption{A sampled output signal $x[n]$ from the charge-sensitive amplifier with a test pulse at its input. The exponential decay of the signal is clearly visible, but the curve fitting of the function ${x}^{*}(t)=A{\mathrm{e}}^{\frac{-t}{\tau}}$ exposes a variation from the ideal shape.}
\label{fig_testpulser}
\end{figure}

For a proof of concept, we will examine the deconvolution of the amplifier signals coming from our charge-sensitive amplifiers. These electronics are designed to operate with a CZT pixel detector~\protect\cite{foedisch_csa}. Further, the amplifiers are equipped with a test pulse input so that pulses with an exponential decay and arbitrary pulse heights can be synthesized. The time constant of the amplifier is in the range of several microseconds, as the feedback resistor is chosen to be $82\,\mathrm{M\Omega}$ and the feedback capacitance is in the range of $100\,\mathrm{fF}$. The exact value of $\tau = R C$ must be experimentally verified, because the electronic components are afflicted with tolerances. An example of a measured pulse from the charge-sensitive amplifier is shown in Fig.~\protect\ref{fig_testpulser}.

\begin{figure}[ht]
\centering
\includegraphics[width=0.50\textwidth]{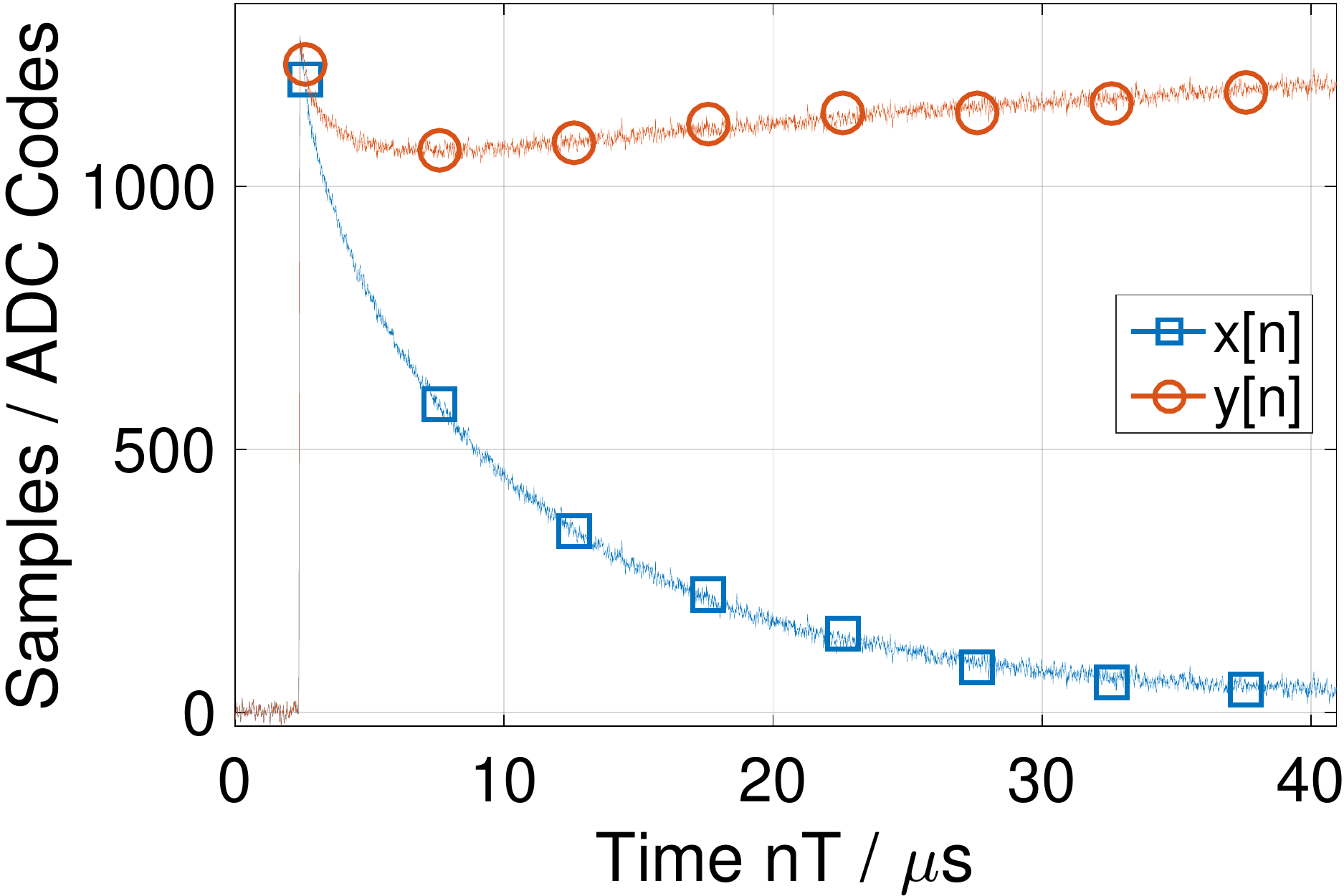}
\caption{The sampled signal $x[n]$ and a deconvolution with the matched-z IIR filter with gain correction. The output of the filter $y[n]$ traces the variations with respect to the ideal exponential decay. The anticipated flat step-like response is distorted.}
\label{fig_testpulse_deconv}
\end{figure}

The values are recorded with a sampling frequency of $100\,\mathrm{MSPS}$ with $14\,\mathrm{bit}$ resolution at an input voltage range of approximately $2.3\,\mathrm{V}$. A curve fit is applied to estimate the time constant $\tau$. For this example, the numerical value of $\tau$ is $8.91\,\mathrm{{\mu}s}$, which corresponds to a $82\,\mathrm{M\Omega}$ resistor in parallel with a $108.7\,\mathrm{fF}$ capacitor. The derived curve fit is in accordance with the expected feedback network, but obviously it does not cover all the features of the pulse shape. Further, a full analysis of the time constant by the best fit function indicates that the parameter $\tau$ is nearly constant over the whole output voltage range. Yet at the same time, a deconvolution of the exponential decay with the estimated $\tau$ by the matched-z IIR filter with gain correction reveals distortion with regard to the predicted flat step-like function. The results of the deconvolution of a measured signal with the inverse high-pass filter are shown in Fig.~\protect\ref{fig_testpulse_deconv}.

\begin{figure}[ht]
\centering
\includegraphics[width=0.50\textwidth]{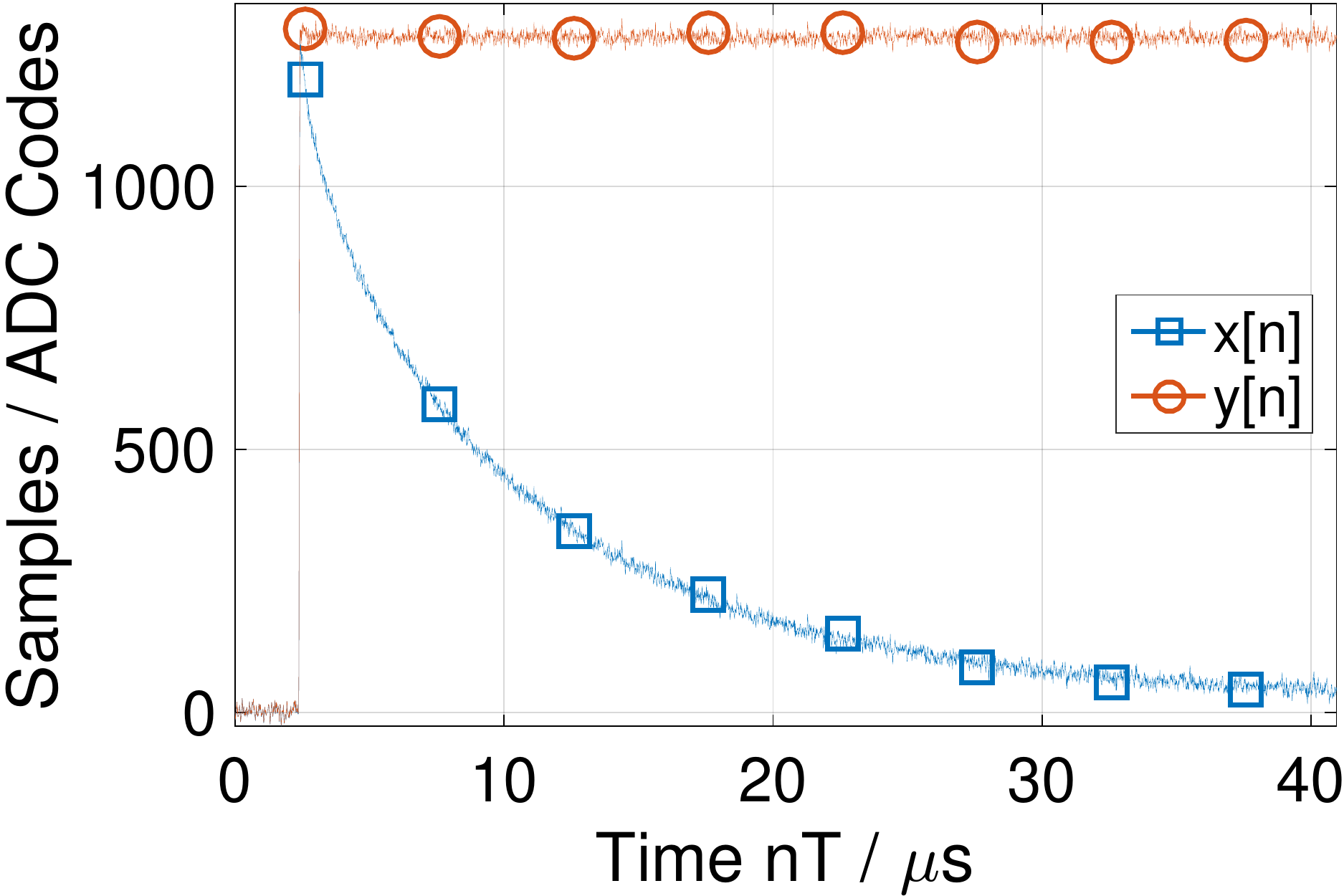}
\caption{The input $x[n]$ and output $y[n]$ of the third-order IIR filter for the deconvolution of the exponential decay with the coefficients quoted in Eqs.~\protect\eqref{eqn_iir_bz} and \protect\eqref{eqn_iir_az}. The output has the expected step-like shape with a flat top.}
\label{fig_testpulse_deconv_advanced}
\end{figure}

In terms of the proposed methods for the deconvolution, the transfer function of the charge-sensitive amplifier cannot be approximated with the inferred model of a first-order high-pass filter from Eq.~\protect\eqref{eqn_hq}. However, the costs of a detailed analysis of all parasitic components of the entire network model are much higher than the benefits. Nevertheless, the transfer function of the charge-sensitive amplifier has been empirically found by the use of the System Identification Toolbox provided by Mathworks~\protect\cite{mathworks}. The tool reports the goodness of fit between the test and reference data to be $93.09\,\%$ (normalized root mean square error) for a continuous-time model with one pole and one zero for the transfer function. This is improved to $96.48\,\%$ by a model with four zeros and four poles. Anticipating the digital implementation of the IIR filter, we chose a model with an accuracy of $96.47\,\%$ and three zeros and three poles (see \protect\ref{sec_app_matlab_fit}). Furthermore, the parameters of the inverse continuous-time transfer function are transformed by the matched-z method and are gain-corrected at the $3\,\mathrm{dB}$ corner frequency. This results in an IIR filter with the coefficients quoted in Eqs.~\protect\eqref{eqn_iir_bz} and \protect\eqref{eqn_iir_az}. The output $y[n]$ from the IIR filter with these coefficients, calculated by a double-precision floating point arithmetic ($64\,\mathrm{bit}$), is shown in Fig.~\protect\ref{fig_testpulse_deconv_advanced}.

In addition to the analysis in the discrete-time domain, the characteristics of the derived IIR filter are shown in \protect\ref{sec_app_iir_three}.

\section{Implementation of a higher order IIR filter}
\begin{figure}[ht]
\centering
\includegraphics[width=0.75\textwidth]{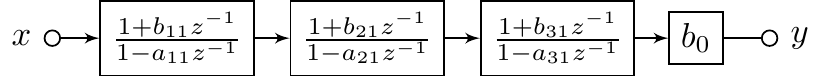}
\caption{A Cascade Form infinite impulse response (IIR) filter of the third-order. Each block is built by a first-order Direct Form IIR filter.}
\label{fig_iir_cascade}
\end{figure}

Nowadays, digital filters are realized with general purpose digital signal processors (DSP) or field-programmable gate arrays (FPGA). We will concentrate our efforts on an implementation of a higher order IIR filter based on an FPGA from Xilinx. The coefficients of the IIR filter are given by Eqs.~\protect\eqref{eqn_iir_bz} and \protect\eqref{eqn_iir_az}. Along with the transfer function described by Eq.~\protect\eqref{eqn_iir_transfer}, the difference equation for the third-order IIR is

\begin{align}
y[n] =  b_0 x[n] + &b_1 x[n-1] + b_2 x[n-2] + b_3 x[n-3] + ...\nonumber\\
                   &a_1 y[n-1] + a_2 y[n-2] + a_3 y[n-3]\label{eqn_diff_iir_three}
\end{align}
The difference equation can be comfortably mapped to the known Direct Form implementation structures of an IIR filter. Still, at the same time, the coefficients need quantization. With a look at the resources provided by a state-of-the-art FPGA from Xilinx, a word length for the coefficients of $25\,\mathrm{bits}$ is easily realizable, since these devices have embedded $18\,\mathrm{bit} \times 25\,\mathrm{bit}$ multipliers \cite{xilinx_dsp}. These come along with a pre-adder and a post-adder in a so-called DSP slice. The built-in logic blocks, with their multiply-accumulate structures, are highly optimized for the requirements of digital filters. But a straightforward implementation of a digital filter with restricted word lengths for the coefficients must be evaluated with regard to quantization errors. We examined the output of the Direct Form IIR filter in the time domain with a variable word length for the coefficient quantization. Even a length of $25\,\mathrm{bits}$ results in clear distortions. Satisfactory results were achieved only with a word length of more than $34\,\mathrm{bits}$. With this implementation on a Xilinx FPGA, one multiplication is mapped to two DSP slices. Thus a Direct Form implementation based on the difference equation (Eq.~\protect\eqref{eqn_diff_iir_three}) requires 14 DSP slices. As we target the processing of highly segmented detectors with more than 65 readout channels, the impact on the quantization error is noticeable in an excessive consumption of slice logic resources.
As already stated by Oppenheim, the Direct Form realization of an IIR filter is very sensitive to quantization errors, because each polynomial root of the transfer function is affected by all of the quantization errors of the coefficients~\protect\cite{oppenheim}. On the whole, the quantization modifies the location of the zeros and poles in the complex $z$-plane, and therefore the response of the filter is changed. This problem is solved by a factorization of the numerator and denominator polynomials of the transfer function in the form

\begin{equation}
H(z) = b_0 \prod_{k=1}^{M}{\frac{1-c_k{z}^{-1}}{1-d_k{z}^{-1}}} \,
\end{equation}
where $M$ is the order of the IIR filter and $c_k$ and $d_k$ are the zeros and poles in the complex $z$-plane. For our derived filter, all zeros and poles are real values where the imaginary part is zero. Thus, the transfer function can be rewritten to a cascade of three Direct Form filters of first-order (Fig.~\protect\ref{fig_iir_cascade}).

The coefficients $-b_{k1}$ and $a_{k1}$ are the roots of the numerator and denominator polynomials and must be quantized. As proven by~\protect{\cite{crochiere} and \protect{\cite{oppenheim}, the Cascade~Form of an IIR filter is generally much less sensitive to coefficient quantization than the equivalent Direct Form implementation. Each first-order filter section of the cascade structure has the form shown in Fig.~\protect\ref{fig_iir_structure}.

\begin{figure}[ht]
\centering
\includegraphics[width=0.75\textwidth]{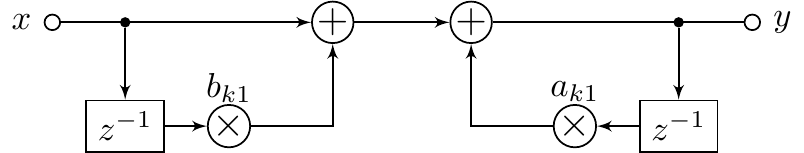}
\caption{A Direct Form implementation of a first-order infinite impulse response filter. The multiply-accumulate architecture from the built-in slices of the field-programmable gate array is adapted to this basic filter structure.}
\label{fig_iir_structure}
\end{figure}
Both the feedforward path (all-zero system) and the feedback path (all-pole system) purely match the multiply-accumulator architecture built into the FPGA. An additional register between the two paths increases the performance in terms of the maximum clock frequency, but also introduces a delay of one clock cycle. 

\section{Results}
We have implemented the presented Cascade Form structure of the third-order IIR filter in VHDL. The design was synthesized and tested with Xilinx FPGAs. Moreover, the tools reported a theoretical maximum frequency of $217\,\mathrm{MHz}$ for the filter running on a Kintex~7 XC7K325T. With the optional register between the feedforward path and feedback path, this value increases to $295\,\mathrm{MHz}$ ($167\,\mathrm{MHz}$ on Spartan~6 XC6SLX45T). For our purposes and for most detector applications, this is sufficient. The fixed-point arithmetic uses the entire bit widths of the built-in $25\,\mathrm{bit} \times 18\,\mathrm{bit}$ multipliers. Even though the fixed-point representation of the poles and zeros is sufficient with a word length of $18\,\mathrm{bits}$, the remaining $25\,\mathrm{bits}$ word length for the sample is absolutely required. As the output of a cascade and the word length of the feedback signal are limited to $25\,\mathrm{bit}$, this truncation propagates a rounding error from each stage to the next. However, the generic VHDL design of the filter is in good agreement with our requirements while running on the Kintex~7 architecture, but on the outdated $18\,\mathrm{bit} \times 18\,\mathrm{bit}$ architecture (e.g.\ Spartan~6), we observed unacceptable distortions due to rounding errors. The design of our third-order IIR filter is mapped to seven DSP slices of the Xilinx FPGA. Each Direct Form module consumes two DSP slices, while the multiplication with the gain correction constant $b_0$ consumes an additional multiplier of a DSP slice.

The step-like output of the IIR filter is turned into a trapezoidal pulse shape by applying a differentiator of the form

\begin{equation}
\label{eqn_fir_diff}
y[n] = x[n] - x[n-M]\,, M\geq1\, ,
\end{equation}
where $M$ is the window length and an average filter with

\begin{equation}
\label{eqn_iir_average}
y[n] = \frac{1}{N}\sum_{k=0}^{N-1}{x[n-k]} \,,
\end{equation}
where $N$ is the number of samples. The implementation of the Eq.~\protect\eqref{eqn_fir_diff} is a straightforward finite impulse response (FIR) filter. Further, the corresponding difference equation of Eq.~\protect\eqref{eqn_iir_average} is another direct form implementation of an IIR filter. Together, the filters implement the MWD algorithm. The results for a recorded sequence with our charge-sensitive amplifier and a CZT detector are shown in Fig.~\protect\ref{fig_results}. This confirms the results of Stein et al. (\cite{stein1}-\cite{stein3}) and of Jordanov et al. (\cite{jordanov1}-\cite{jordanov3}}).

\begin{figure}[ht]
\centering
\includegraphics[width=0.5\textwidth]{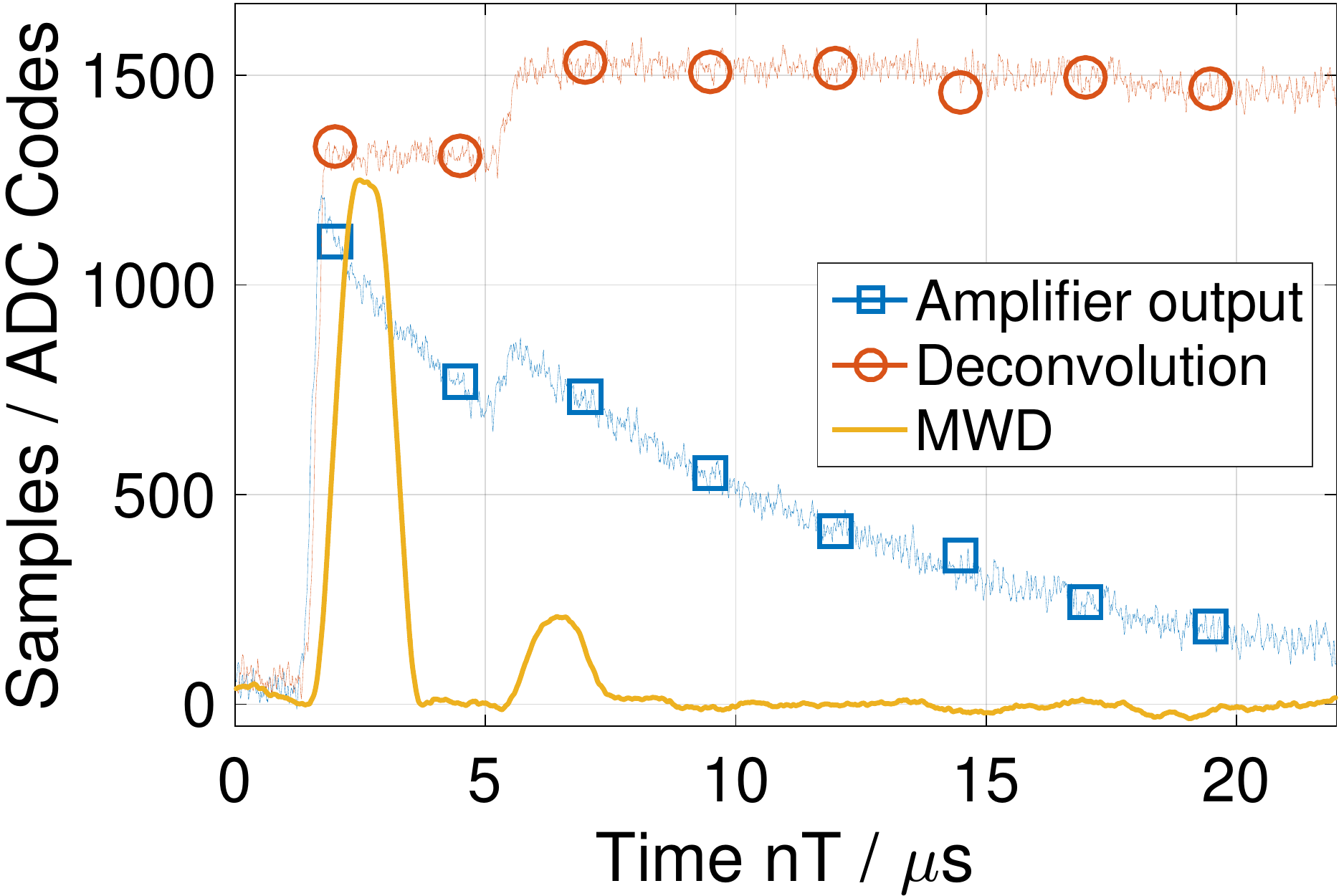}
\caption{An example of the deconvolution with the obtained inverse amplifier transfer function. Advanced pulse processing (MWD) with a differentiator ($M=128$) and an average filter ($N=64$) accomplishes the pulse shaping with respect to the well-known trapezoidal shaper from~\protect\cite{stein1} or \protect\cite{jordanov1}.}
\label{fig_results}
\end{figure}
This example shows that the obtained transfer function of the charge-sensitive amplifier is sufficiently robust to be applied to an identical amplifier because the recorded sequence was captured from an arbitrarily selected pixel with a dedicated amplifier. A single pixel calibration further improves the results, since the absolute values of the electronic components vary.

\section{Summary}
Based on the fundamental description from Stein et al.~\protect\cite{stein1} of the Moving Window Deconvolution, we investigated the deconvolution of an exponential decay related to an output signal of a charge-sensitive amplifier. We showed that the approach of Stein et al. for signal processing relies on an inversion of a first-order high-pass filter representing the amplifier transfer function. Thus the deconvolution can be realized with an infinite impulse response (IIR) filter. In the context of established design methods for digital filters, we derived the optimum coefficients for the demanded IIR filter. In conclusion, the matched-z transformation with a gain correction at the $3\,\mathrm{dB}$ corner frequency performed best while regarding the output in the time and frequency domain. Moreover, we examined the signal processing with measured signals from a charge-sensitive amplifier. We revised the apparent model of a first-order high-pass filter for the transfer function of the amplifier because it fails in terms of accuracy. After an estimation of the transfer function by experimental data, we enhanced this model to the third-order. Finally, we implemented the inverse amplifier transfer function by means of an IIR filter with an adapted Cascade Form structure. The results showed, that the proposed method for the deconvolution is realizable with little computational effort in a Xilinx Kintex~7 FPGA. The presented method and implementation of the signal processing are essential for our advanced pulse processing of the CZT detector signals with high resolution and high throughput.

\clearpage
\singlespacing
\appendix
\section{Continuous-time inverse amplifier transfer function}
\subsection{First-order model}
The transfer function of the amplifier is modeled with a first-order high-pass filter. Therefore, the inversion of this transfer function (Eq.~\protect\eqref{eqn_hq}) is used for the deconvolution of the exponential decay. The characteristics in the continuous-time frequency domain ($s$-domain) are illustrated in Fig.~\protect\ref{fig_app_sdomain_mag} and ~\protect\ref{fig_app_sdomain_phase}.

\begin{figure}[h]
\centering
\includegraphics[width=0.40\textwidth]{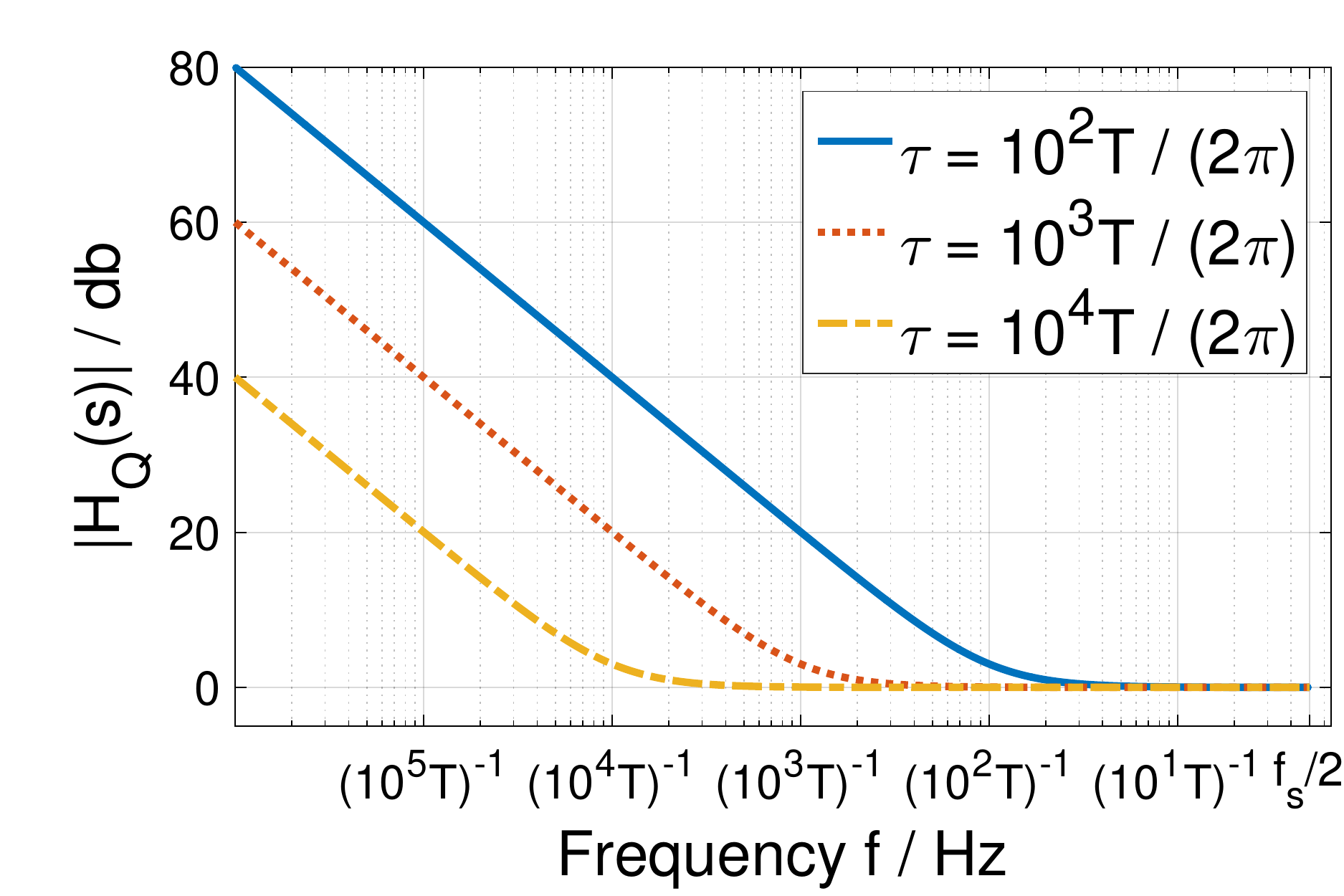}
\caption{Magnitude of the frequency response of the inverse amplifier transfer function in the $s$-domain.}
\label{fig_app_sdomain_mag}
\end{figure}

\begin{figure}[h]
\centering
\includegraphics[width=0.40\textwidth]{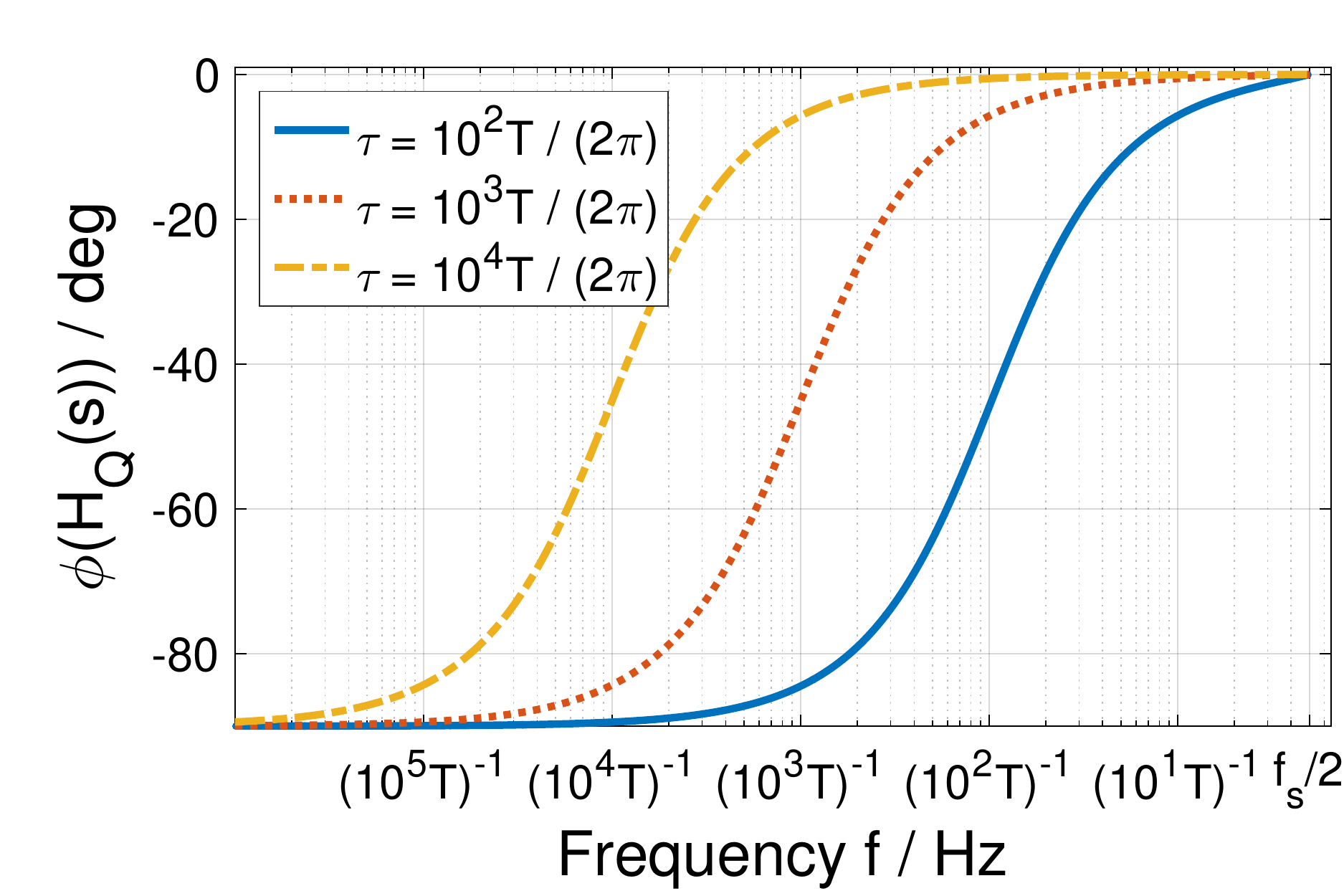}
\caption{Phase shift of the frequency response of the inverse amplifier transfer function in the $s$-domain.}
\label{fig_app_sdomain_phase}
\end{figure}
\subsection{Third-order model}
\label{sec_app_matlab_fit}
The coefficients for the continuous-time third-order filter (inverse amplifier transfer function in the $s$-domain), obtained by measured data and the Matlab System Identification Toolbox~\protect\cite{mathworks}.

\begin{align} 
{b_s}_n=\{&1.003296462624417, 3.287812476298027 {e}{06},\nonumber\\
      &1.440835556293589 {e}{12}, 1.141346517666954 {e}{17}\}\label{eqn_iir_bs}\\
{a_s}_n=\{&1.000000000000000, 2.849177008886374 {e}{06},\nonumber\\
      &8.633921892399874 {e}{11}, 2.337183622326430 {e}{15}\}\label{eqn_iir_as}\, .
\end{align}

\clearpage
\section{Discrete-time inverse amplifier transfer functions}
\label{sec_app_iir_comparison}
\subsection{IIR filter with forward difference method}
\label{sec_app_forward}

\begin{figure}[h]
\centering
\includegraphics[width=0.40\textwidth]{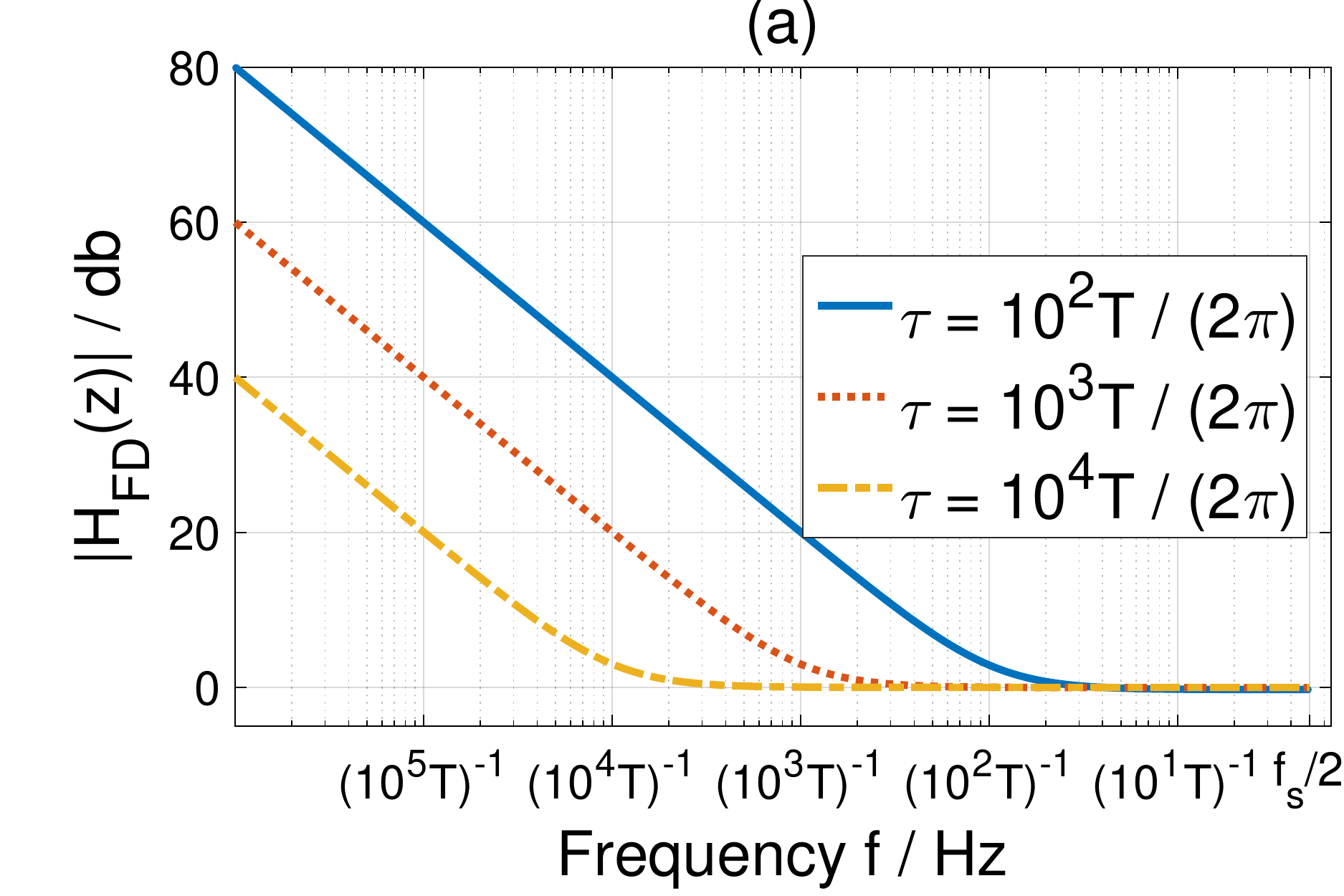}
\includegraphics[width=0.40\textwidth]{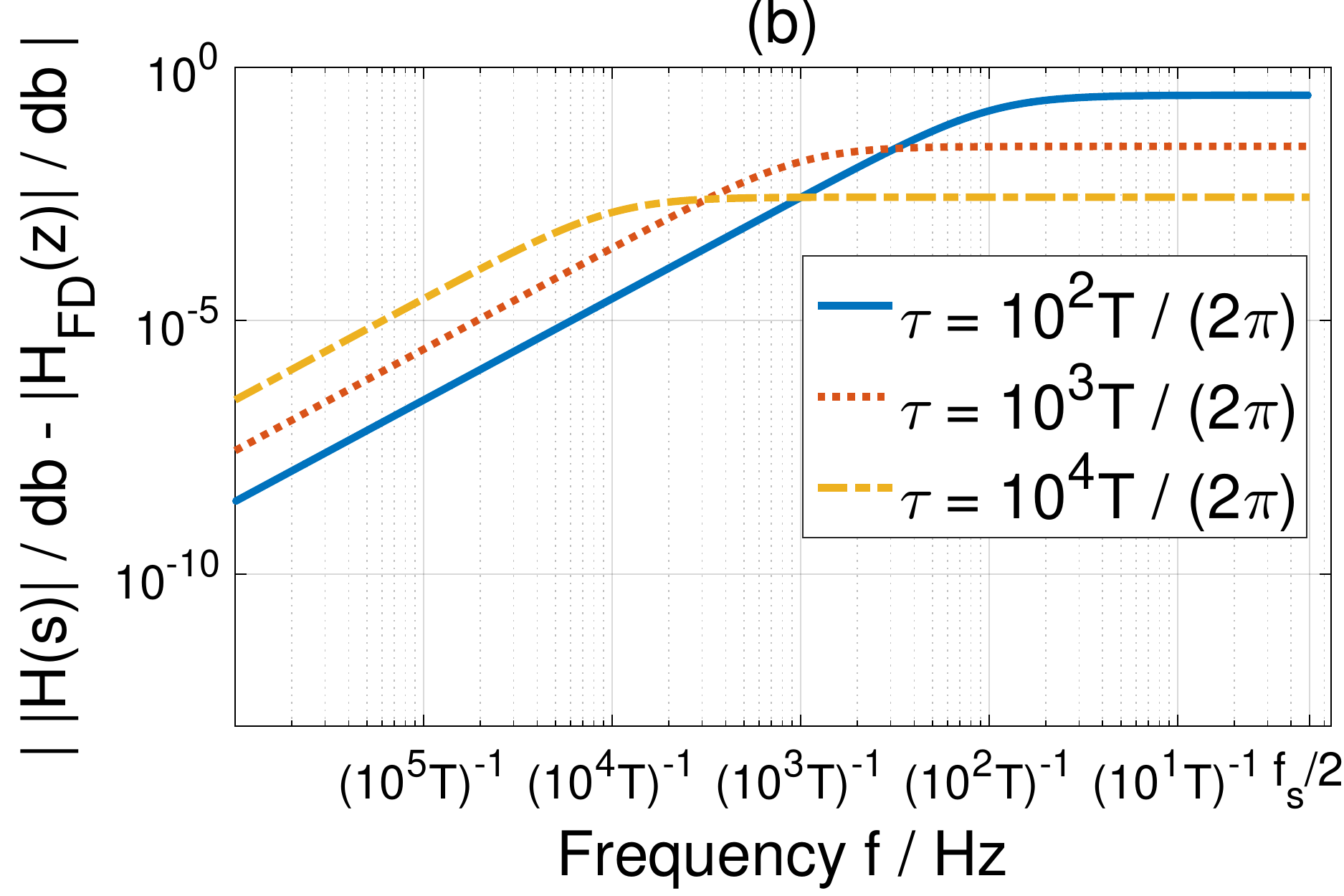}
\caption{Magnitude of the frequency response of the IIR filter (a) and the corresponding difference from the ideal filter (b).}
\label{fig_app_forward_mag}
\end{figure}

\begin{figure}[h]
\centering
\includegraphics[width=0.40\textwidth]{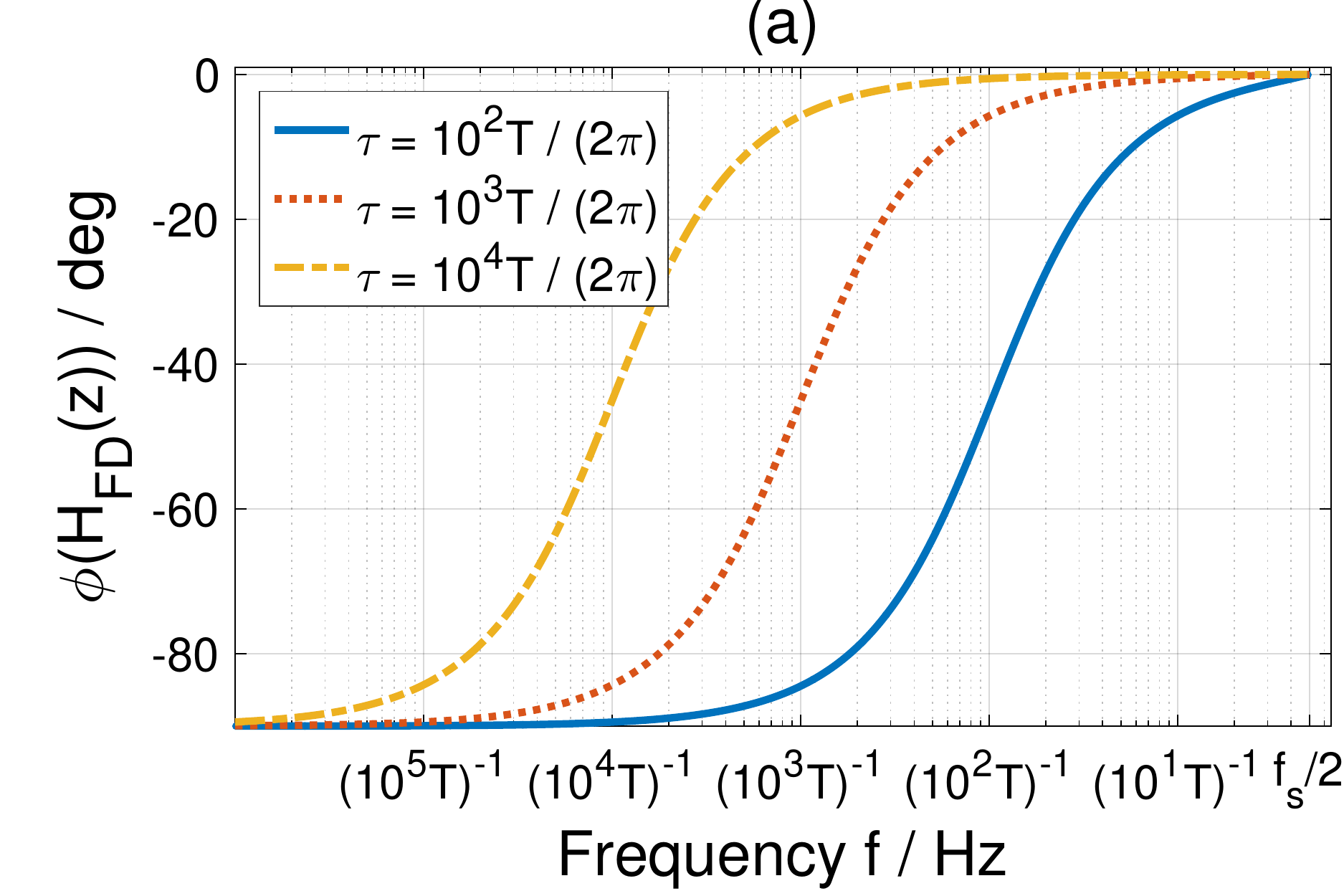}
\includegraphics[width=0.40\textwidth]{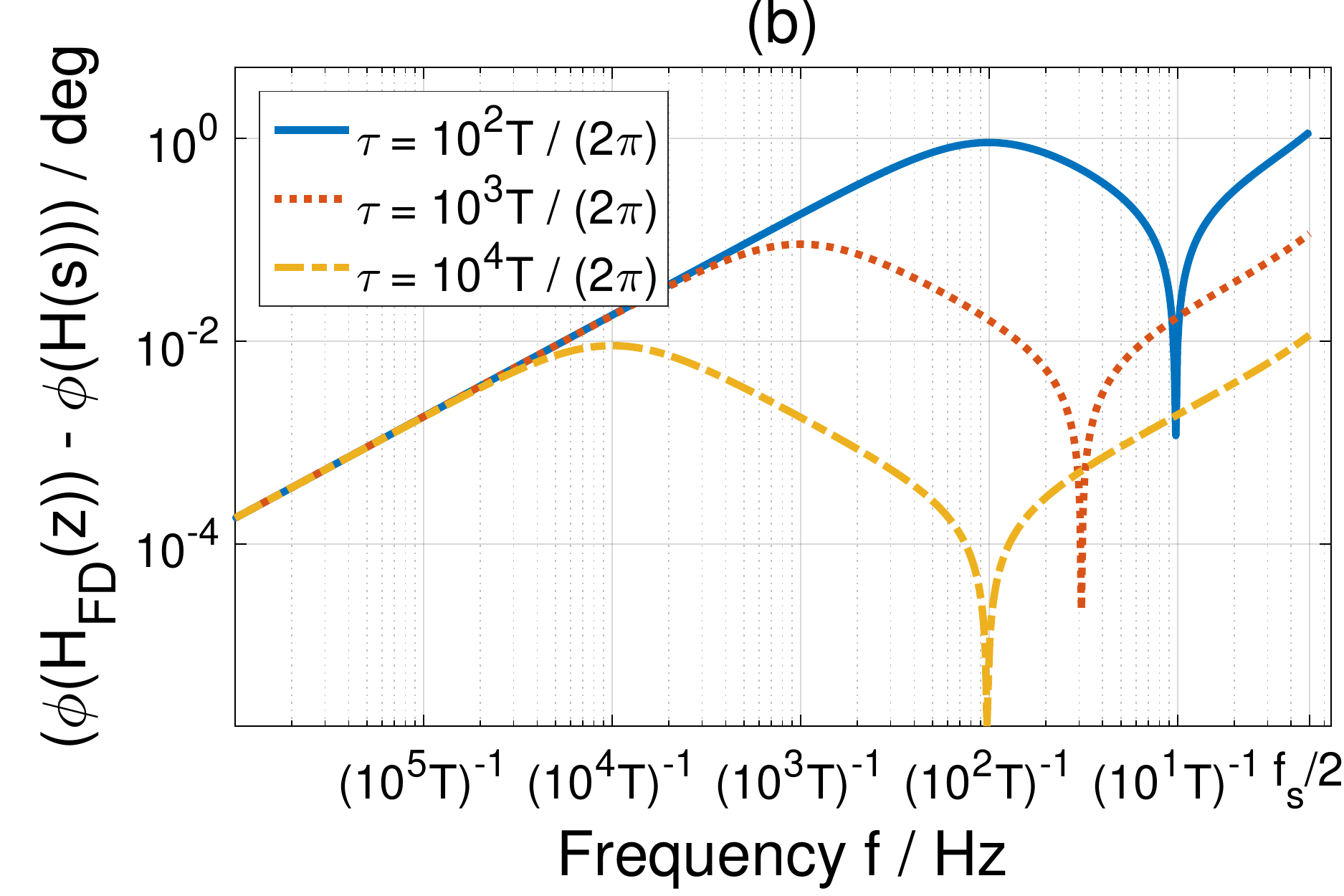}
\caption{Phase shift of the frequency response of the IIR filter (a) and the corresponding difference from the ideal filter (b).}
\label{fig_app_forward_phase}
\end{figure}

\begin{figure}[h]
\centering
\includegraphics[width=0.40\textwidth]{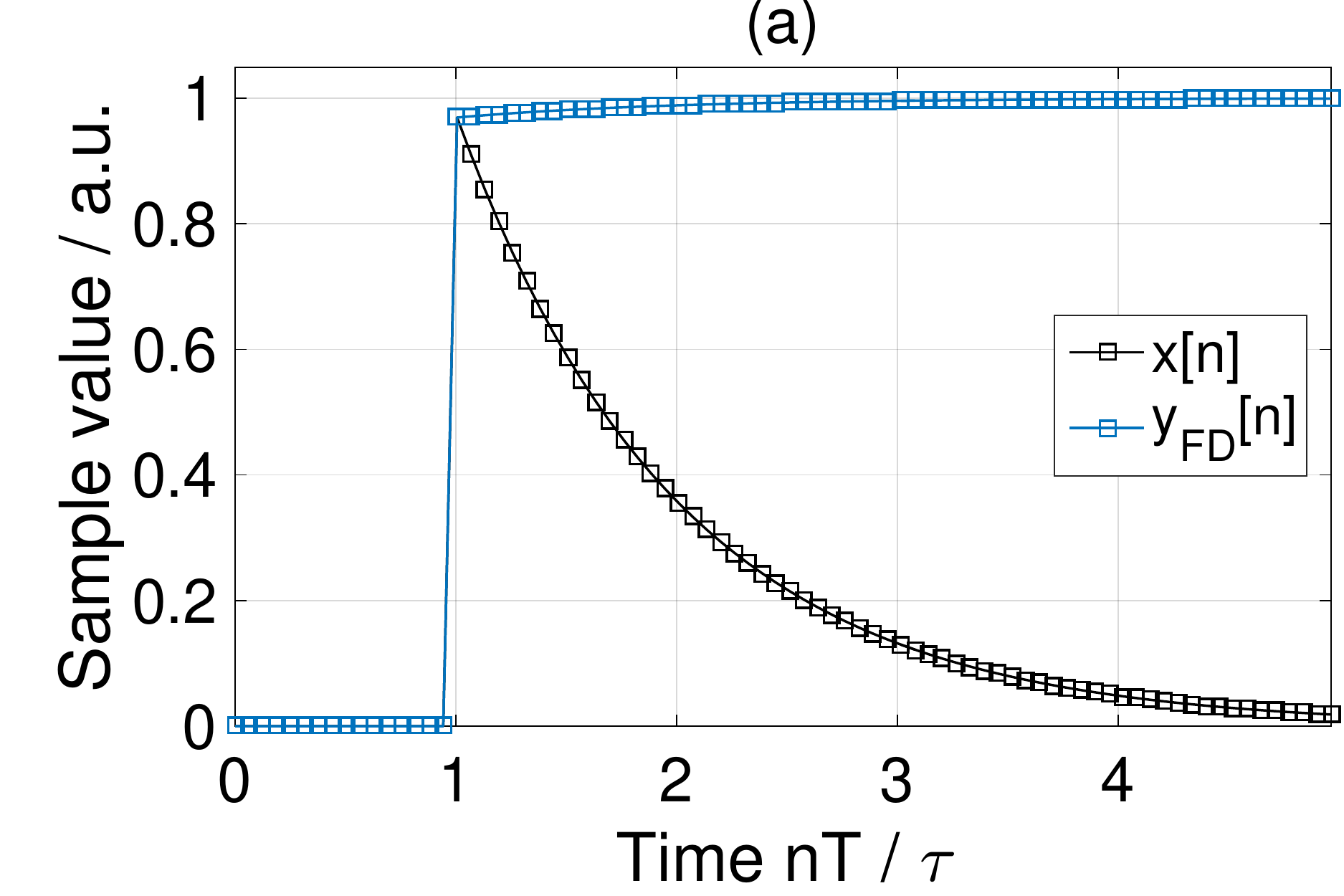}
\includegraphics[width=0.40\textwidth]{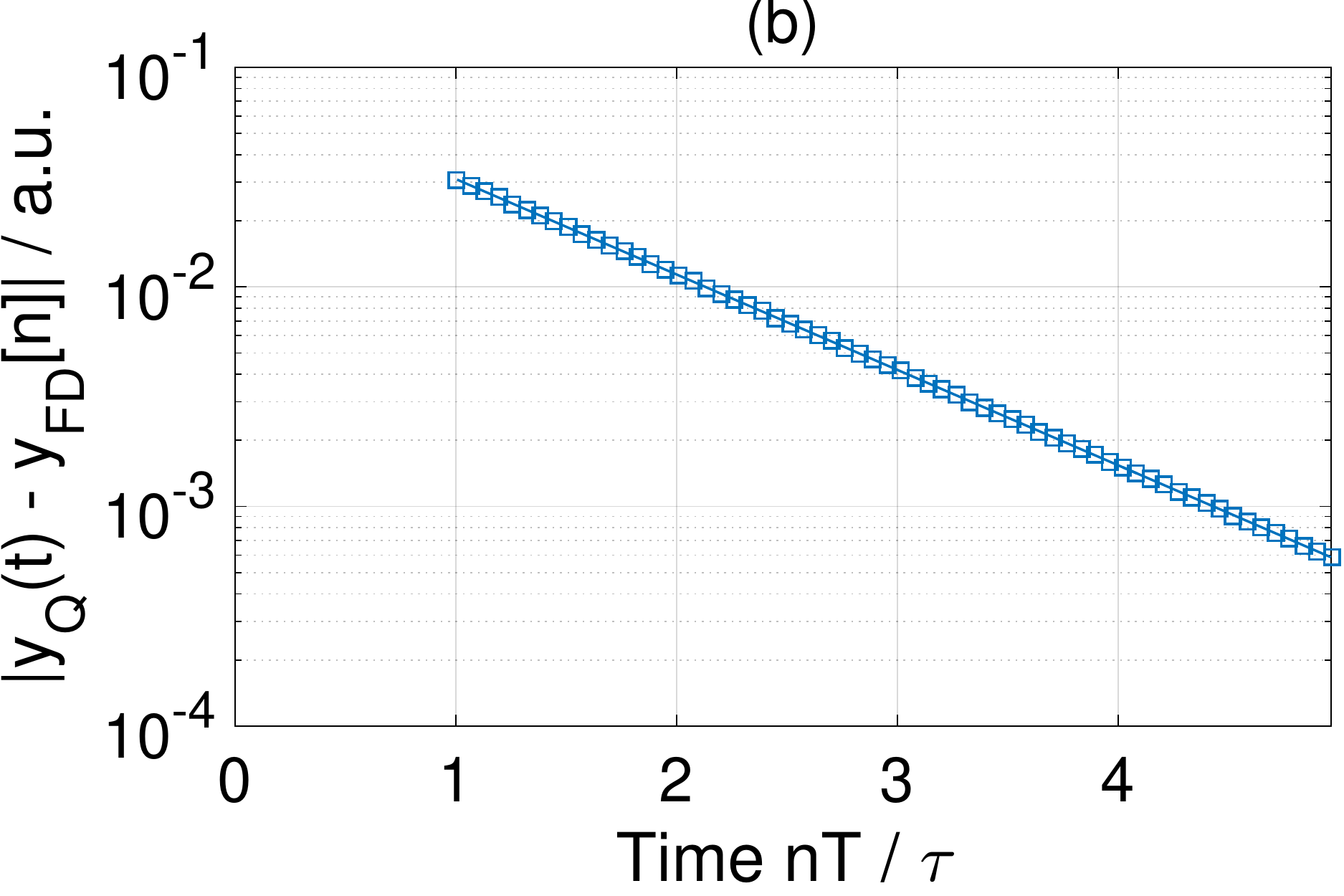}
\caption{Simulated amplifier signal with $\tau = \frac{100T}{2\pi}$ and the corresponding deconvolution with the IIR filter (left) and the difference from the ideal output.}
\label{fig_app_forward_time}
\end{figure}

\clearpage
\subsection{IIR filter with backward difference method}
\label{sec_app_backward}

\begin{figure}[h]
\centering
\includegraphics[width=0.40\textwidth]{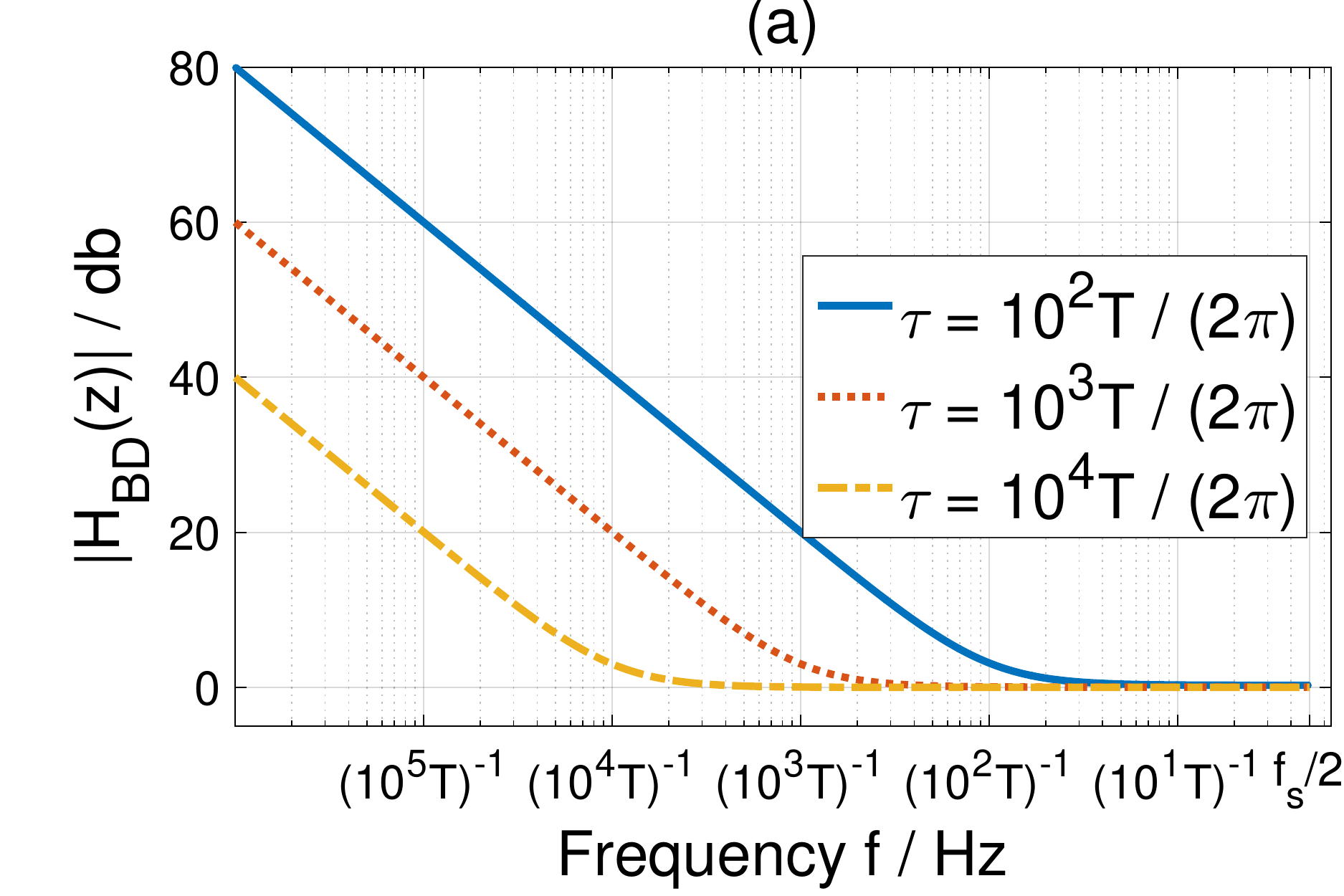}
\includegraphics[width=0.40\textwidth]{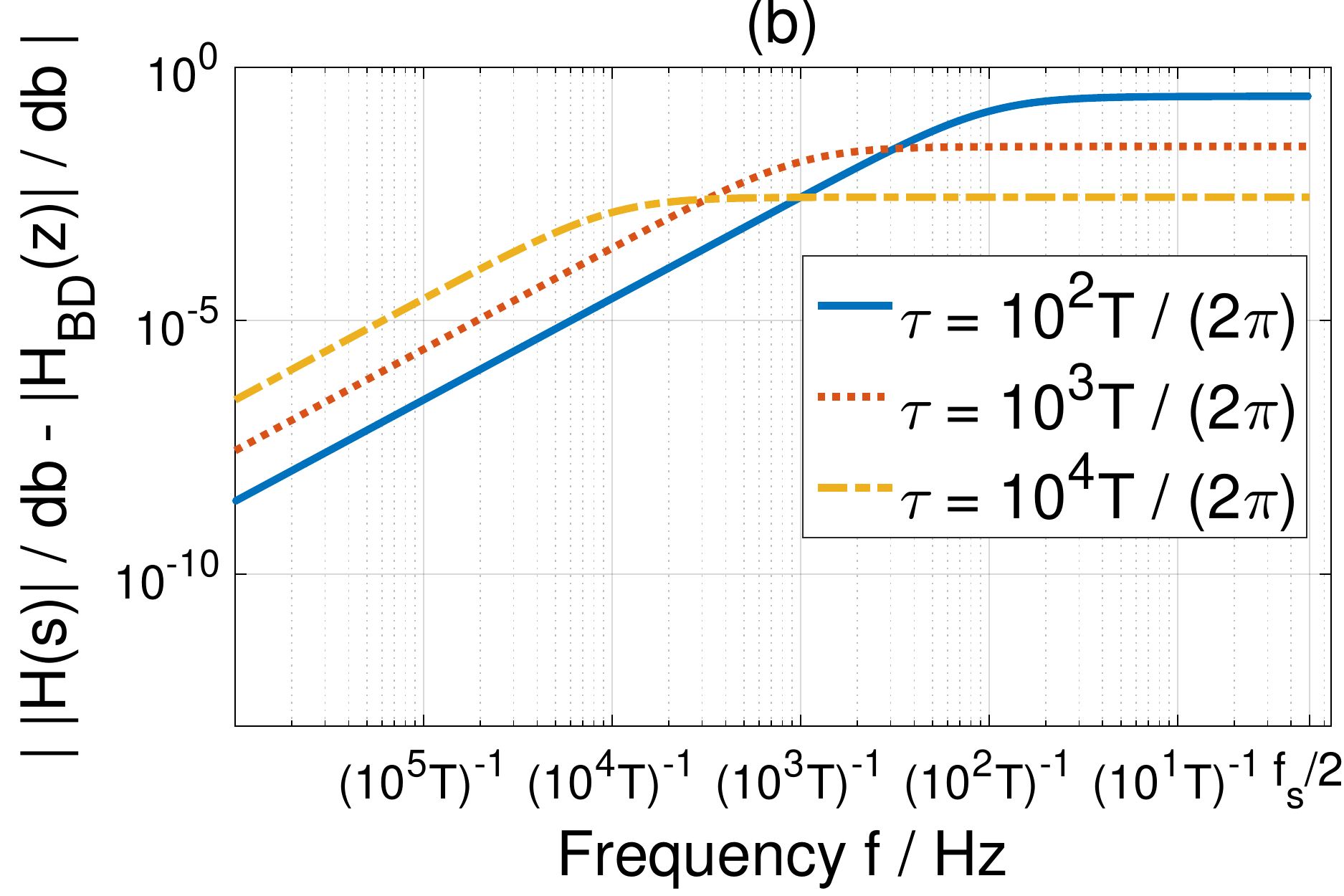}
\caption{Magnitude of the frequency response of the IIR filter (left) and the corresponding difference from the ideal filter (right).}
\label{fig_app_backward_mag}
\end{figure}

\begin{figure}[h]
\centering
\includegraphics[width=0.40\textwidth]{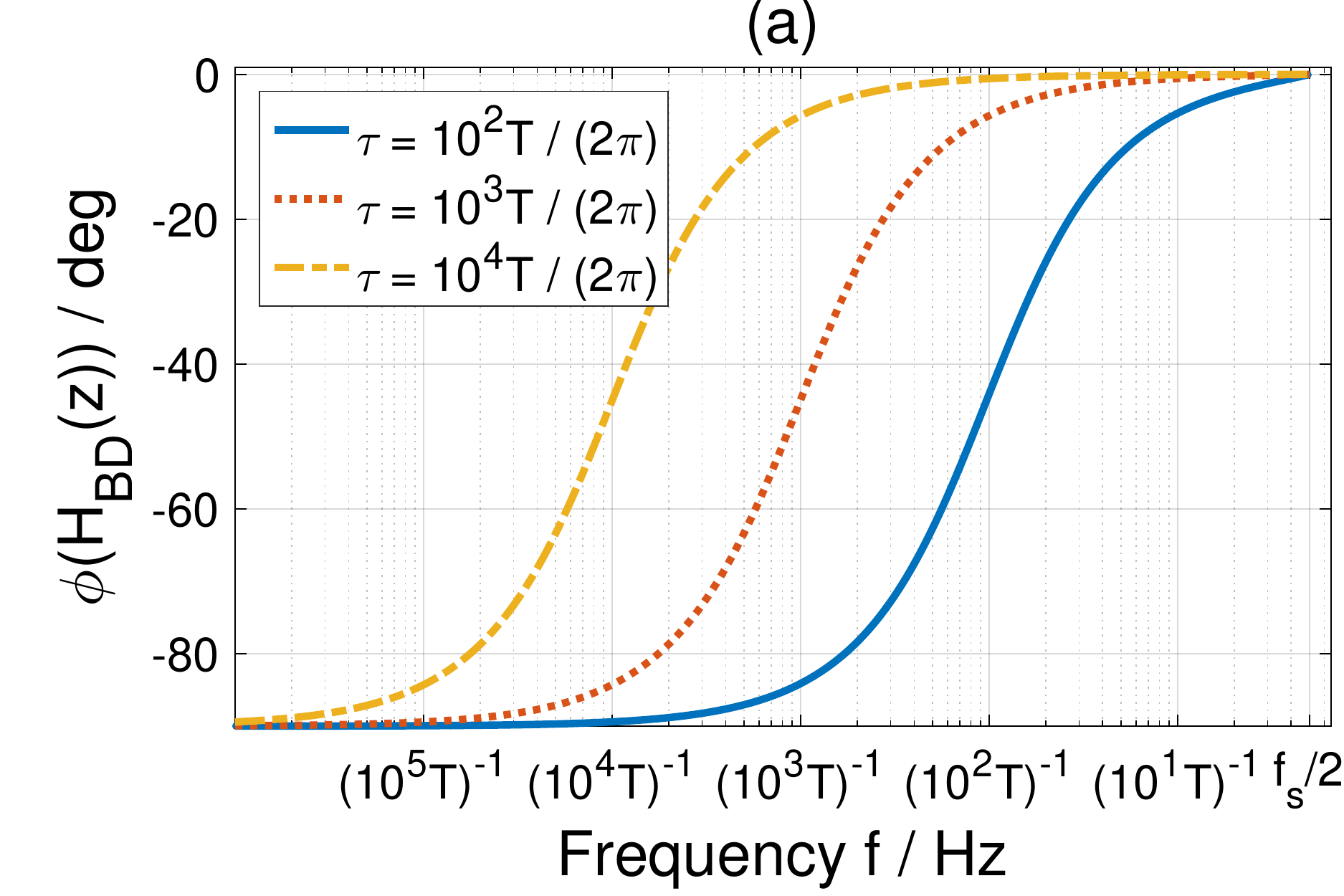}
\includegraphics[width=0.40\textwidth]{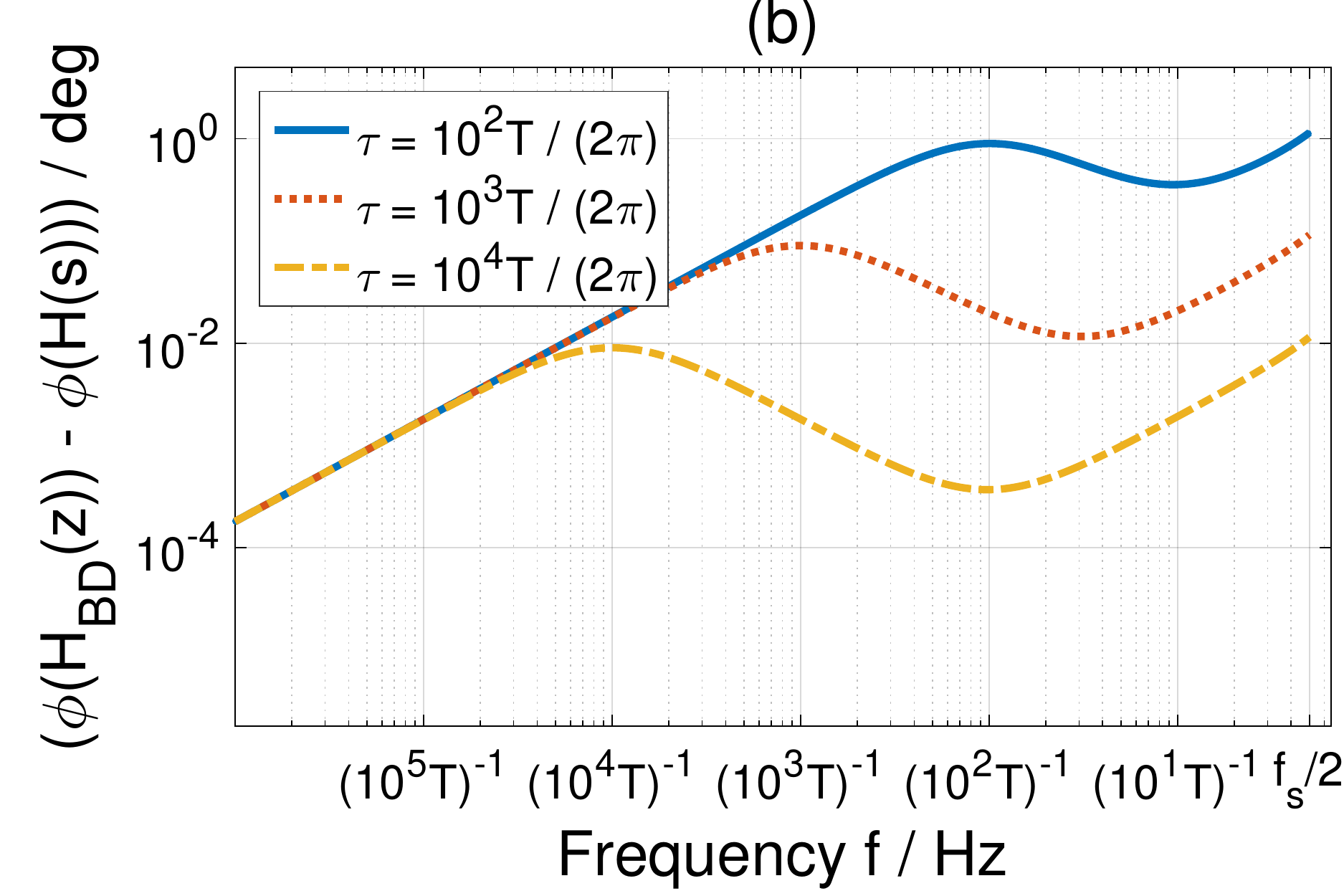}
\caption{Phase shift of the frequency response of the IIR filter (left) and the corresponding difference from the ideal filter (right).}
\label{fig_app_backward_phase}
\end{figure}

\begin{figure}[h]
\centering
\includegraphics[width=0.40\textwidth]{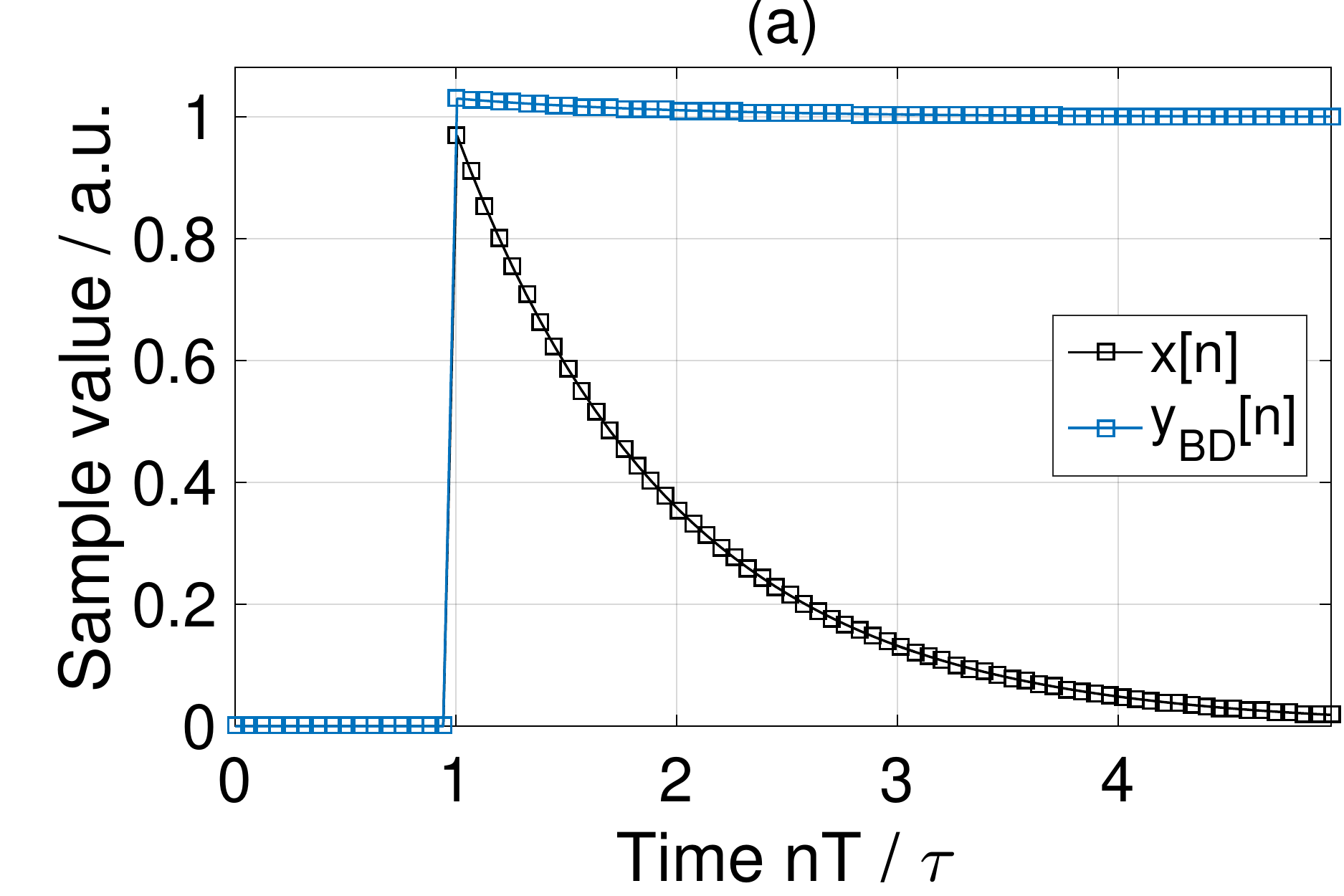}
\includegraphics[width=0.40\textwidth]{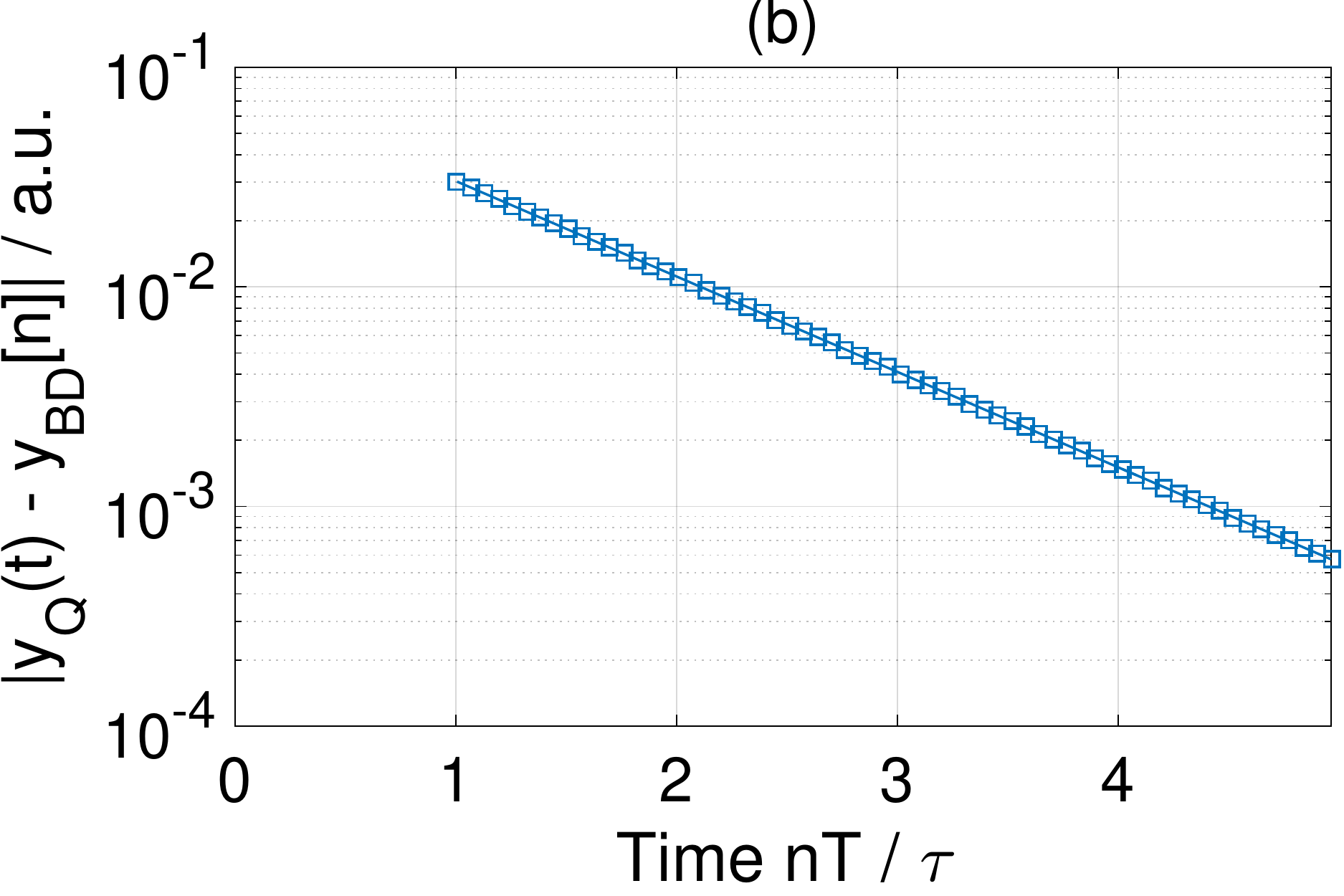}
\caption{Simulated amplifier signal with $\tau = \frac{100T}{2\pi}$ and the corresponding deconvolution with the IIR filter (left) and the difference from the ideal output.}
\label{fig_app_backward_time}
\end{figure}

\clearpage
\subsection{IIR filter with bilinear transformation}
\label{sec_app_bilinear}

\begin{figure}[h]
\centering
\includegraphics[width=0.40\textwidth]{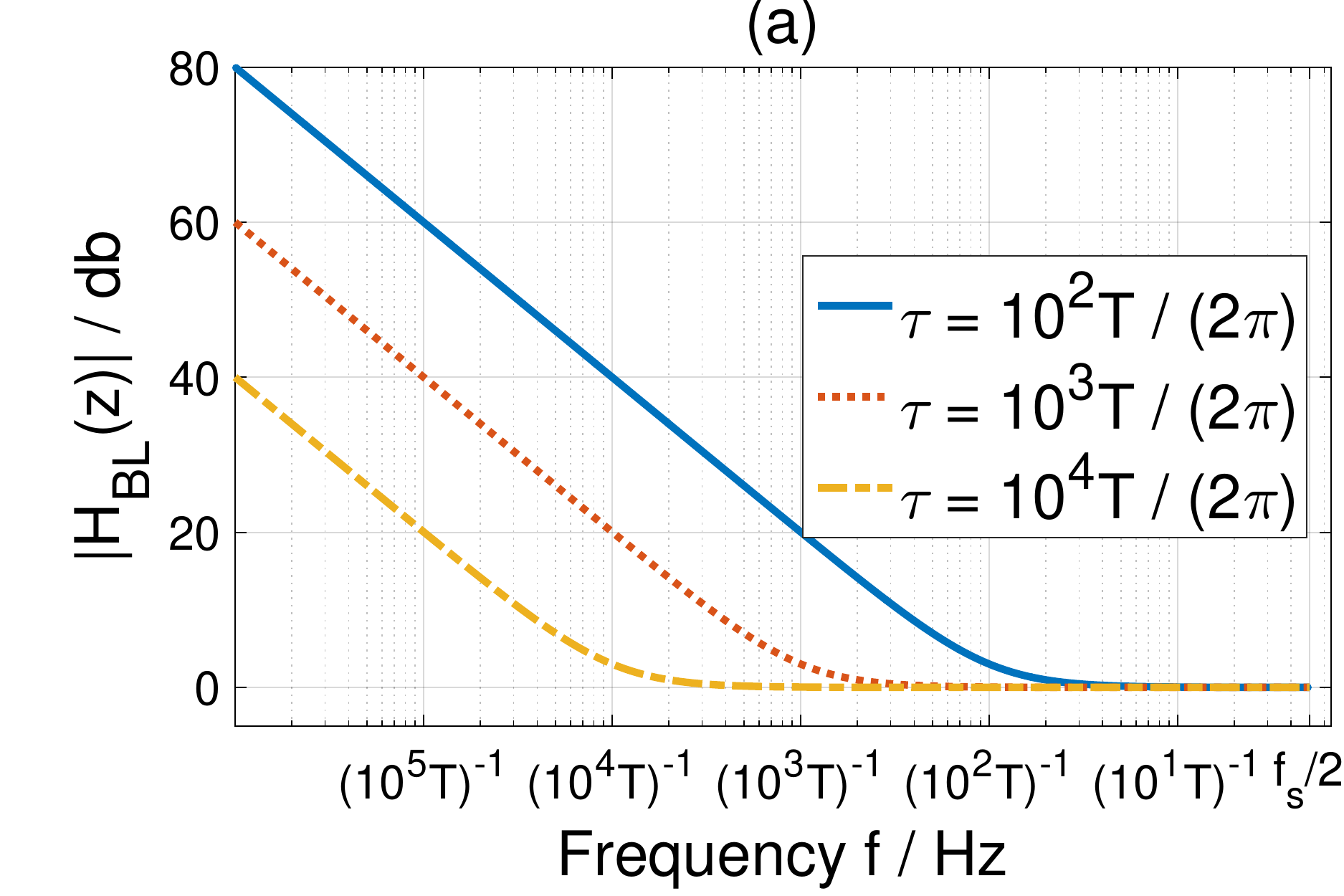}
\includegraphics[width=0.40\textwidth]{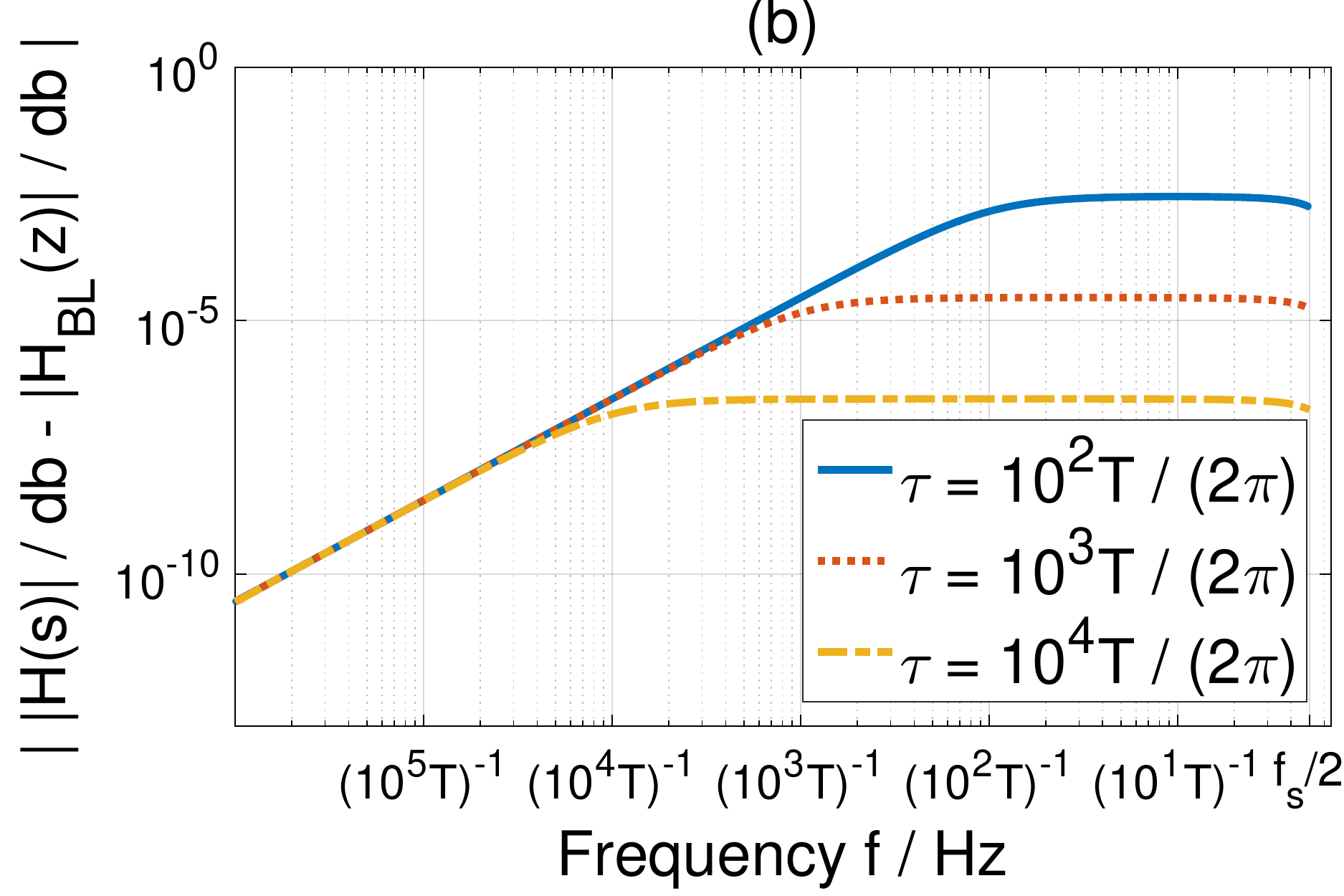}
\caption{Magnitude of the frequency response of the IIR filter (left) and the corresponding difference from the ideal filter (right).}
\label{fig_app_bilinear_mag}
\end{figure}

\begin{figure}[h]
\centering
\includegraphics[width=0.40\textwidth]{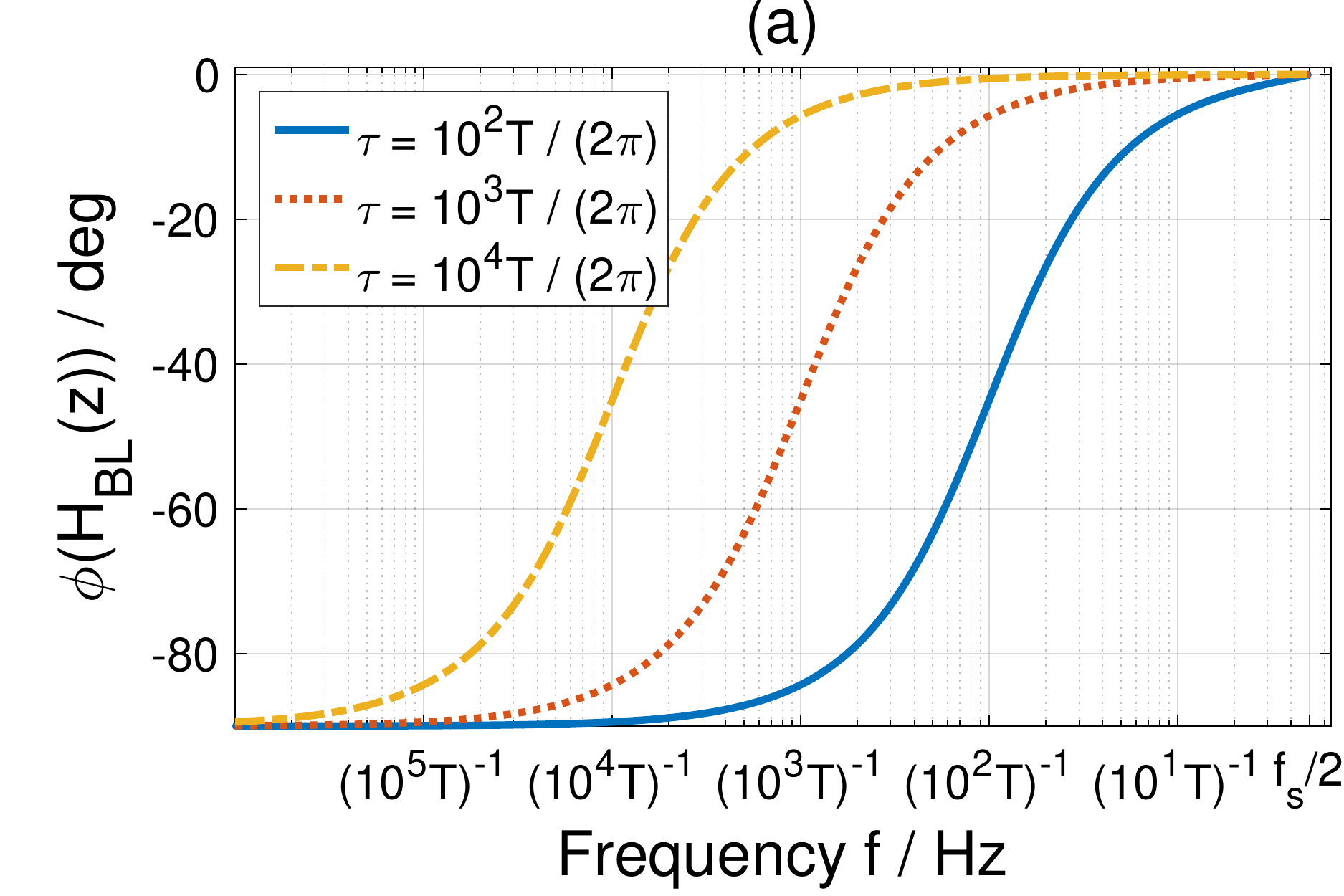}
\includegraphics[width=0.40\textwidth]{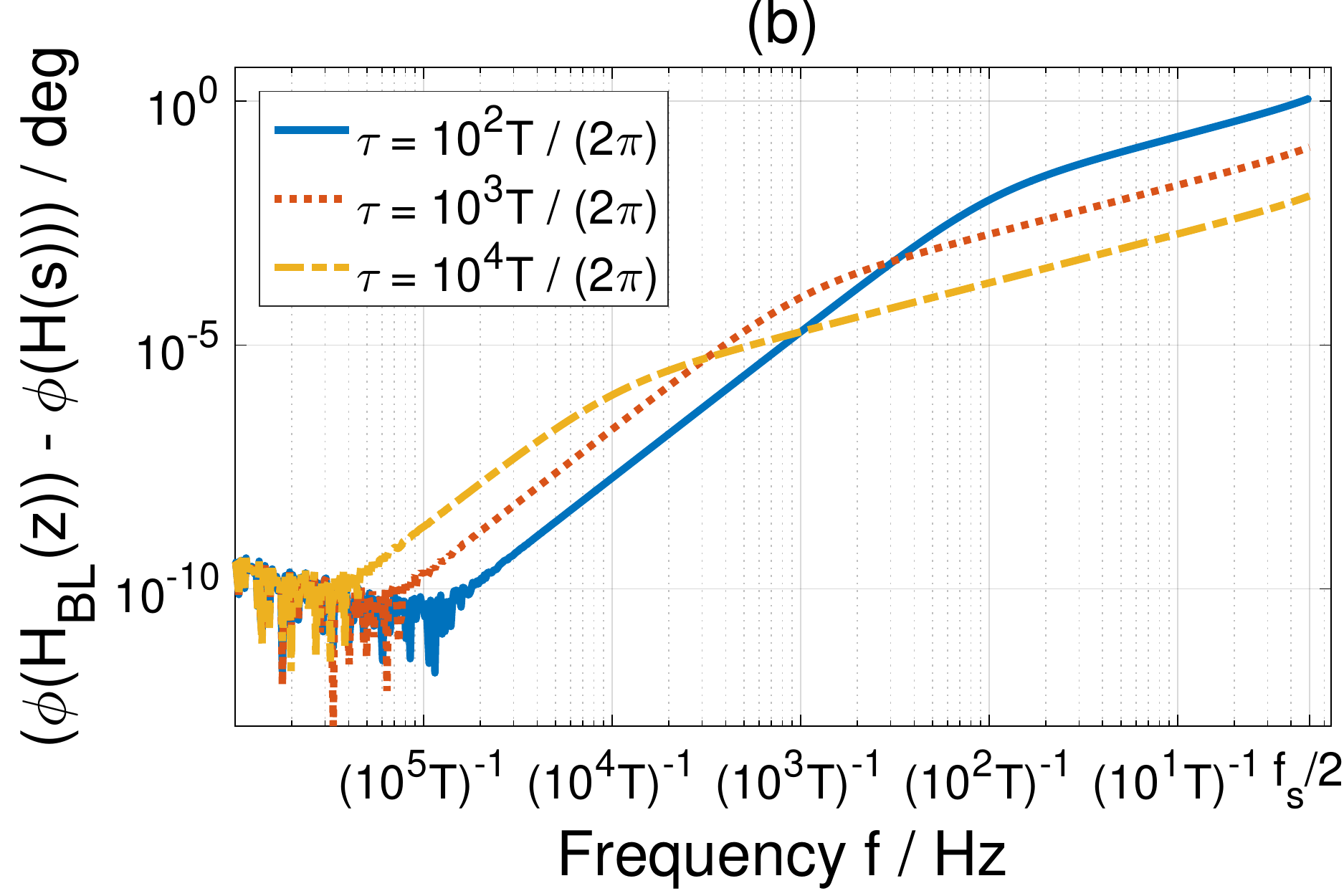}
\caption{Phase shift of the frequency response of the IIR filter (left) and the corresponding difference from the ideal filter (right).}
\label{fig_app_bilinear_phase}
\end{figure}

\begin{figure}[h]
\centering
\includegraphics[width=0.40\textwidth]{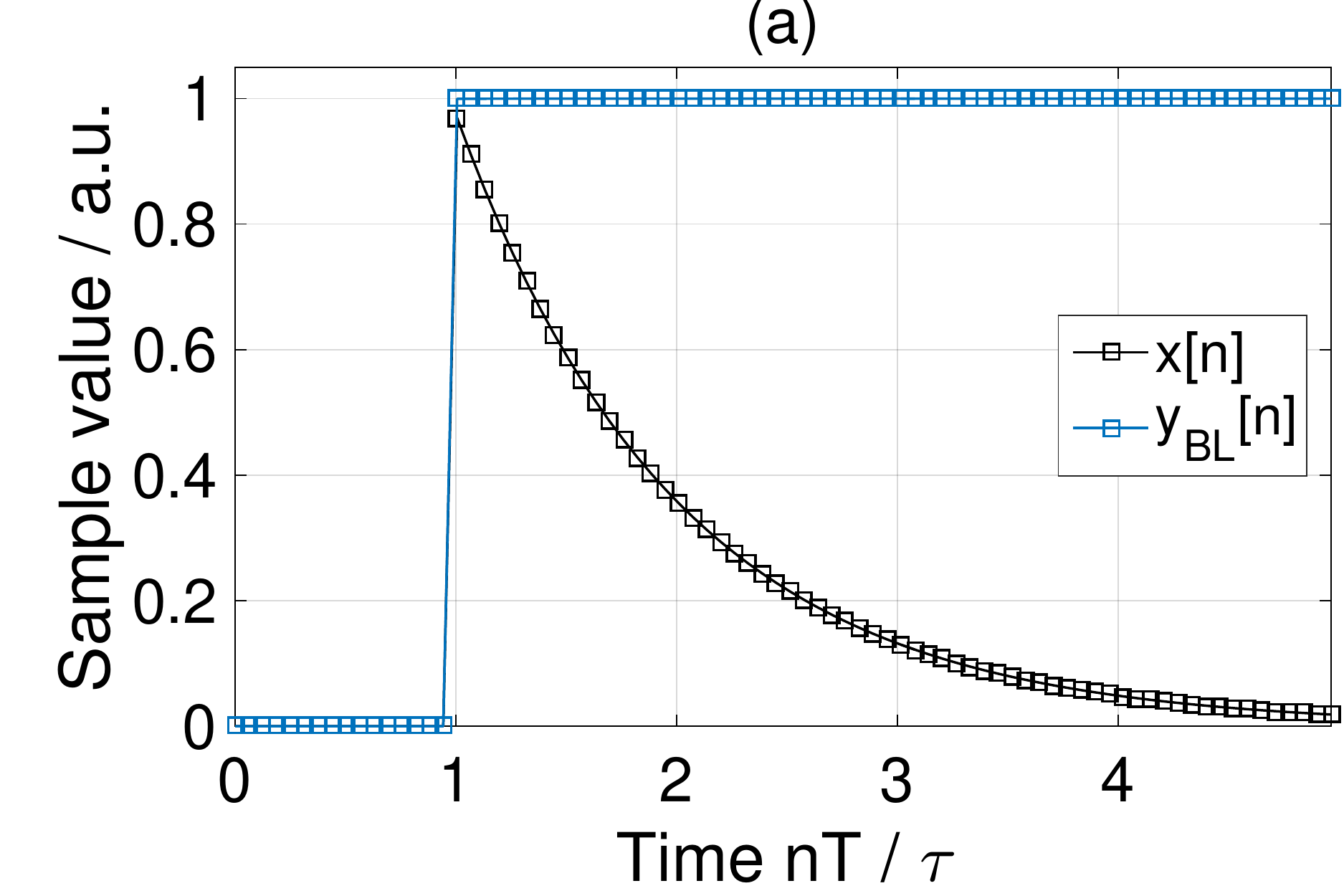}
\includegraphics[width=0.40\textwidth]{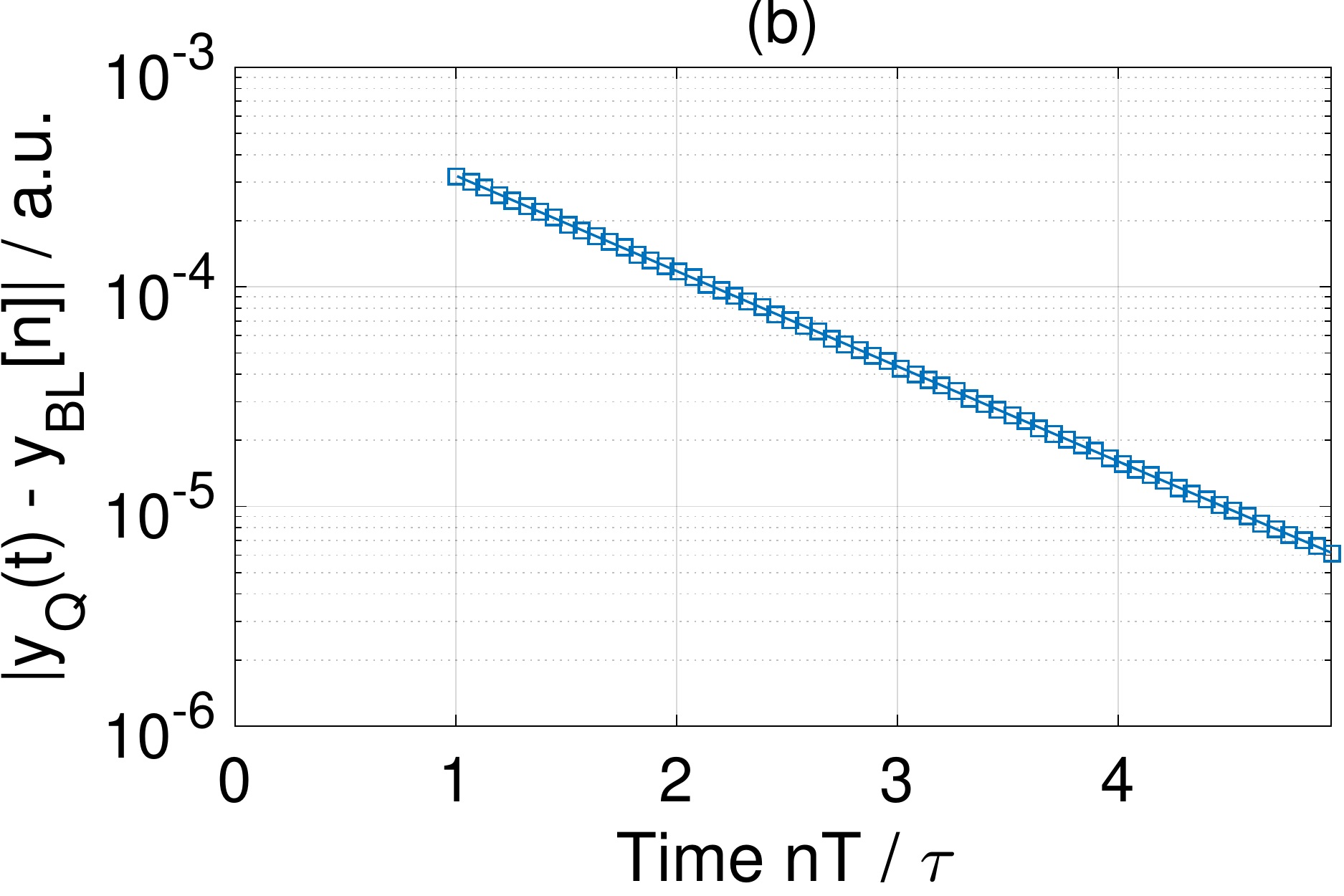}
\caption{Simulated amplifier signal with $\tau = \frac{100T}{2\pi}$ and the corresponding deconvolution with the IIR filter (left) and the difference from the ideal output.}
\label{fig_app_bilinear_time}
\end{figure}

\clearpage
\subsection{IIR filter with matched-z transformation}
\label{sec_app_matchedz}

\begin{figure}[h]
\centering
\includegraphics[width=0.40\textwidth]{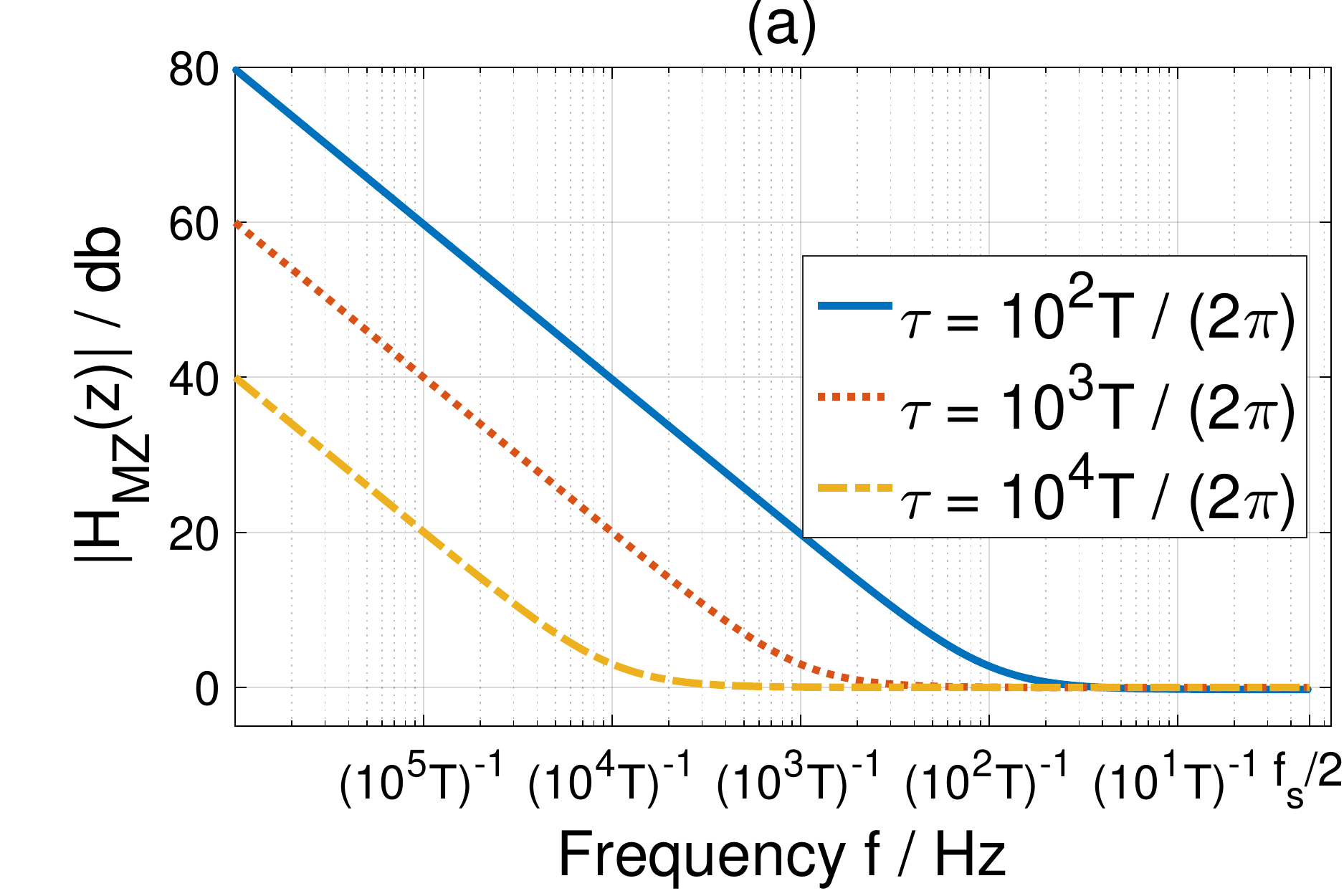}
\includegraphics[width=0.40\textwidth]{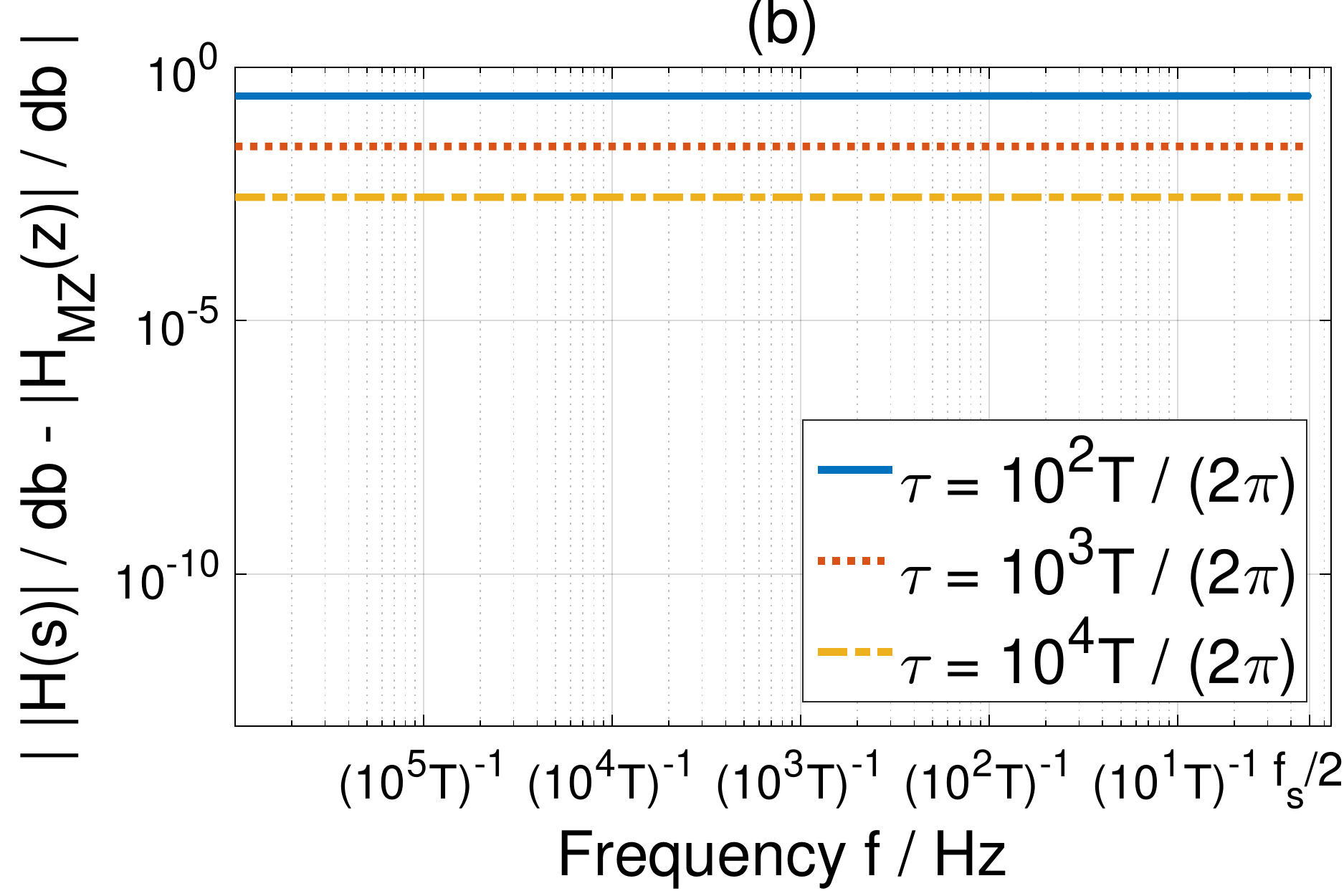}
\caption{Magnitude of the frequency response of the IIR filter (left) and the corresponding difference from the ideal filter (right).}
\label{fig_app_matchedz_mag}
\end{figure}

\begin{figure}[h]
\centering
\includegraphics[width=0.40\textwidth]{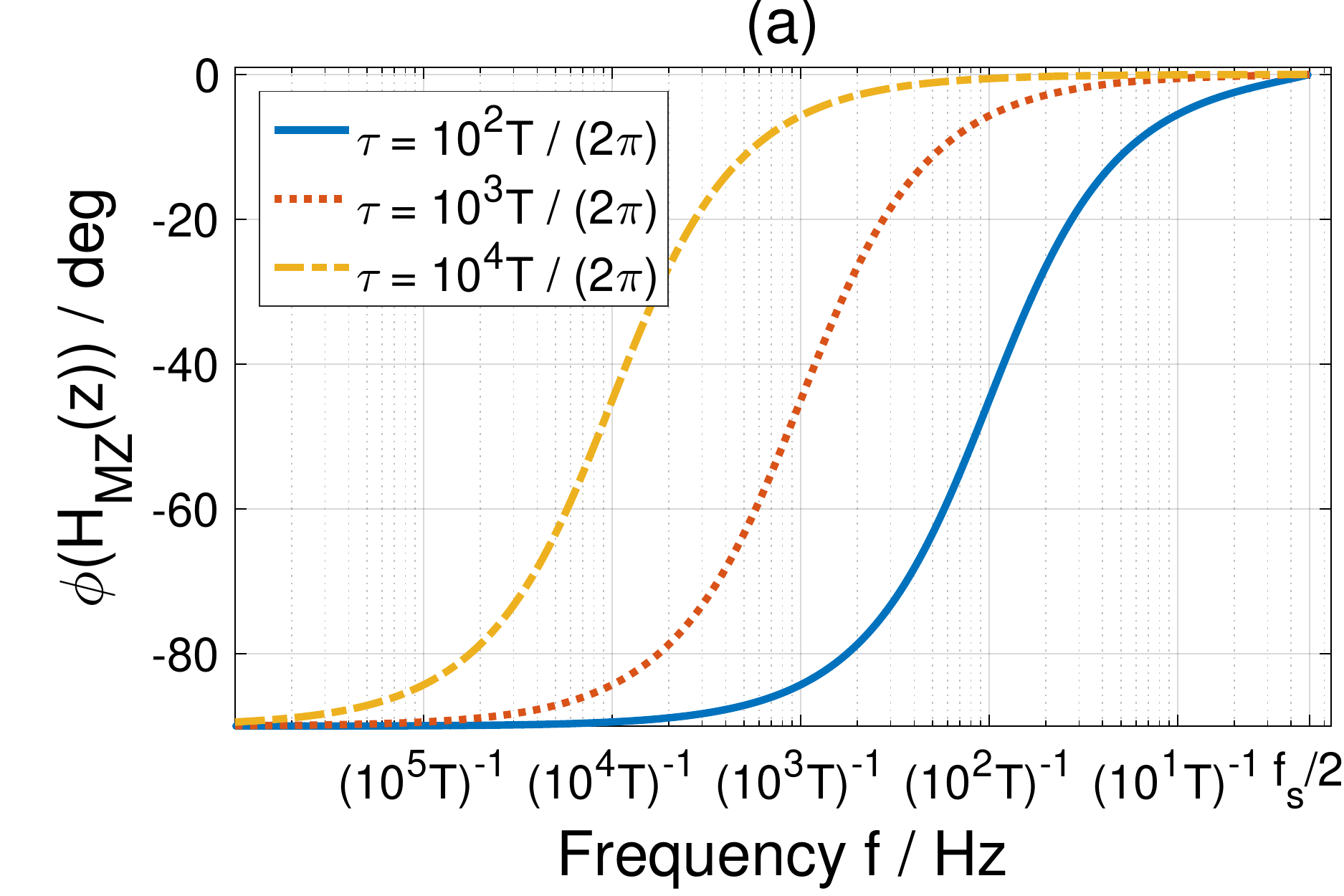}
\includegraphics[width=0.40\textwidth]{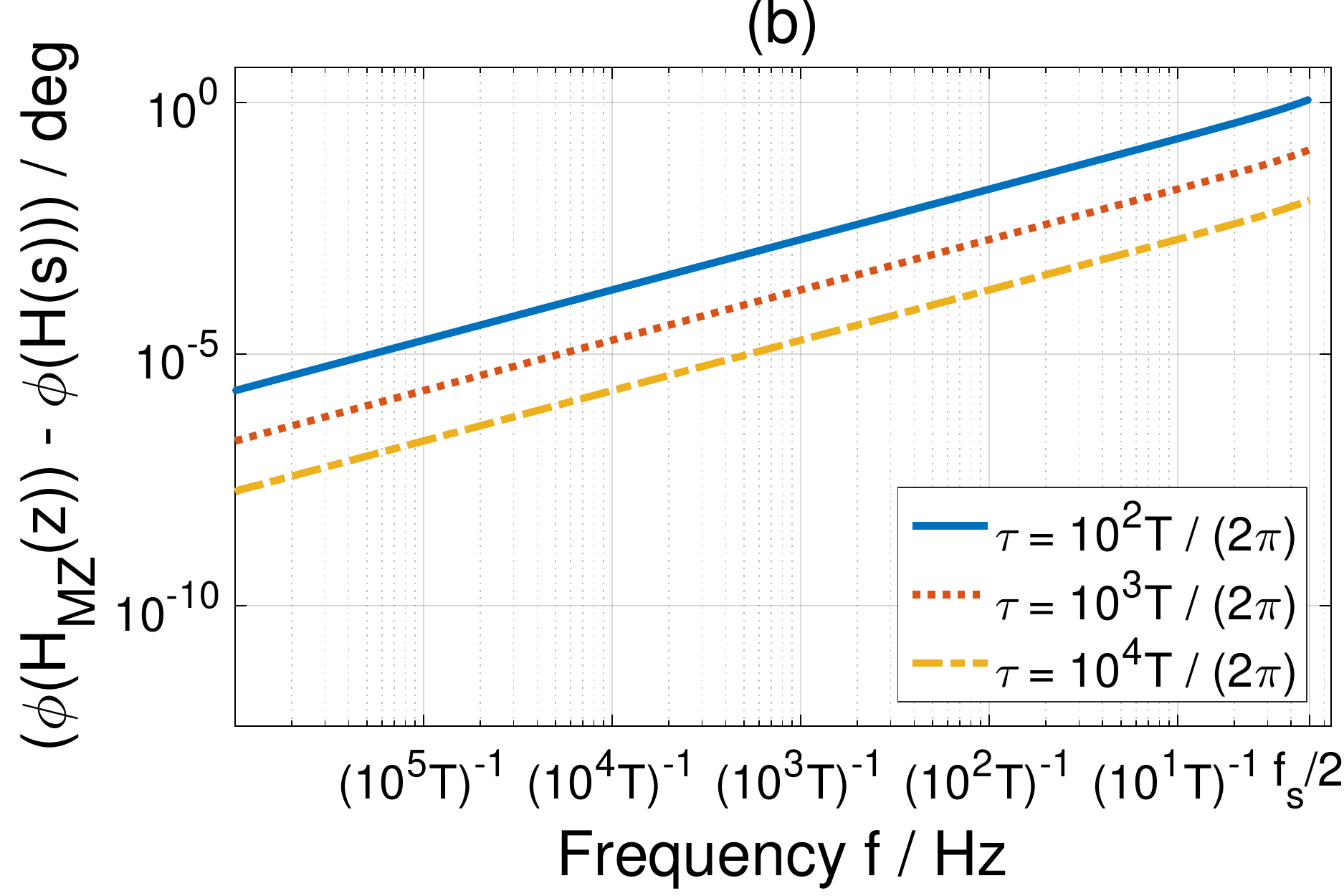}
\caption{Phase shift of the frequency response of the IIR filter (left) and the corresponding difference from the ideal filter (right).}
\label{fig_app_matchedz_phase}
\end{figure}

\begin{figure}[h]
\centering
\includegraphics[width=0.40\textwidth]{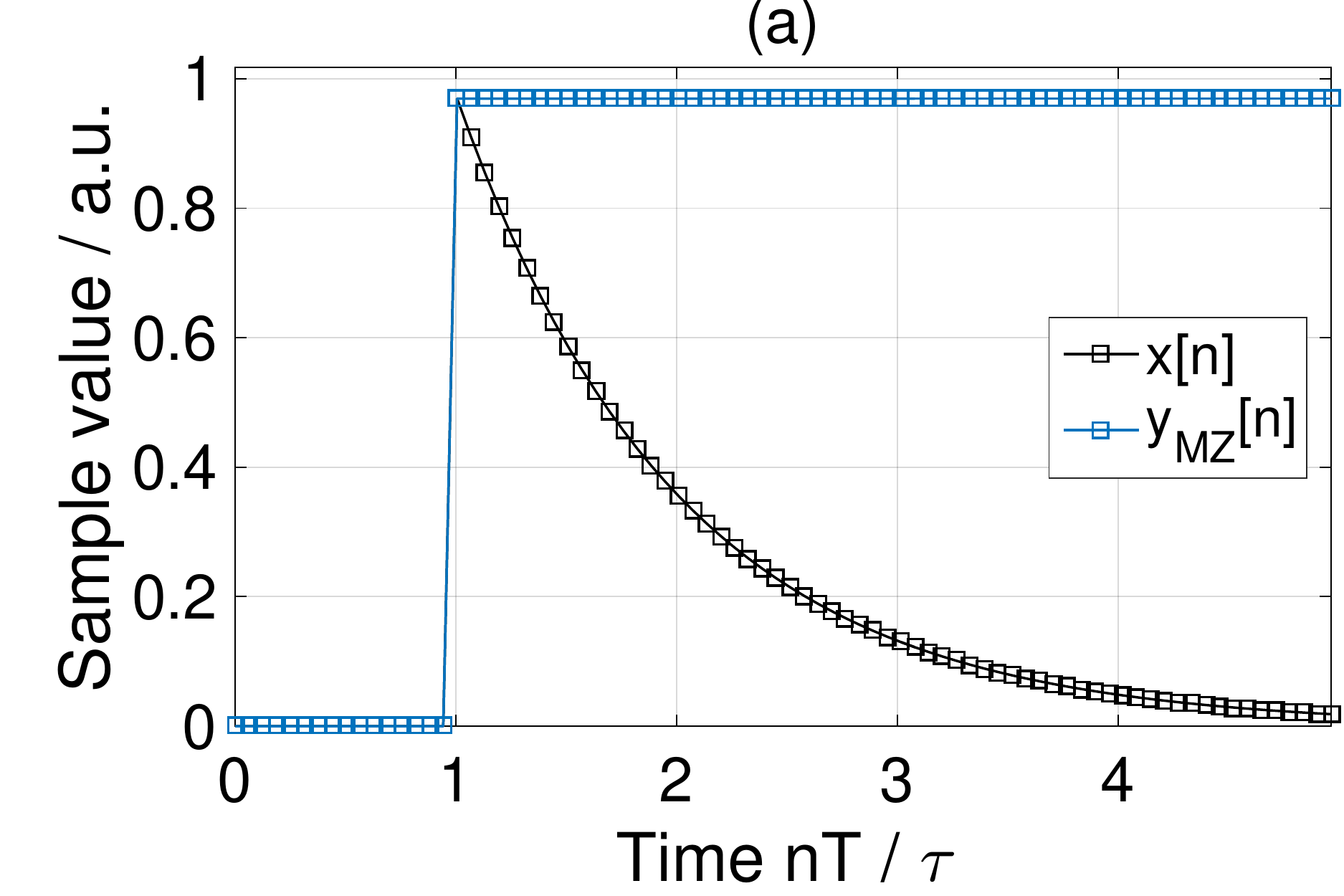}
\includegraphics[width=0.40\textwidth]{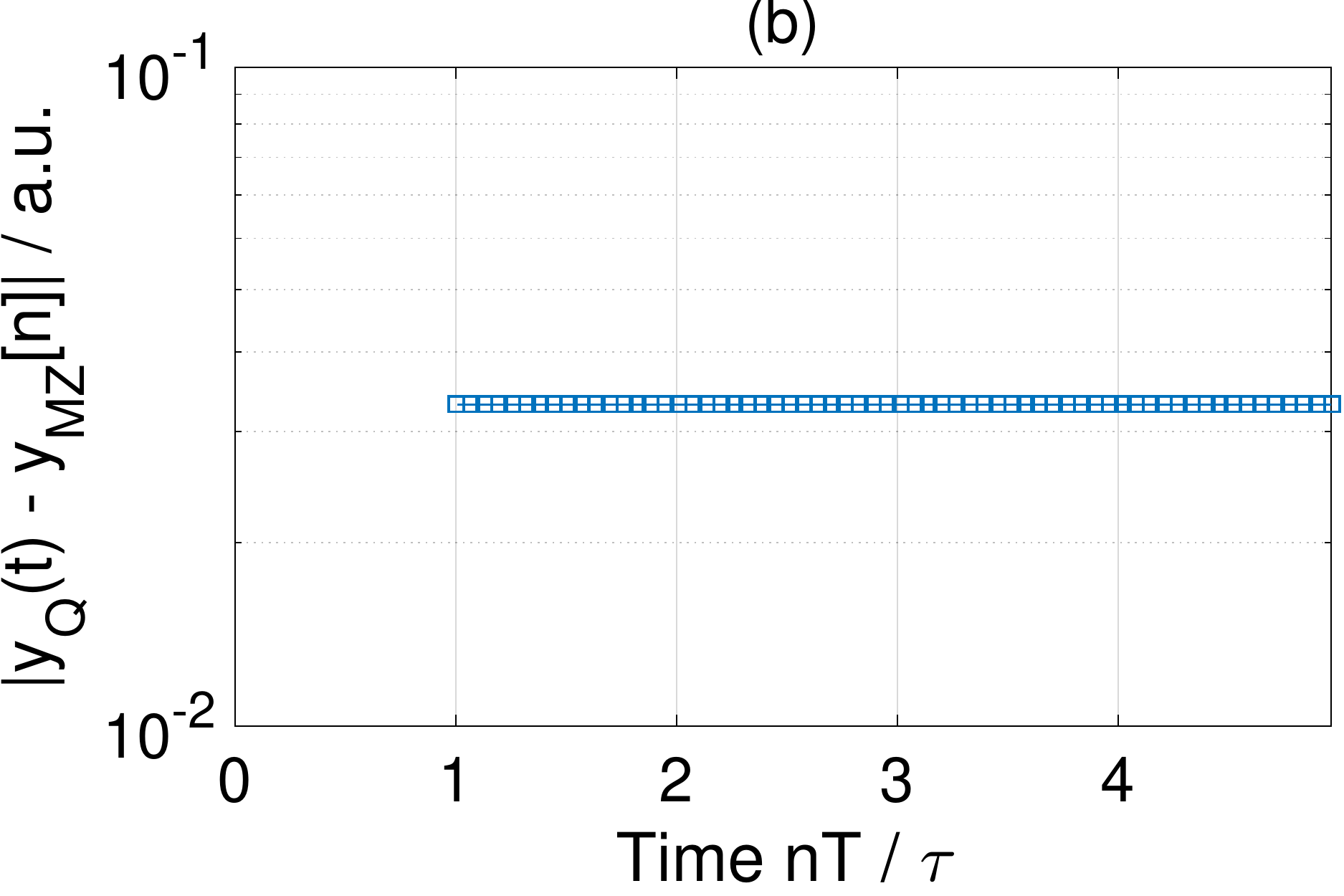}
\caption{Simulated amplifier signal with $\tau = \frac{100T}{2\pi}$ and the corresponding deconvolution with the IIR filter (left) and the difference from the ideal output.}
\label{fig_app_matchedz_time}
\end{figure}

\clearpage
\subsection{IIR filter with matched-z transformation and gain correction at frequency ($\omega = \frac{\pi}{T}$)}
\label{sec_app_gainT}

\begin{figure}[h]
\centering
\includegraphics[width=0.40\textwidth]{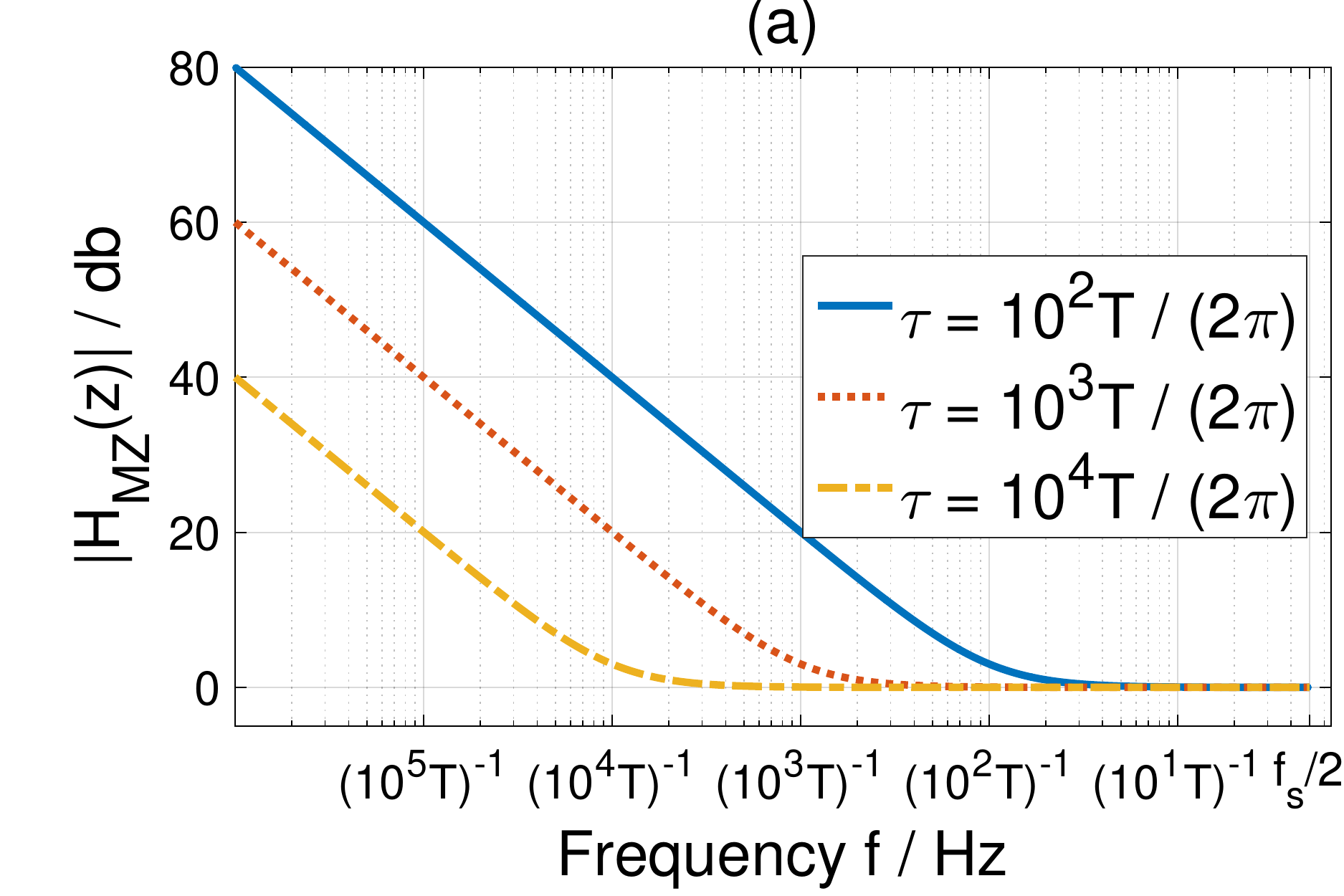}
\includegraphics[width=0.40\textwidth]{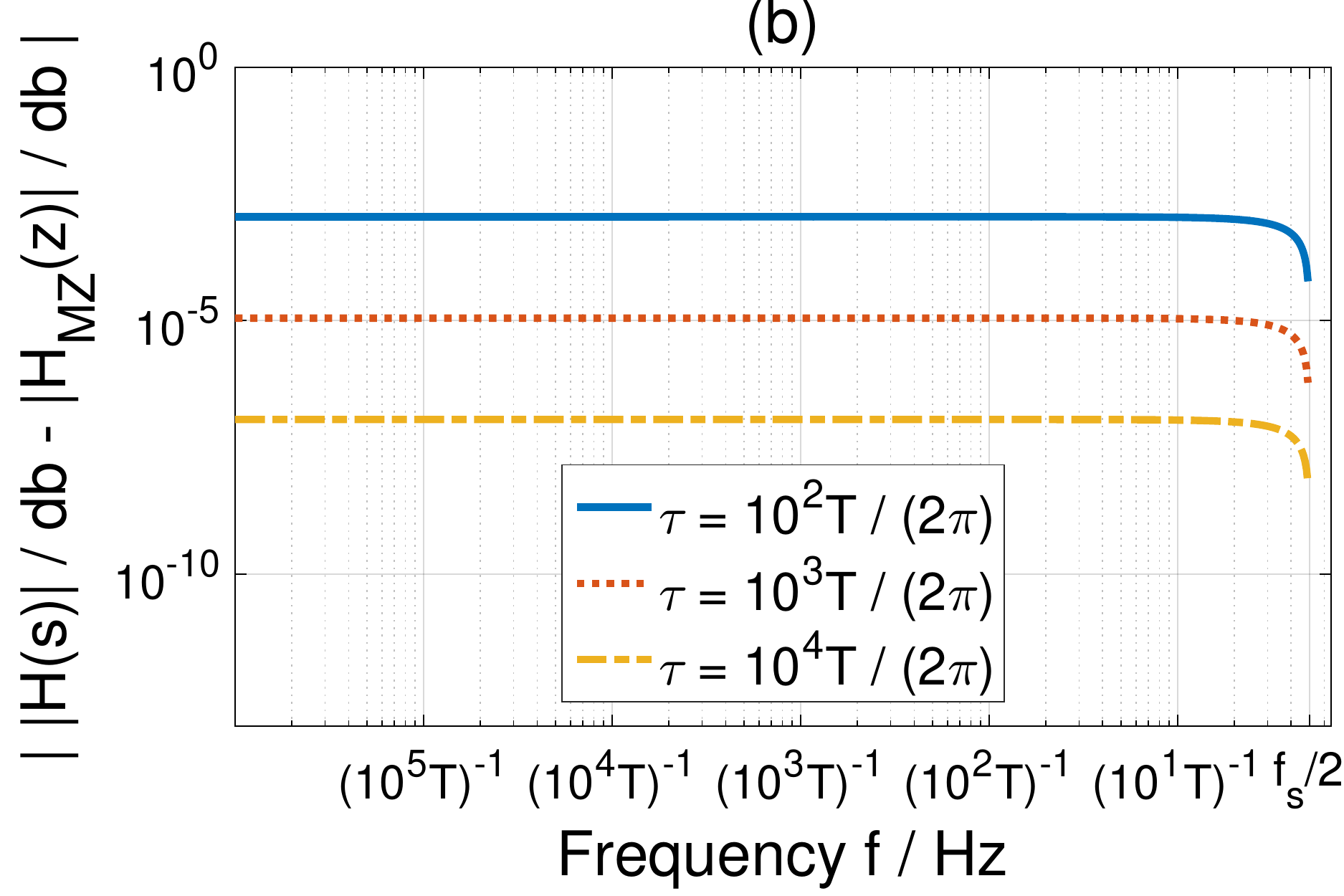}
\caption{Magnitude of the frequency response of the IIR filter (left) and the corresponding difference from the ideal filter (right).}
\label{fig_app_gainT_mag}
\end{figure}
\begin{figure}[h]
\centering
\includegraphics[width=0.40\textwidth]{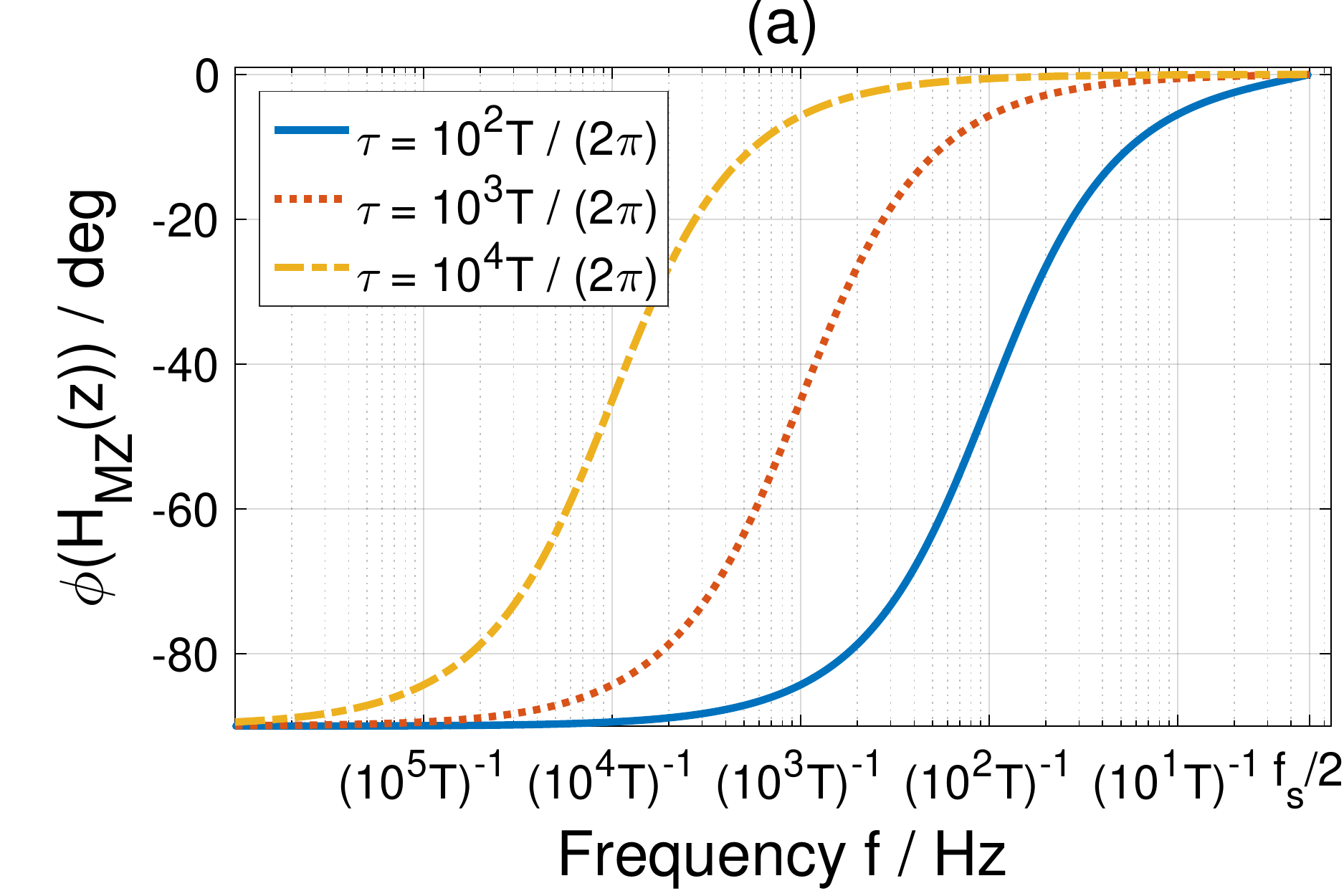}
\includegraphics[width=0.40\textwidth]{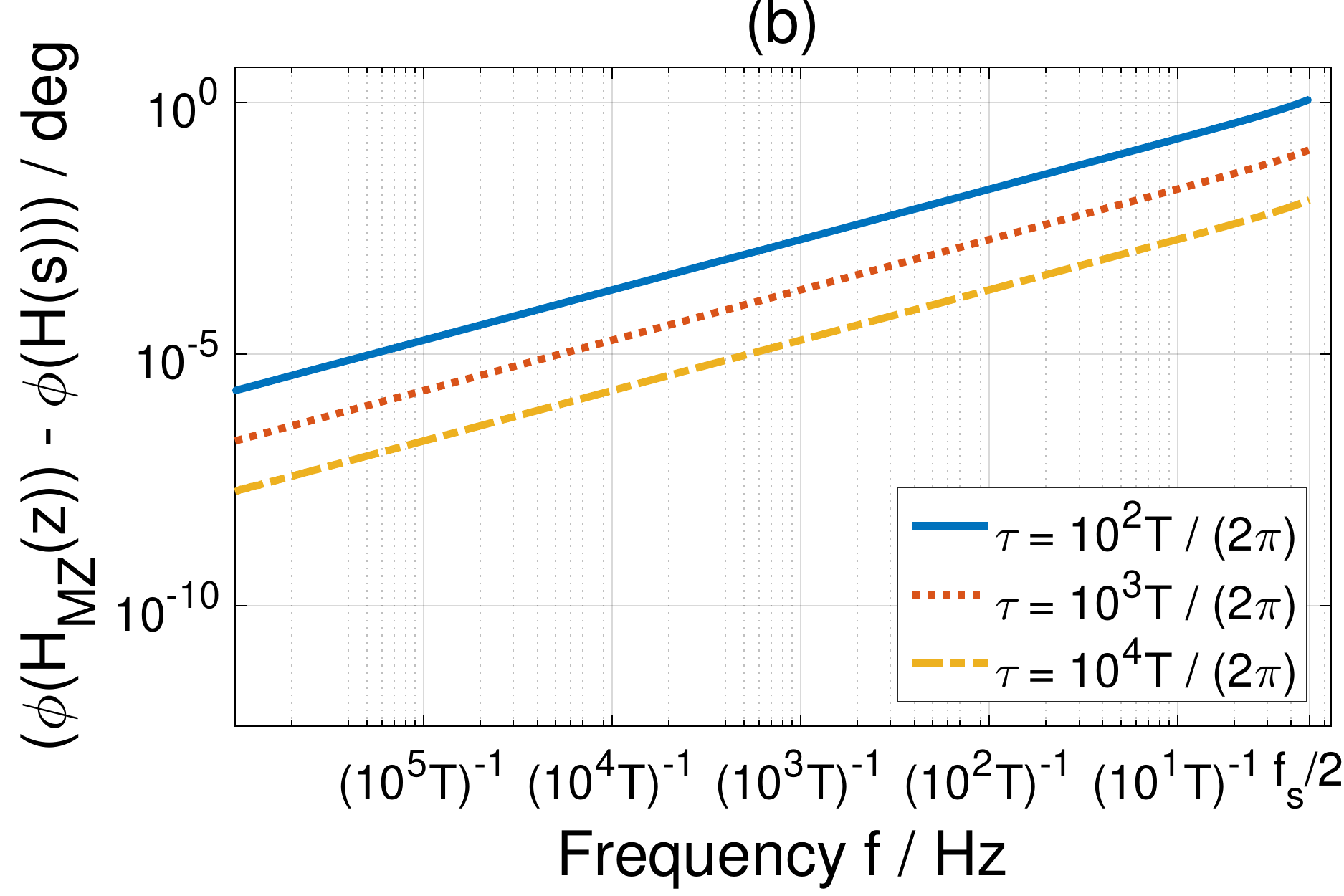}
\caption{Phase shift of the frequency response of the IIR filter (left) and the corresponding difference from the ideal filter (right).}
\label{fig_app_gainT_phase}
\end{figure}

\begin{figure}[h]
\centering
\includegraphics[width=0.40\textwidth]{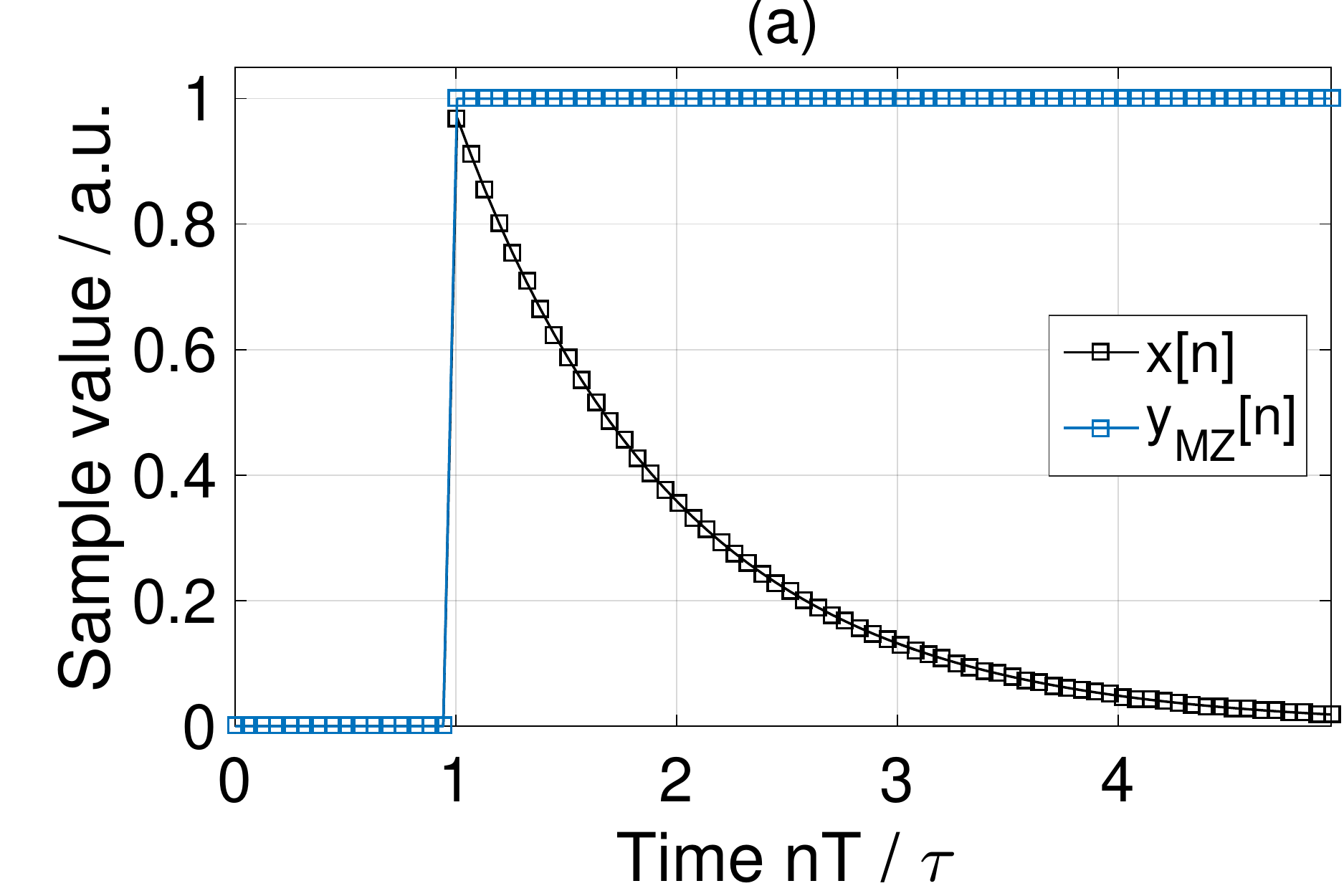}
\includegraphics[width=0.40\textwidth]{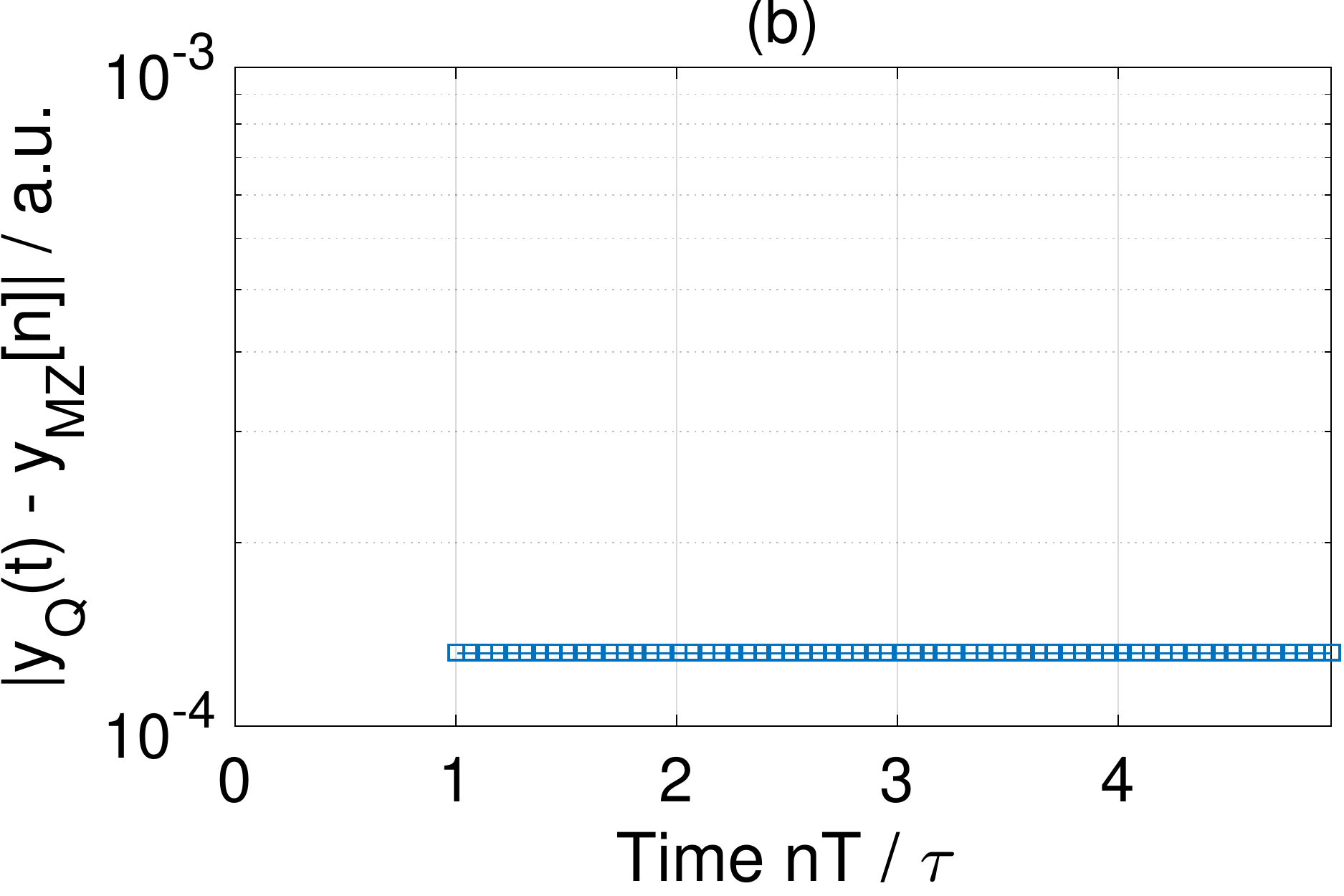}
\caption{Simulated amplifier signal with $\tau = \frac{100T}{2\pi}$ and the corresponding deconvolution with the IIR filter (left) and the difference from the ideal output.}
\label{fig_app_gainT_time}
\end{figure}

\clearpage
\subsection{IIR filter with matched-z transformation and gain correction at frequency ($\omega = \frac{1}{\tau}$)}
\label{sec_app_gainTAU}

\begin{figure}[h]
\centering
\includegraphics[width=0.40\textwidth]{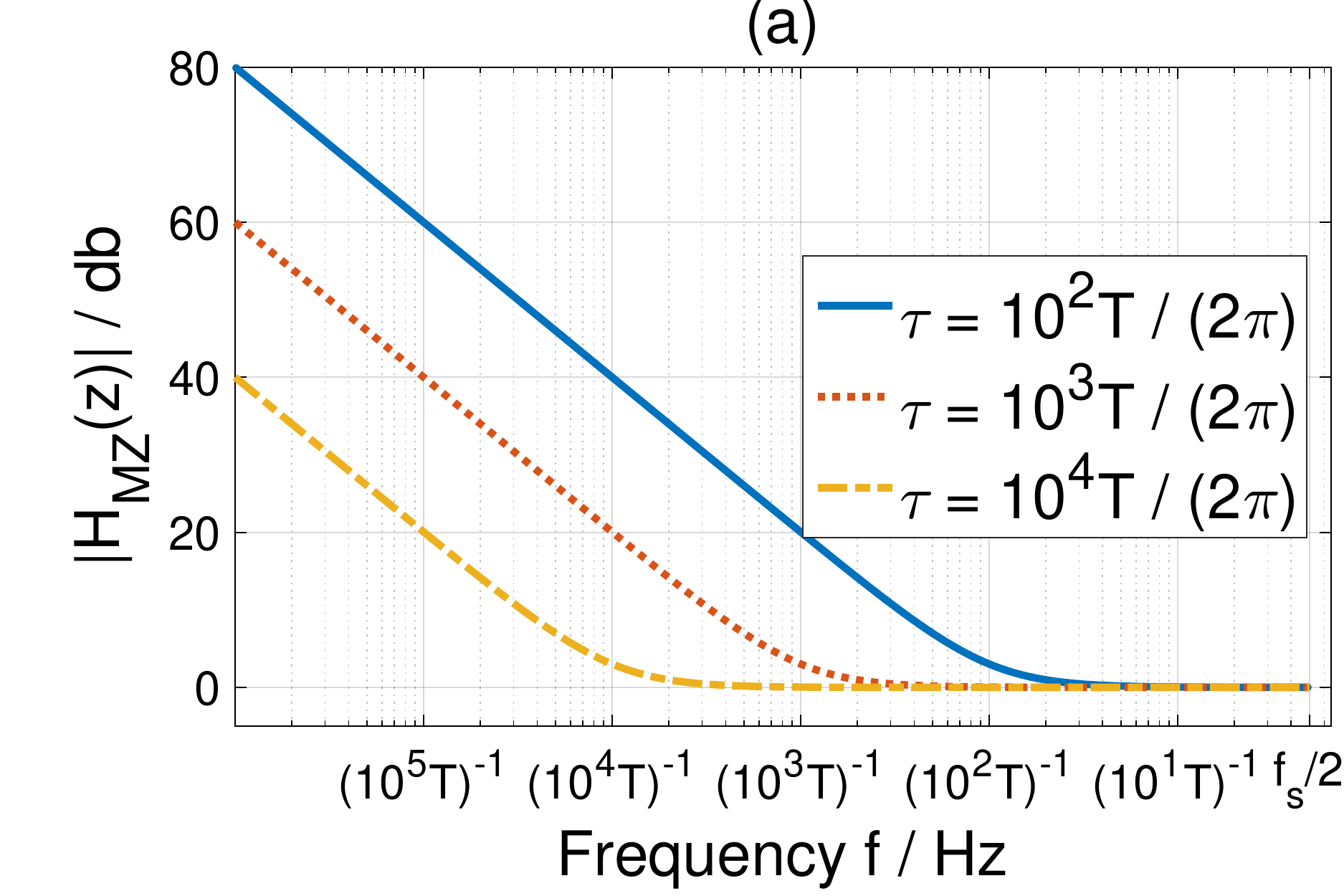}
\includegraphics[width=0.40\textwidth]{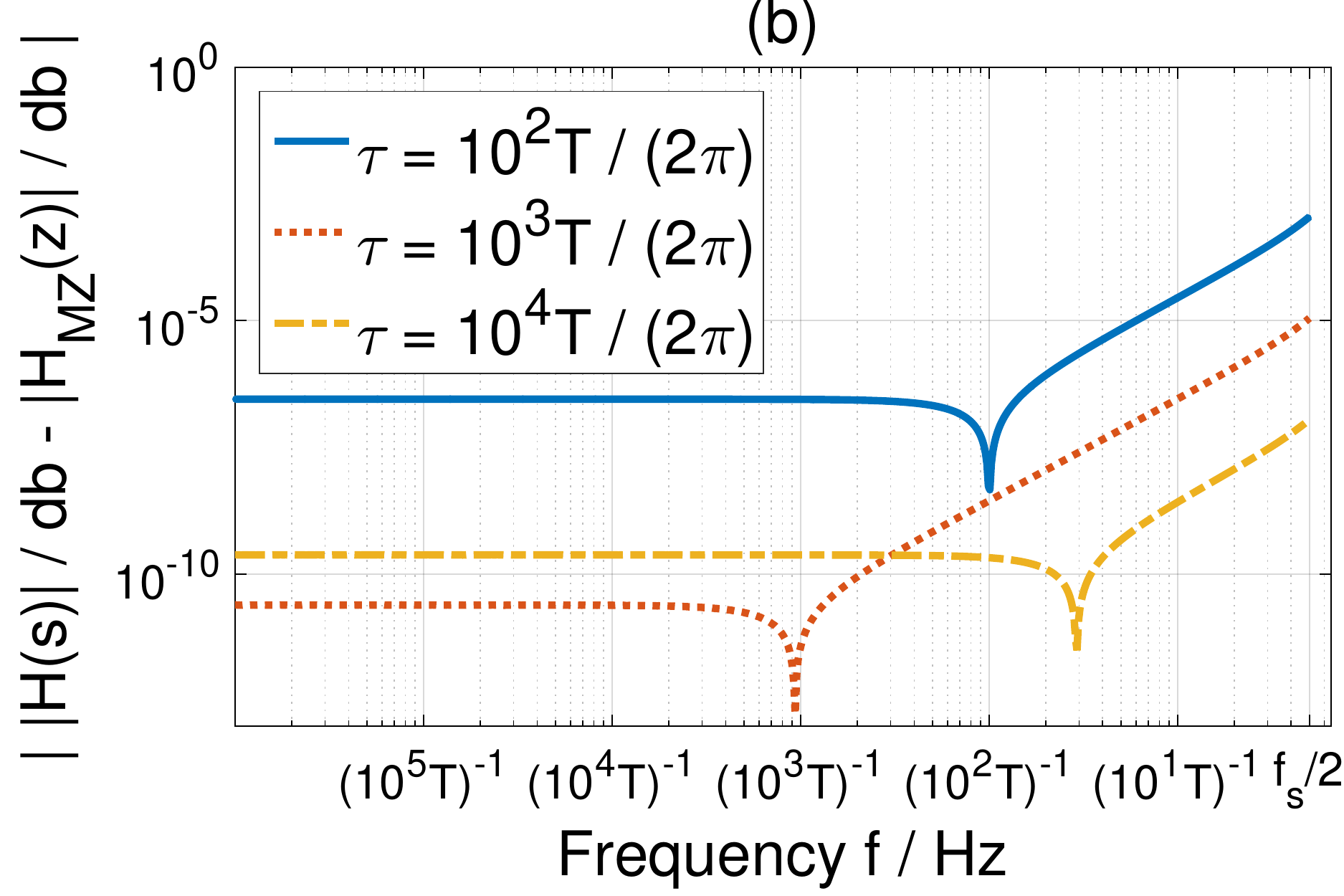}
\caption{Magnitude of the frequency response of the IIR filter (left) and the corresponding difference from the ideal filter (right).}
\label{fig_app_gainTAU_mag}
\end{figure}

\begin{figure}[h]
\centering
\includegraphics[width=0.40\textwidth]{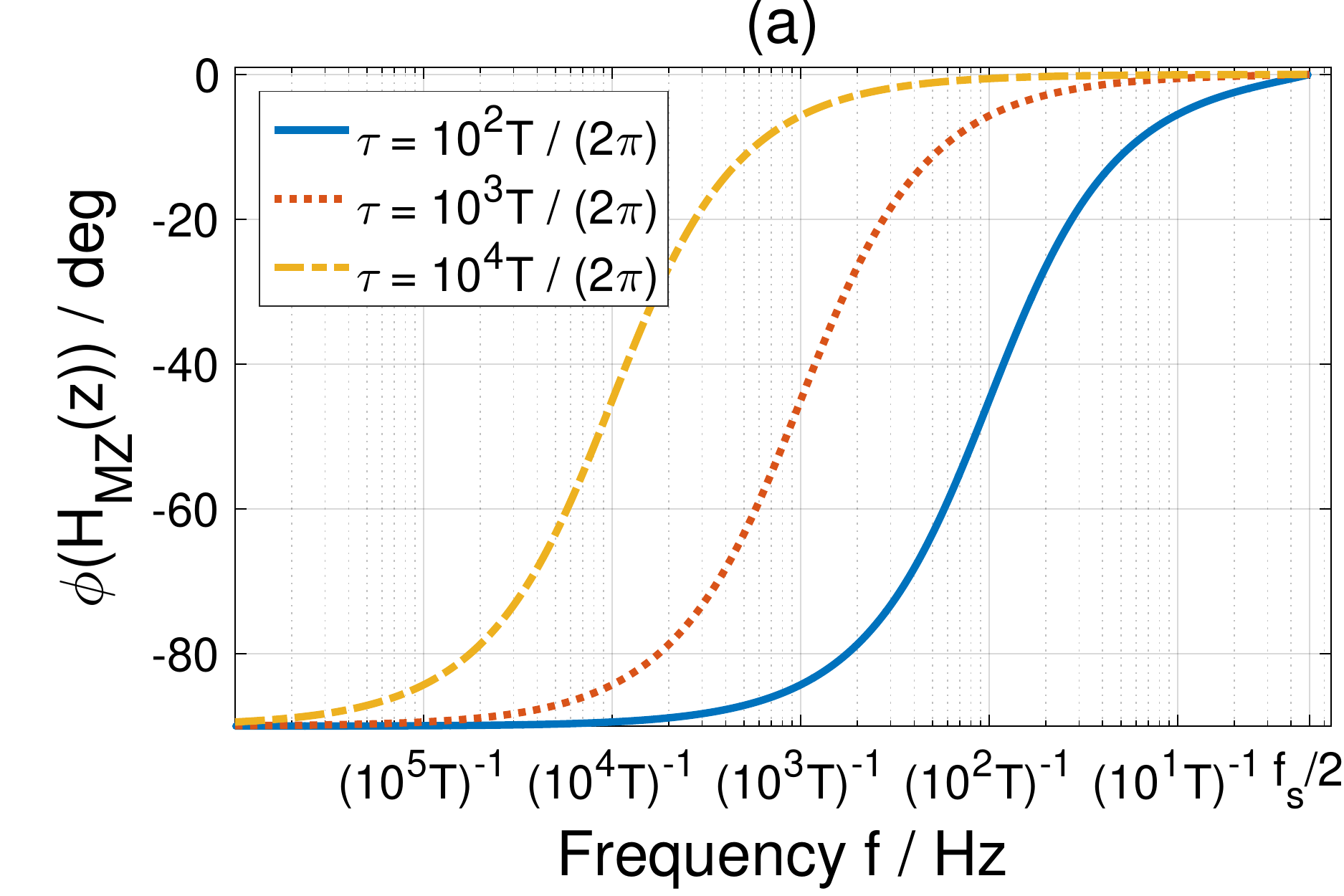}
\includegraphics[width=0.40\textwidth]{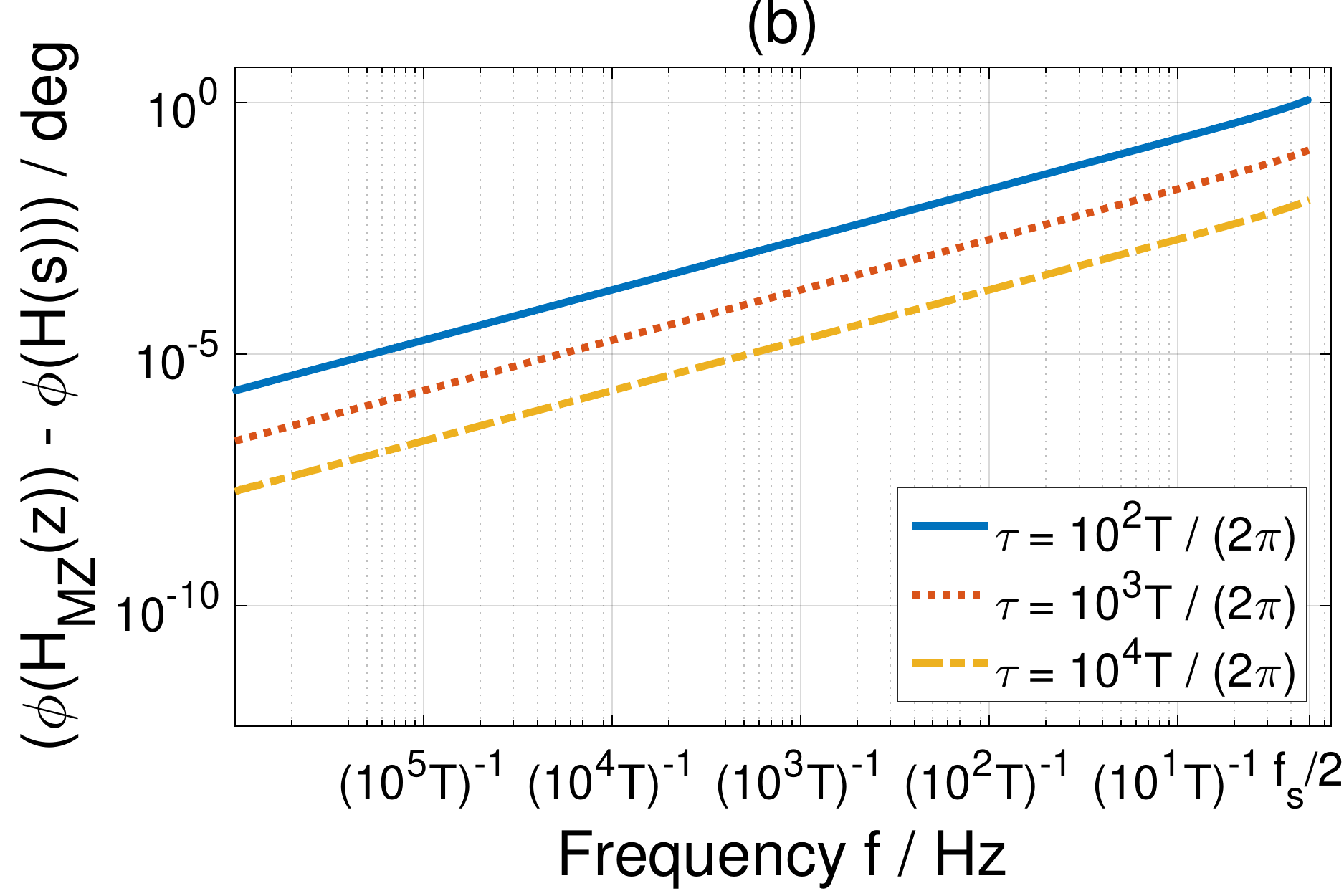}
\caption{Phase shift of the frequency response of the IIR filter (left) and the corresponding difference from the ideal filter (right).}
\label{fig_app_gainTAU_phase}
\end{figure}

\begin{figure}[h]
\centering
\includegraphics[width=0.40\textwidth]{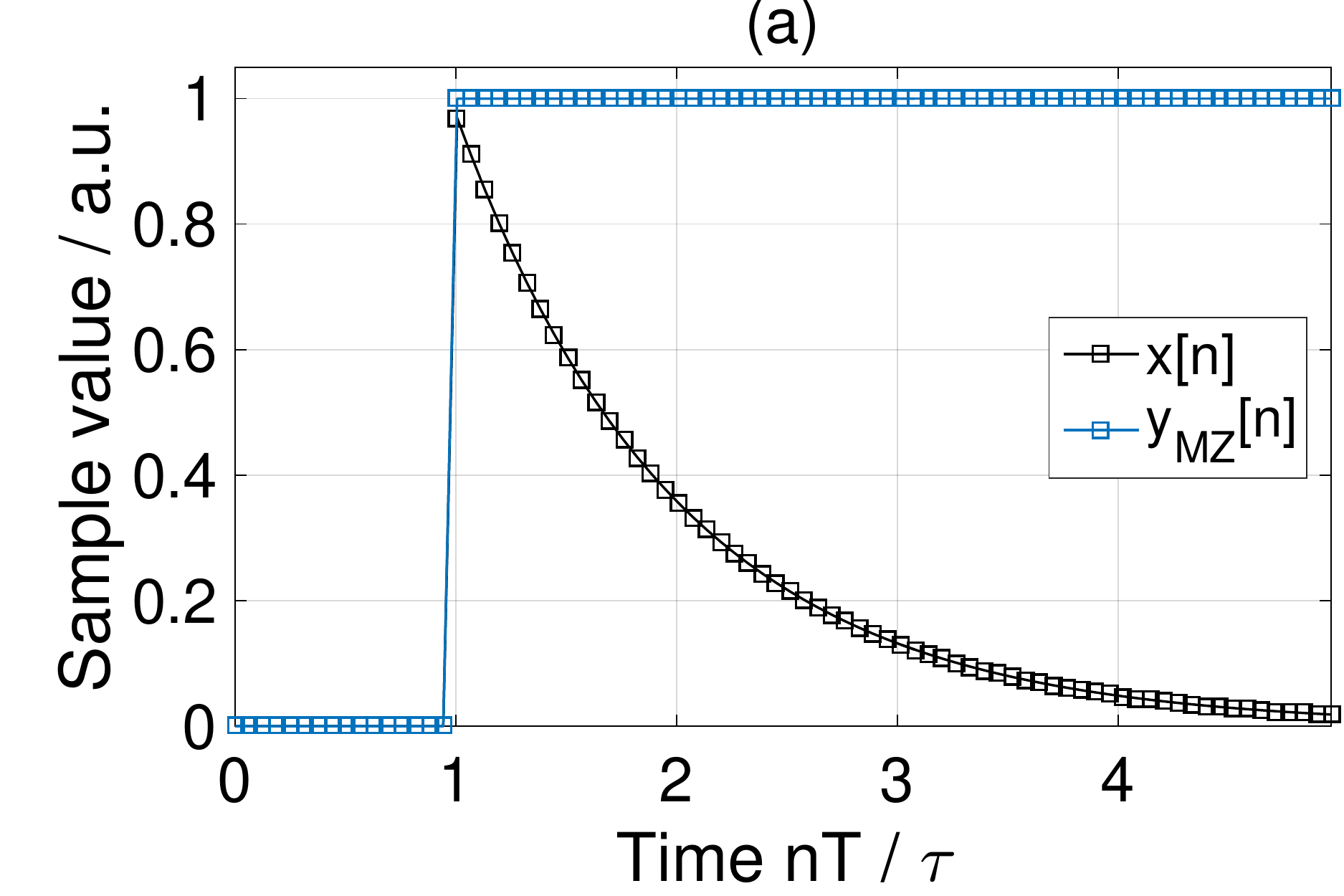}
\includegraphics[width=0.40\textwidth]{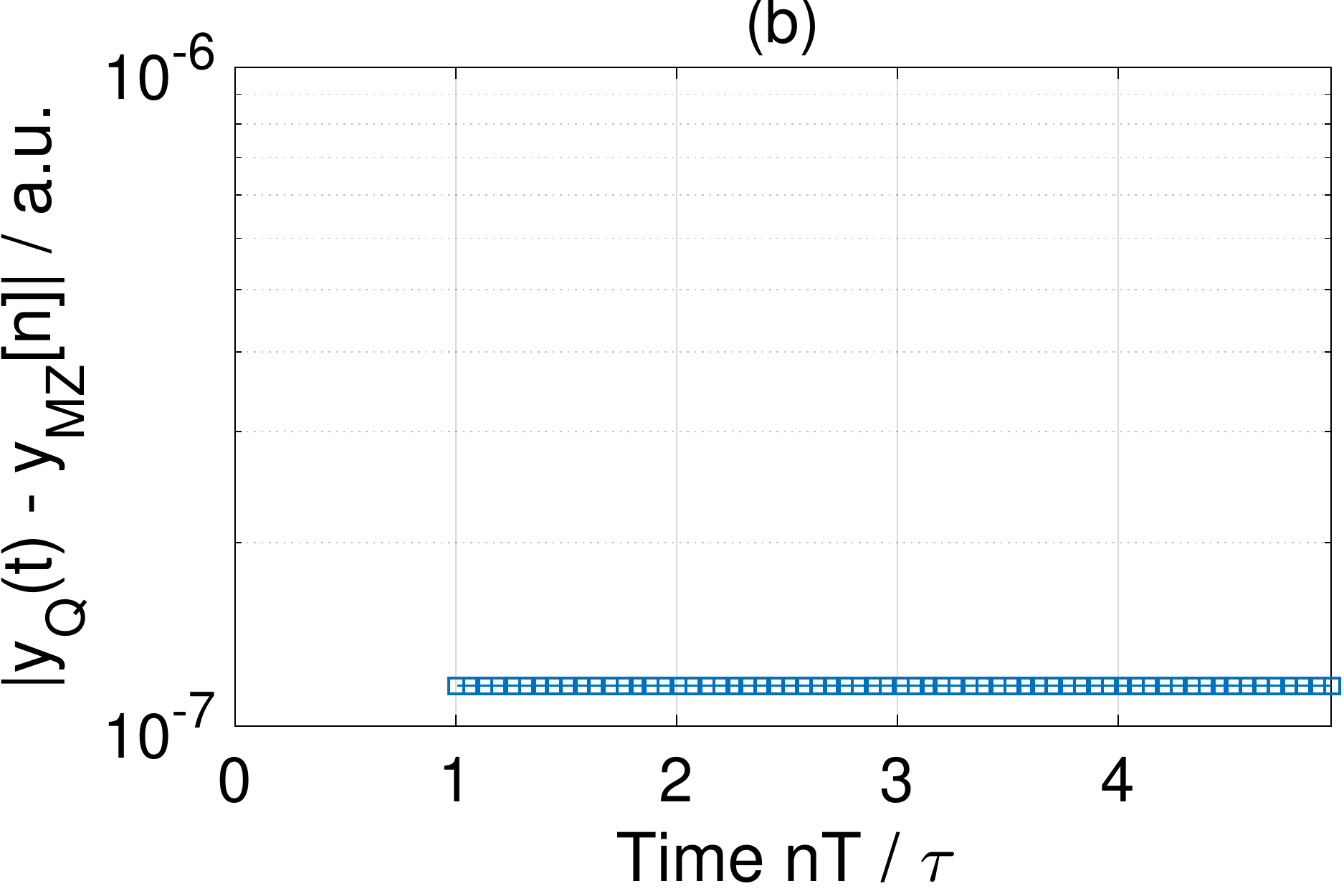}
\caption{Simulated amplifier signal with $\tau = \frac{100T}{2\pi}$ and the corresponding deconvolution with the IIR filter (left) and the difference from the ideal output.}
\label{fig_app_gainTAU_time}
\end{figure}

\clearpage
\subsection{IIR filter with matched-z transformation and gain correction at $3\,\mathrm{dB}$ corner frequency (model with three zeros and three poles)}
\label{sec_app_iir_three}
Coefficients for the continuous-time filter in frequency domain (inverse amplifier transfer function in $z$-domain):

\begin{align} 
{b_z}_n=\{&1.002135293208258, -2.973956423901655,\nonumber\\
      &2.941648719285271, -0.969827476438357\}\label{eqn_iir_bz}\\
{a_z}_n=\{&1.000000000000000, -2.971825171464407,\nonumber\\
      &2.943735466617156, -0.971910292848565\}\label{eqn_iir_az}\, .
\end{align}

\begin{figure}[h]
\centering
\includegraphics[width=0.40\textwidth]{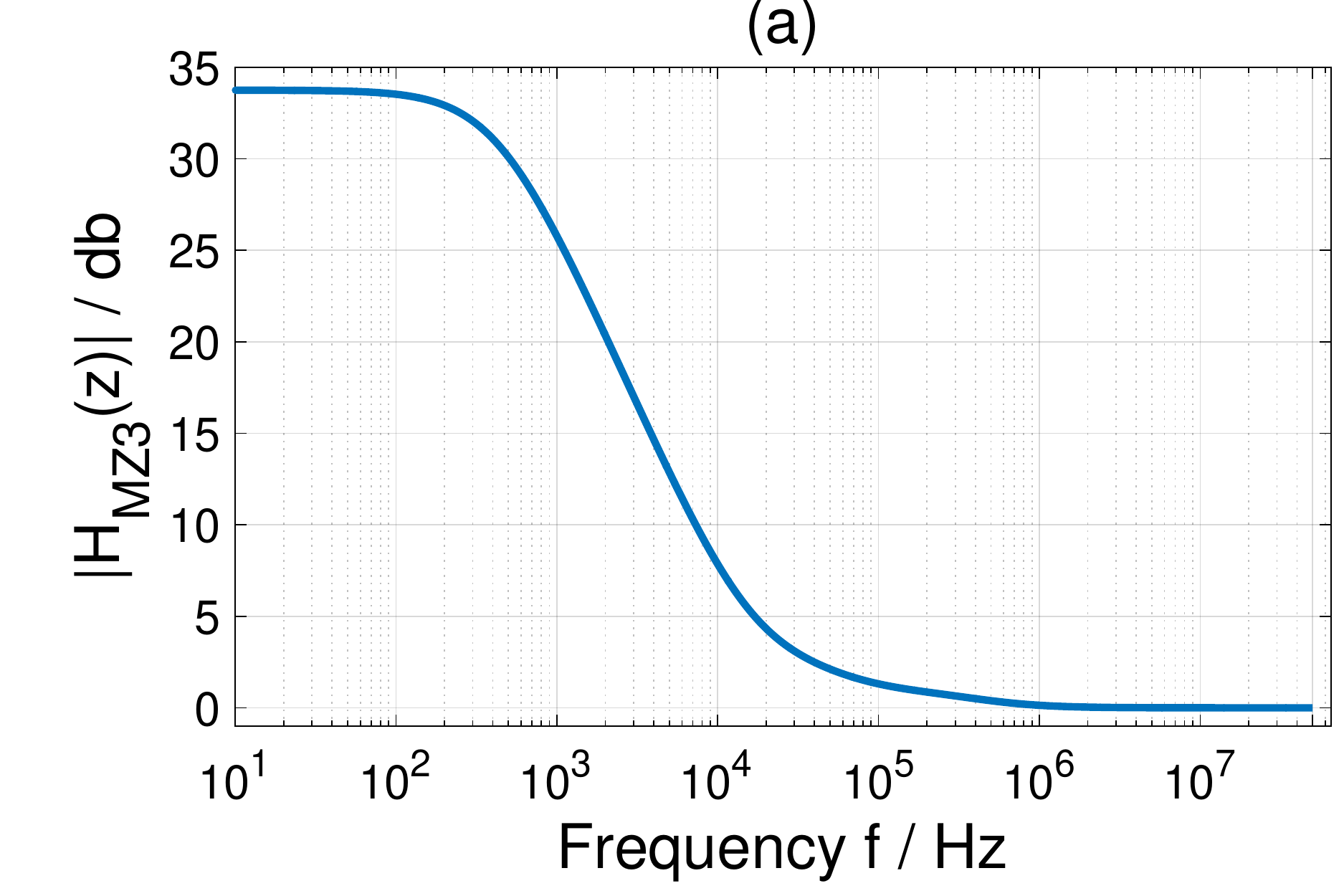}
\includegraphics[width=0.40\textwidth]{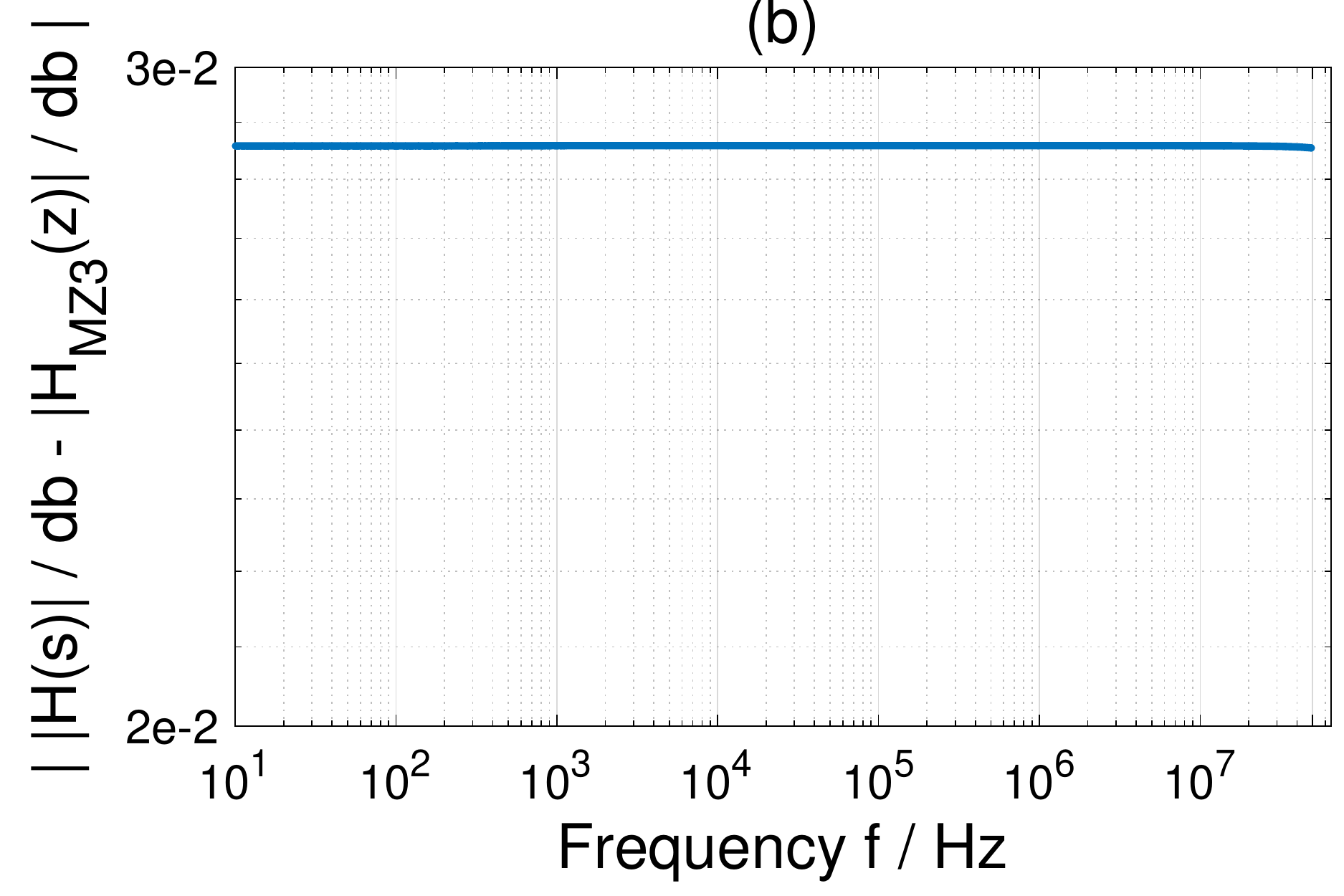}
\caption{Magnitude of the frequency response of the IIR filter (left) and the corresponding difference from the ideal filter (right).}
\label{fig_app_gainTAU_mag}
\end{figure}

\begin{figure}[h]
\centering
\includegraphics[width=0.40\textwidth]{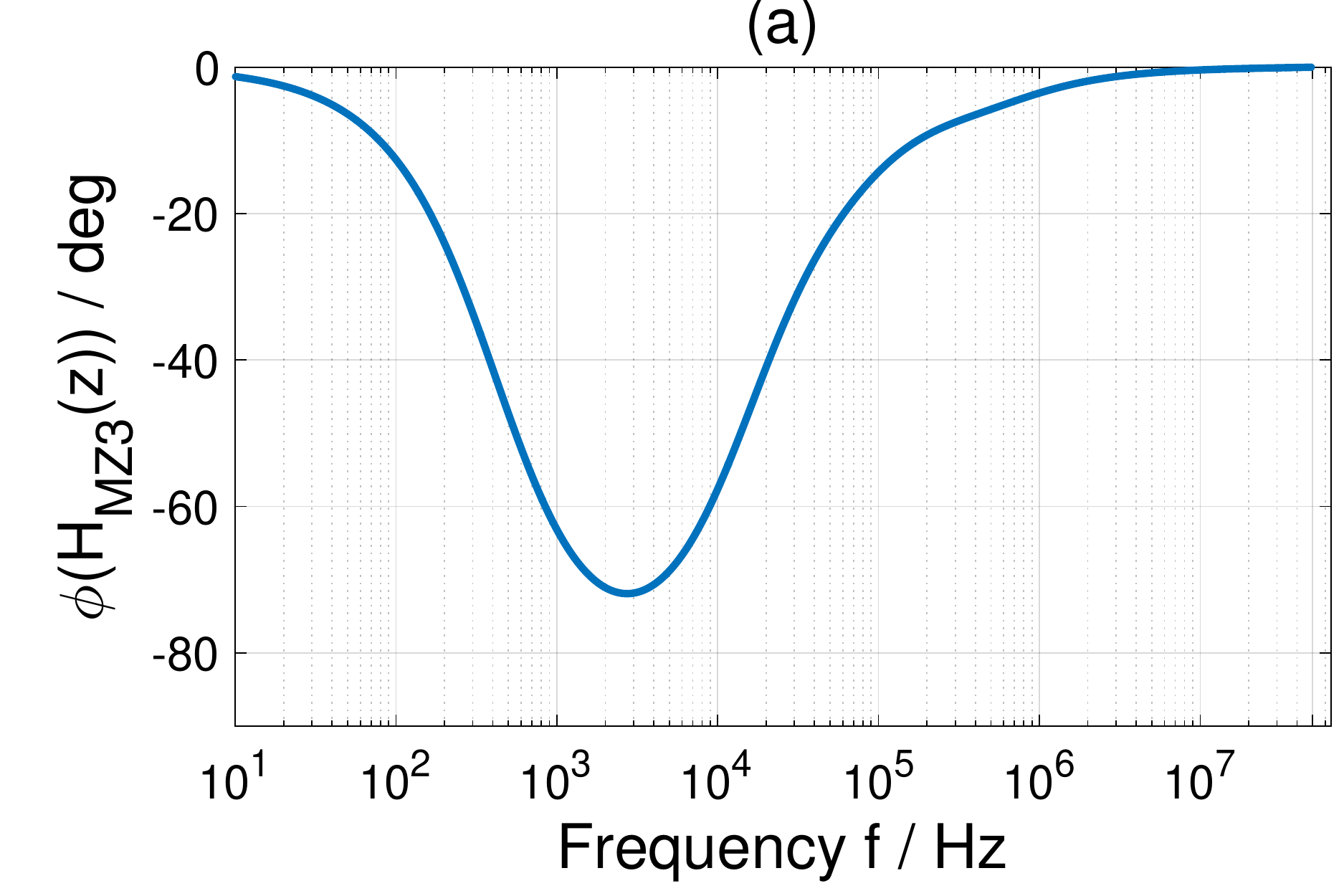}
\includegraphics[width=0.40\textwidth]{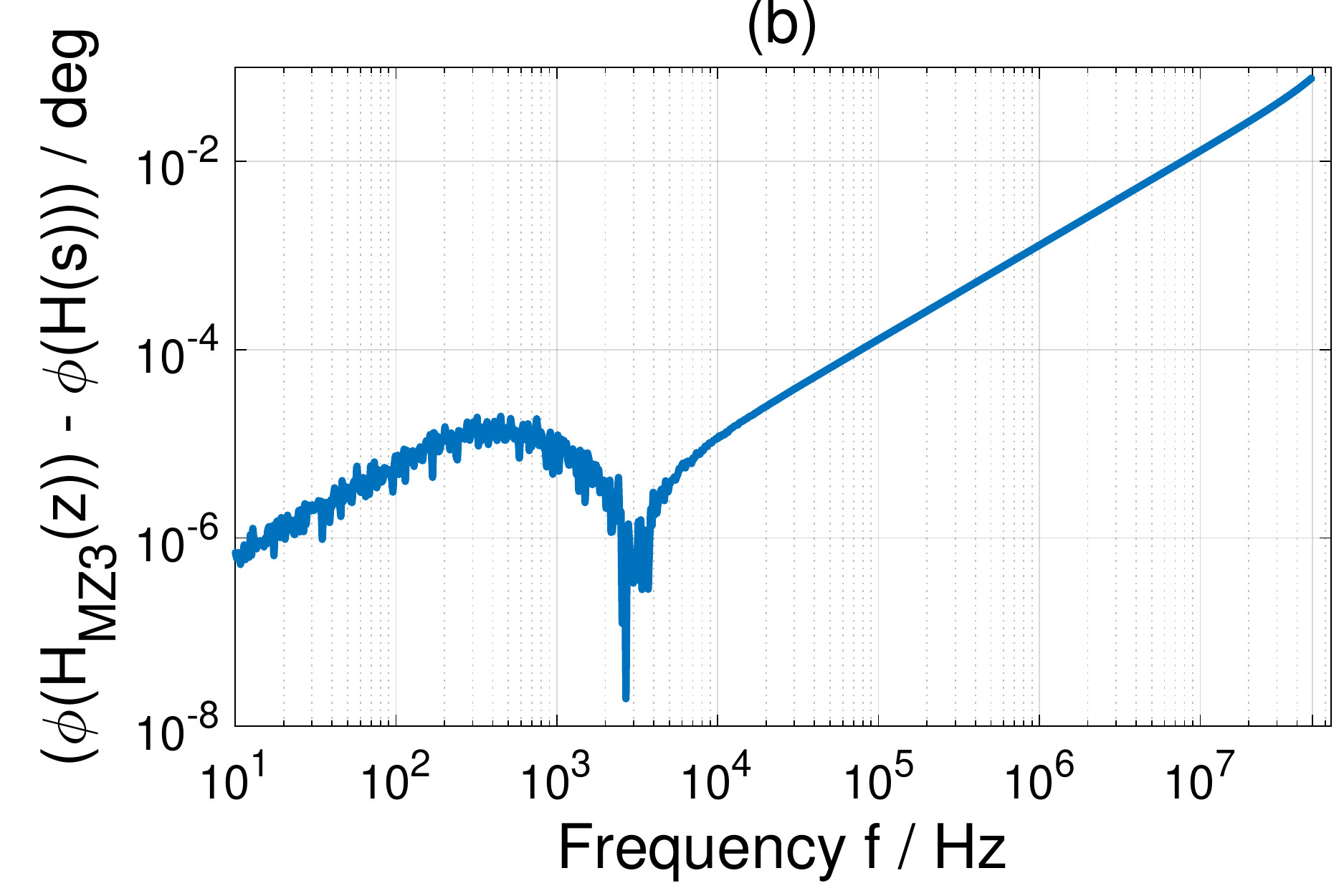}
\caption{Phase shift of the frequency response of the IIR filter (left) and the corresponding difference from the ideal filter (right).}
\label{fig_app_gainTAU_phase}
\end{figure}

\subsubsection{Examples of the pulse processing}
Some examples (Fig.~\protect\ref{fig_results_tespulse55}, Fig.~\protect\ref{fig_results_tespulse250}) of the deconvolution with the IIR filter with the coefficients \protect\eqref{eqn_iir_bz}, \protect\eqref{eqn_iir_az}. The differentiator has the length $M=128$ and the average filter is of length $N=64$.

\begin{figure}[ht]
\centering
\includegraphics[width=0.5\textwidth]{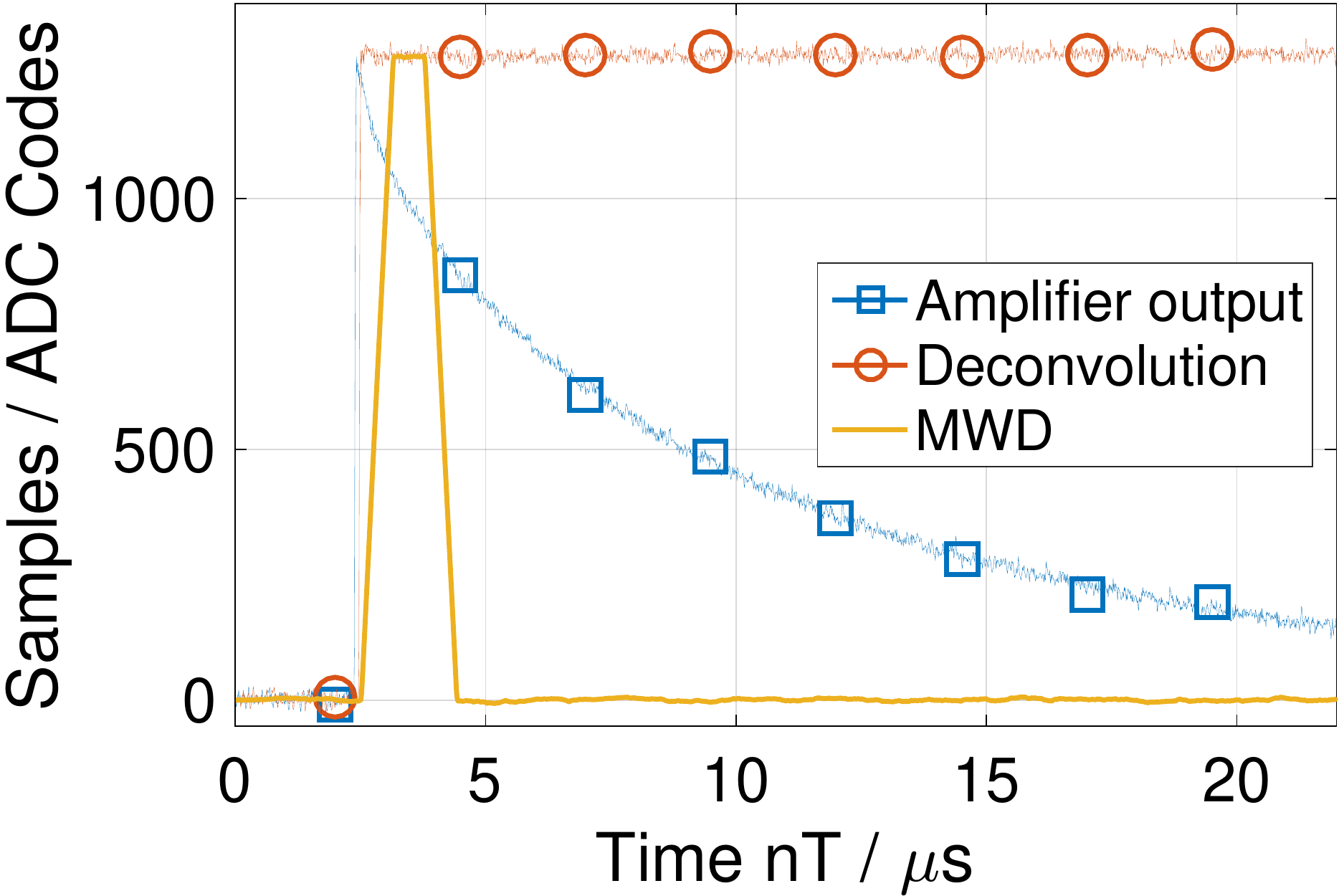}
\caption{Derived output signals from the charge-sensitive amplifier ($55\,\mathrm{mV}$ test input voltage) and the presented pulse shaping with deconvolution and advanced pulse processing (MWD).}
\label{fig_results_tespulse55}
\end{figure}

\begin{figure}[ht]
\centering
\includegraphics[width=0.5\textwidth]{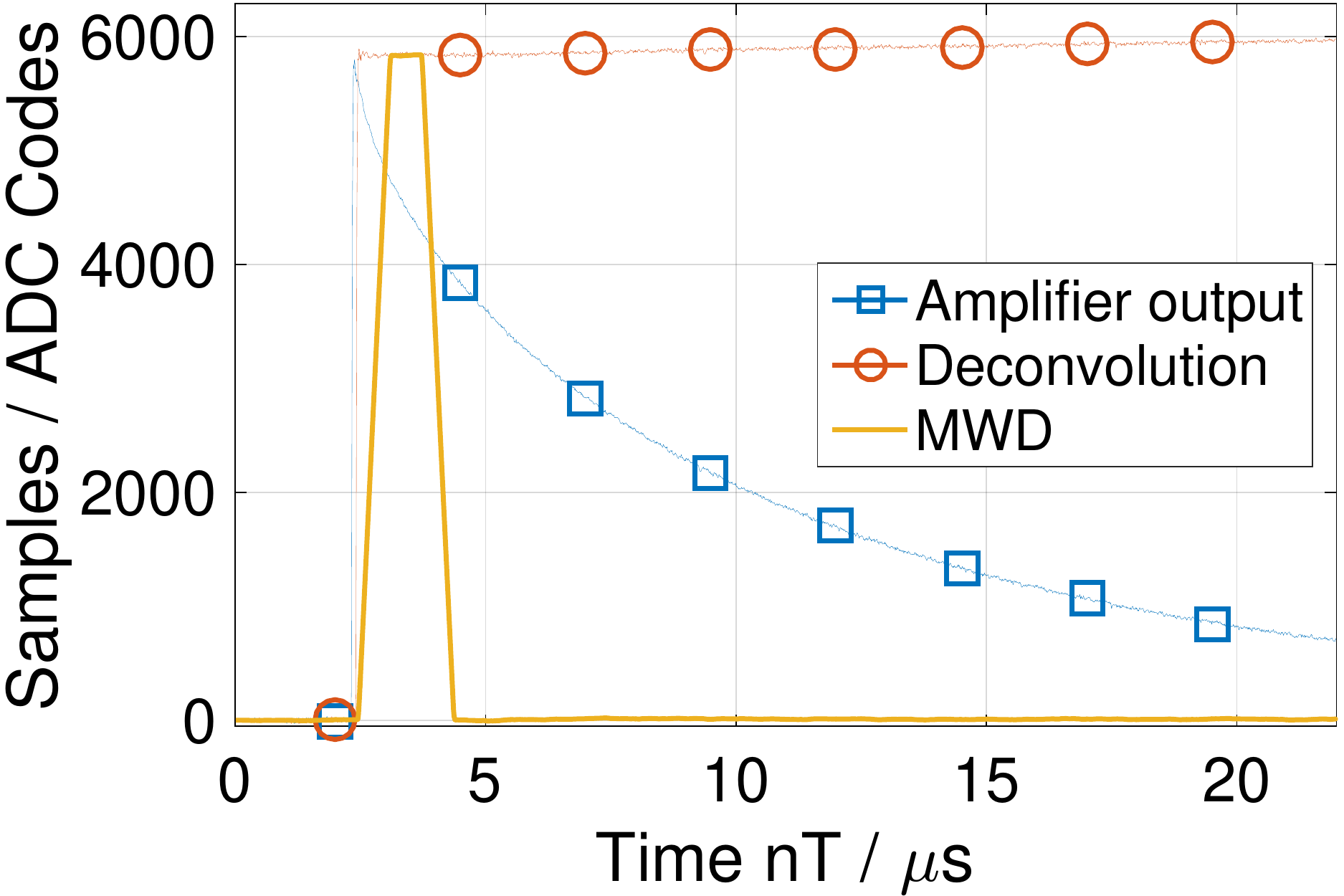}
\caption{Derived output signals from the charge-sensitive amplifier ($250\,\mathrm{mV}$ test input voltage) and the presented pulse shaping with deconvolution and advanced pulse processing (MWD). The step-like response of the deconvolution filter is not maximally flat due to rounding errors in the Cascade Form structure.}
\label{fig_results_tespulse250}
\end{figure}

\clearpage


\begin{thebibliography}{00}
\bibitem{foedisch_csa}

P. F\"odisch et al., \emph{Charge-sensitive front-end electronics with operational amplifiers for CdZnTe detectors}, \href{http://arxiv.org/abs/1603.05098}{arXiv:1603.05098}, submitted to J. Instrum.

\bibitem{stein1}
J. Stein et al., \emph{Circuit arrangement for the digital processing of semiconductor detector signals}, US Patent No. 5\,307\,299, 1994

\bibitem{stein2}
J. Stein et al., \emph{Schaltungsanordnung f\"ur die digitale Verarbeitung von Halbleiterdetektorsignalen}, Europ\"aische Patentschrift EP\,0\,550\,830\,B1, 1997

\bibitem{georgiev1}
A. Georgiev et al., \emph{Digital Pulse Processing in High Resolution, High Throughput Gamma-Ray Spectroscopy}, \href{http://dx.doi.org/10.1109/23.256659}{IEEE Trans. Nucl. Sci. {\bf{40}} (1993) 770}

\bibitem{georgiev2}
A. Georgiev et al., \emph{An Analog-to-Digital Conversion Based on a Moving Window Deconvolution}, \href{http://dx.doi.org/10.1109/23.322868}{IEEE Trans. Nucl. Sci. {\bf{41}} (1994) 1116}

\bibitem{stein3}
J. Stein et al., \emph{X-ray detectors with digitized preamplifiers}, \href{http://dx.doi.org/10.1016/0168-583X(95)01417-9}{Nucl. Instr. and Meth. B {\bf{113}} (1996) 141}

\bibitem{jordanov1}
V. T. Jordanov et al., \emph{Digital synthesis of pulse shapes in real time for high resolution radiation spectroscopy}, \href{http://dx.doi.org/10.1016/0168-9002(94)91011-1}{Nucl. Instr. and Meth. A {\bf{345}} (1994) 337}

\bibitem{jordanov2}
V. T. Jordanov et al., \emph{Deconvolution of pulses from a detector-amplifier configuration}, \href{http://dx.doi.org/doi:10.1016/0168-9002(94)91394-3}{Nucl. Instr. and Meth. A {\bf{351}} (1994) 592}

\bibitem{jordanov3}
V. T. Jordanov et al., \emph{Digital techniques for real-time pulse shaping in radiation measurements}, \href{http://dx.doi.org/10.1016/0168-9002(94)91652-7}{Nucl. Instr. and Meth. A {\bf{353}} (1994) 261}

\bibitem{jordanov4}
V. T. Jordanov, \emph{Unfolding-synthesis technique for digital pulse processing. Part 1: Unfolding}, \href{http://dx.doi.org/10.1016/j.nima.2015.07.040}{Nucl. Instr. and Meth. A {\bf{805}} (2016) 63}

\bibitem{oppenheim}
A. V. Oppenheim et al., \emph{Discrete-Time Signal Processing}, Third Editon, Prentice Hall, 2010

\bibitem{proakis}
J. G. Proakis et al., \emph{Digital Signal Processing: Principles, Algorithms and Applications}, Third Edition, Prentice Hall, 1996

\bibitem{mathworks}
The MathWorks, Inc., \emph{MATLAB and System Identification Toolbox Release 2016a}

\bibitem{xilinx_dsp}
Xilinx, Inc., \emph{7 Series DSP48E1 Slice}, UG479

\bibitem{crochiere}
R. E. Crochiere et al., \emph{Analysis of linear digital networks}, \href{http://dx.doi.org/10.1109/PROC.1975.9793}{Proc. IEEE {\bf{63}} (1975) 581}

\end{thebibliography}
\end{document}